\documentclass[prb,twocolumn,superscriptaddress,eqsecnum]{revtex4}
\usepackage[dvips]{graphicx}
\usepackage{latexsym}
\usepackage{amsmath}
\begin{document}
\title{Quantum criticality beyond the Landau-Ginzburg-Wilson paradigm}

\author{T. Senthil}
\affiliation{Department of Physics, Massachusetts Institute of
Technology, Cambridge MA 02139}

\author{Leon Balents}
\affiliation{Department of Physics, University of California,
Santa Barbara, CA 93106-4030}

\author{Subir Sachdev}
\affiliation{Department of Physics, Yale University, P.O. Box
208120, New Haven, CT 06520-8120}

\author{Ashvin Vishwanath}
\affiliation{Department of Physics, Massachusetts Institute of
Technology, Cambridge MA 02139}

\author{Matthew P. A. Fisher}
\affiliation{Kavli Institute for Theoretical Physics, University
of California, Santa Barbara, CA 93106-4030}

\date{\today}

\begin{abstract}

We present the critical theory of a number of zero temperature
phase transitions of quantum antiferromagnets and interacting
boson systems in two dimensions. The most important example is the
transition of the $S=1/2$ square lattice antiferromagnet between
the N\'eel state (which breaks spin rotation invariance) and the
paramagnetic valence bond solid (which preserves spin rotation
invariance but breaks lattice symmetries). We show that these two
states are separated by a second-order quantum phase transition.
The critical theory is not expressed in terms of the order
parameters characterizing either state (as would be the case in
Landau-Ginzburg-Wilson theory), but involves fractionalized
degrees of freedom and an emergent, topological, global
conservation law. A closely related theory describes the
superfluid-insulator transition of bosons at half-filling on a
square lattice, in which the insulator has a bond density wave
order. Similar considerations are shown to apply to transitions of
antiferromagnets between the valence bond solid and the $Z_2$ spin
liquid: the critical theory has deconfined excitations interacting
with an emergent U(1) gauge force. We comment on the broader
implications of our results for the study of quantum criticality
in correlated electron systems.
\end{abstract}

\maketitle

\tableofcontents

\section{Introduction and motivation}
\label{sec:intro}

A central concept in the theory of phase transitions is that of
the `order parameter', which expresses the different symmetries of
the phases on either side of the critical point.  If the
transition is second order, there is interesting universal
singular behavior that is manifested in many physical quantities.
According to the prevalent paradigm largely due to Landau and
Ginzburg\cite{landau}, these universal critical singularities are
associated with long wavelength low energy fluctuations of the
order parameter degree of freedom. When combined with general
renormalization group ideas \cite{wilson}, this notion provides
the sophisticated Landau-Ginzburg-Wilson (LGW) theoretical
framework for thinking about critical phenomena in various diverse
contexts. Specifically, static critical properties at non-zero
temperature are supposed to be determined from effective models in
which all modes other than the order parameter have been
eliminated. Similarly, for dynamical critical properties, the only
degrees of freedom that purportedly need be retained are the order
parameter and at most a few additional ``hydrodynamic'' modes
having slow relaxation times due to conservation laws.

Recent years have seen much interest in the study of zero
temperature phase transition phenomena in correlated many body
systems.  Unlike their thermal counterpart, such transitions are
often driven by quantum fluctuation effects and are hence known as
`quantum phase transitions'\cite{subirbook}.  Indeed, it has been
proposed that proximity to quantum critical points (QCPs)
separating two distinct phases is responsible for the anomalous
properties of some interesting correlated materials such as, for
instance, the cuprate superconductors.  Theoretically, the LGW
paradigm has thus far provided the basic framework to examine
quantum critical phenomena as well.  In particular, the critical
modes specific to a quantum critical point are presumed to be the
long distance, long time fluctuations of the order parameter,
described in a continuum field theory.

In the last few years some interesting and tantalizing evidence
has emerged that points toward the failure of the LGW paradigm at
certain quantum phase transitions. First, there are numerical
calculations \cite{assaad,sandvik} that see a direct second order
quantum phase transition between two phases with different broken
symmetry characterized by two apparently independent order
parameters.  A LGW description of the competition between such two
kinds of orders would then generically predict either a
first-order transition, or an intermediate region of coexistence
where both orders simultaneously exist, or an intermediate region
with neither order. A direct second order transition between these
two broken symmetry phases would seem to require fine-tuning to a
`multicritical' point. Are the numerics managing to achieve this
`fine-tuning' or is the LGW paradigm simply invalid?

Second, there have been a number of fascinating experiments
probing the onset of magnetic long range order in a class of
rare-earth inter-metallics known as the heavy fermion metals
\cite{piers1,stewart}. Remarkably, the behavior right at the
quantum transition between the magnetic and non-magnetic metallic
phases is usually very different from that of a Fermi liquid.
Furthermore, such behavior is in severe disagreement with
expectations based on LGW analyses. Specifically, theories
associating the critical singularities with fluctuations of the
natural magnetic order parameter in a metallic environment seem to
have a hard time explaining the observed non-Fermi liquid
phenomena. Once again it appears that more than the obvious
possibly happens at some quantum critical points.

In this paper we demonstrate and study various specific examples
of quantum phase transitions which violate the LGW paradigm. We
will show that in a number of different quantum transitions, the
natural field theoretic description of the critical singularities
is not in terms of the order parameter field(s) that describe the
bulk phases but in terms of some new `emergent' degrees of freedom
that are specific to the critical point. These new degrees of freedom may be
thought of
`fractional' quantum number particles that interact with each other through an
emergent gauge force in a sense made
precise below. Laughlin has previously
argued for fractionalization at quantum critical points on
phenomenological grounds \cite{rbl}. A non-technical overview of
our results has appeared previously.\cite{shortpap}

We note, in passing, that there are already numerous
well-documented examples of the breakdown of the LGW paradigm in
quantum systems in one dimension.\cite{oned} However, these rely
rather crucially on the description of various states in terms of
the harmonic phase degrees of freedom of the Tomonaga-Luttinger
liquid, and do not have any direct generalization to higher
dimensions.

For the most part in this paper, we will study phase transitions
in two dimensional quantum magnetism. These may also be fruitfully
viewed from a different point of view as representing transitions
of interacting bosons on a lattice at commensurate density.
Quantum magnets provide a particularly useful laboratory to
develop and test ideas on the theory of quantum phase transitions.
Consider a quantum system of spin $S = 1/2$ moments $\vec S_r$ on
a two dimensional square lattice ($r=(x,y)$) with the Hamiltonian
\begin{equation}
H = J\sum_{\langle rr'\rangle}\vec S_r\cdot \vec S_{r'} + \ldots
\end{equation}
The ellipses represent other short ranged interactions that may be tuned
to drive various zero temperature phase transitions.  We assume $J > 0$,
{\em i.e} antiferromagnetic interactions. Later we will consider various
generalizations to other lattices, higher spins, etc.
\begin{figure}[t]
\centerline{\includegraphics[width=3.2in]{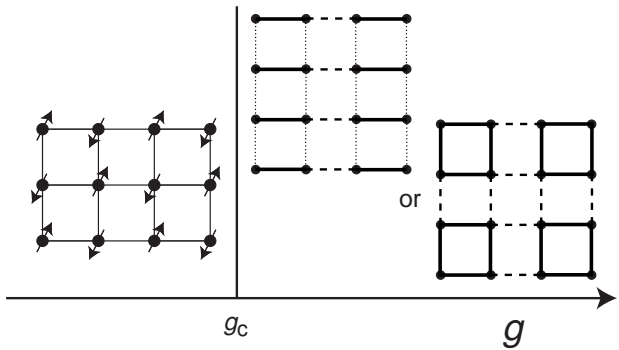}}
\caption{Ground states of the square lattice $S=1/2$
antiferromagnet studied in this paper. The coupling $g$ controls
the strength of quantum spin fluctuations about a magnetically
ordered state, and appears in Eq.~(\ref{eq:nlsm}) (the classical
limit is $g=0$). There is broken spin rotation invariance in the
N\'eel state for $g<g_c$, described by the order parameter
$\vec{N}_r$ in Eq.~(\ref{eq:staggered}). The VBS ground state
appears for $g>g_c$, and is characterized by the order parameter
$\psi_{\rm VBS}$ in Eq.~(\ref{eq:VBSorder}) -- the distinct lines
represent distinct values of $\langle \vec S_r \cdot \vec S_{r'}
\rangle$ on each link. The VBS state on the left has ``columnar''
bond order, while that on the right has ``plaquette'' order. The
theory $\mathcal{L}_z$ in Eq.~(\ref{sz}) applies only at the QCP
$g=g_c$ at its critical point obtained at $s=s_c$.} \label{pdiag}
\end{figure}

The nature of some of the various possible ground states of such a
Hamiltonian are quite well understood.  First, there are states
that develop magnetic long range order and break the spin rotation
symmetry. The simplest example (and the one that we will focus on)
are collinear antiferromagnets where the order parameter is a
single vector $\vec{N}_r$ (the N\'eel vector), defined to describe
a state of staggered magnetization,
\begin{equation}
  \label{eq:staggered}
  \vec{S}_r = \epsilon_r \vec{N}_r ,
\end{equation}
where
\begin{equation}
\epsilon_r \equiv (-1)^{x+y} \label{defeps}
\end{equation}
is $+1$ on one checkerboard sublattice and $-1$ on the other. The
N\'eel state has $\langle \vec{N}_r \rangle \neq 0$ and
independent of $r$ (see Fig~\ref{pdiag}), but more generally
$\vec{N}_r$ is presumed to vary ``slowly'' on the lattice scale
over at least most of space. The low energy excitations of the
N\'eel state are simply linear dispersing spin waves.

It is now recognized that a variety of quantum paramagnetic ground
states are also possible where quantum fluctuations have prevented
the spins from developing magnetic long range order, and so
$\langle \vec{S}_r \rangle = 0$. Such paramagnetic states can be
broadly divided into two groups. First, there are states that can
be described as `Valence Bond Solid' (VBS) states \cite{ReSaSuN}.
In a simple caricature of such a state, each spin forms a singlet
with one particular other spin resulting in an ordered pattern of
`valence bonds'.  For spin-$1/2$ systems on a square lattice, such
states necessarily break lattice translational symmetry. The
so-called ``columnar'' and ``plaquette'' ordering patterns (see
Fig.~\ref{pdiag}) are described by a complex VBS order parameter
$\psi_{\rm VBS}$, where
\begin{eqnarray}
  \vec{S}_{r}\cdot\vec{S}_{r+\hat{x}} & \sim & {\rm Re}[\psi_{\rm VBS}]
  (-1)^x, \nonumber \\
  \vec{S}_{r}\cdot\vec{S}_{r+\hat{y}} & \sim & {\rm Im}[\psi_{\rm VBS}]
  (-1)^y,   \label{eq:VBSorder}
\end{eqnarray}
and $r=(x,y)$ (here columnar states have $\psi_{\rm VBS}^4$ real
and positive, while plaquette states have $\psi_{\rm VBS}^4$ real
and negative). In these states there is an energy gap for
spin-carrying $S=1$ quasi-particle excitations; these `triplons'
\cite{triplon} are quite distinct from spin waves, and are instead
adiabatically connected to spin excitons in band insulators. A
second class of more exotic paramagnetic states are also
possible\cite{rvb,krs,ReSaSpN,Wen,sf} in principle: in these
states the valence bond configurations resonate amongst each other
and form a `liquid'. The resulting state has been argued to
possess excitations with fractional spin $1/2$ and interesting
topological structure.

Our focus will be on the nature of the evolution of the ground
state between these various phases. Our primary example is that between the
ordered magnet and a valence bond solid. We also discuss the phase
transitions between valence bond solid and `spin' liquid phases (see
Section~\ref{sec:vbssl}). Qualitatively similar phenomena will be ahown to
obtain at both these transitions.

Both the magnetic N\'eel
state, and the valence bond solid are states of broken symmetry.
The former breaks spin rotation symmetry, and the latter the
symmetry of lattice translations. The order parameters $\vec{N}$
and $\psi_{\rm VBS}$ associated with these two different broken
symmetries are very different.  A LGW picture of the evolution
between these two distinct ground states would be formulated in
terms of an effective action that is a functional of $\vec{N}$ and
$\psi_{\rm VBS}$.  Such a construction would suggest either a
first order transition, or passage through an intermediate phase
which breaks both kinds of symmetry. (Actually, the general LGW
analysis also allows an intermediate `disordered' state with {\em
neither} order, but this possibility was excluded in early
analyses \cite{ReSaSuN}; this exclusion was already an indication
that LGW theory did not apply here.) A direct second order
transition would be expected only by further fine-tuning to
special multicritical points.  Our central thesis is that this
expectation is wrong. A generic second order transition is
possible between these two phases with different broken
symmetries. The resulting critical theory is however unusual and
{\em not} naturally described in terms of the order parameter
fields of either phase. Instead, the natural description is in
terms of spin-$1/2$ ``spinon'' or CP$^1$ fields $z_{\alpha}$
($\alpha=1,2$ is a spinor index).  The N\'eel order parameter is
bilinear in the spinons:
\begin{equation}
  \label{eq:cp1}
  \vec{N} \sim z^\dagger \vec\sigma z^{\vphantom\dagger}.
\end{equation}
Here $\vec\sigma$ is the usual vector of Pauli matrices and
multiplication of the spinor index is implied.  The fields $z_\alpha$
create single spin-$1/2$ quanta, ``half'' that of the spin-$1$ quanta
created by the N\'eel field $\vec{N}$.

The spinon fields $z_\alpha$ so defined have a U(1) ``gauge''
redundancy.  Specifically the {\em local} phase rotation
\begin{equation}
  z \rightarrow e^{i{\gamma}(r, \tau)}z
\end{equation}
leaves the N\'eel vector invariant and hence is a gauge degree of
freedom.  Here $\tau$ is the imaginary time coordinate.  Thus the
spinons are coupled to a U(1) gauge field $a_{\mu}( r, \tau)$ (we
will use the Greek indices $\mu,\nu,\ldots$ to represent the three
spacetime indices $x,y,\tau$). Our central thesis -- substantiated
by a variety of arguments to follow -- is that the critical field
theory for the N\'eel-VBS transition is just the simple continuum
action $\mathcal{S}_z = \int d^2 r d \tau \mathcal{L}_z$, and
\begin{eqnarray}
\mathcal{L}_{z} &=&  \sum_{a = 1}^N |\left(\partial_{\mu} -
ia_{\mu}\right) z_{a}|^2 + s |z|^2 +
u\left(|z|^2 \right)^2 \nonumber \\
&~&~~~~~~~ +
\kappa\left(\epsilon_{\mu\nu\kappa}\partial_{\nu}a_{\kappa}\right)^2,
\label{sz}
\end{eqnarray}
where $N=2$ is the number of $z$ components (later we will
consider the case of general $N$), $|z|^2 \equiv \sum_{a=1}^N |z_a
|^2 $, and the value of $s$ is to be tuned to a critical value
$s=s_c$ so that $\mathcal{L}_z$ is at its scale-invariant critical
point. The same action with a simple modification also describes
the critical field theory for systems with easy-plane anisotropy,
with the addition of the simple term
\begin{equation}
  \label{eq:epterm}
  \mathcal{L}_{\rm ep} = w |z_1|^2 |z_2|^2,
\end{equation}

with $w<0$. We will discuss in more detail later why these would
describe stable critical points - perhaps the most direct evidence
comes from the numerical simulations reported in
Ref.~\onlinecite{mv} of a lattice model of a CP$^1$ field coupled
to a noncompact gauge field (a lattice version of Eqn. \ref{sz}),
where a continuous transition was found in both the isotropic and
easy plane cases.

How can this action describe the onset of VBS order when it does
not contain $\psi_{\rm VBS}$, and the $z_\alpha$ are closely
related to the N\'eel order parameter?  In writing Eq.~(\ref{sz}),
we have tacitly assumed $a_\mu$ to be a single-valued continuous
field.  In a more careful lattice implementation of
Eq.~(\ref{eq:cp1}), however, the resulting gauge field that
appears is {\em compact}, {\em i.e.\/} defined only modulo $2\pi$.
This allows for the presence of topological defects occurring at a
single instant of space-time (``instantons'') called monopoles, at
which magnetic flux $\partial_x a_y-\partial_y a_x$ is created or
destroyed in integer multiples of $2\pi$.  In general,
Eq.~(\ref{sz}) should thus be supplemented by terms which create
or destroy such $2\pi$ fluxes, or equivalently insert monopoles
into the partition function:
\begin{equation}
  \label{eq:smono}
  \mathcal{L}_{\rm mp} = \sum_{n=1}^\infty \lambda_n(r) \left(
    [v_{r\tau}^{\vphantom{\dagger}}]^n + [v_{r\tau}^\dagger]^n \right),
\end{equation}
where $v_{r\tau}^\dagger$ and $v_{r\tau}^{\vphantom{\dagger}}$
insert monopoles of strength $2\pi$ and $-2\pi$ at the space-time
point $(r,\tau)$, respectively.
Remarkably, it
has been shown by Read and Sachdev \cite{ReSaSuN,ReSa} that this
operator may be identified with the VBS order parameter, {\em
i.e.\/}
\begin{equation}
  \label{eq:veqvbs}
  v_{r\tau} \sim \psi_{\rm VBS}(r,\tau).
\end{equation}
A simple argument to this effect will be given in
Sec.~\ref{sec:models}. Thus VBS physics is implicitly (albeit
highly non-linearly) contained in the gauge theory of
Eqs.~(\ref{sz}) and (\ref{eq:smono}).  Our claim that
Eq.~(\ref{sz}) {\em without the monopole creation terms of
Eq.~(\ref{eq:smono})\ } describes the critical properties of the
N\'eel-VBS transition requires that the monopole ``fugacities''
$\lambda_n$ are {\em irrelevant} in the renormalization group
sense at the QCP.  Later sections of this paper will give a
variety of compelling arguments, relying upon destructive quantum
interference between different monopole events, for this
irrelevance for spin-$1/2$ antiferromagnets. The arguments are
based on quantum Berry phase effects described first by
Haldane\cite{hald88}, which render $\lambda_n(r)$ oscillatory and
negligible for $n \neq 0 \, ({\rm mod}\, 4)$ for spin $S=1/2$ (a different
derivation appears in Appendix~\ref{app:SJB} and in the review in
Ref.~\onlinecite{srev}).

Although monopoles can be neglected at
the QCP, this is not true at low energies in the VBS phase.
Indeed, it is well-known from studies of pure compact U(1) gauge
theories, that the fugacities $\lambda_n$ are always relevant in
the absence of gapless ``matter fields'' ({\em i.e.\/} the
$z_\alpha$), so that monopoles inevitably proliferate in this
case. This proliferation leads to a ``condensation'' of the
monopole operator, $\langle v_{r\tau}\rangle \sim \langle
\psi_{\rm VBS}\rangle \neq 0$, hence VBS order
\cite{ReSaSuN,ReSa}.  At the same time it generates a gap for the
gauge ``photon''.  In renormalization group terminology, the
monopole condensation in the VBS phase -- despite the fact that
the $\lambda_n$ are negligible at the QCP -- indicates that (some)
$\lambda_n$ are ``dangerously irrelevant''.

\begin{figure}
\centerline{\includegraphics[width=3in]{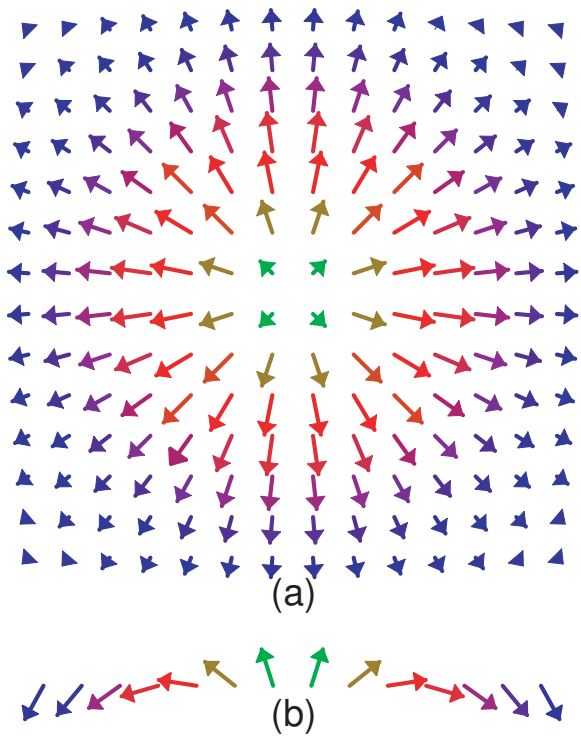}} \caption{A skyrmion
configuration of the field $\hat{n}_r$. In (a) we show the vector
$(n^x,n^y)$ at different points in the XY plane; note that
$\hat{n} \propto (-1)^{x+y} \vec S_r$, and so the underlying spins
have a rapid sublattice oscillation which is not shown. In (b) we
show the vector $(n^x, n^z)$ along a section of (a) on the $x$
axis. Along any other section of (a), a picture similar to (b)
pertains, as the former is invariant under rotations about the $z$
axis. The skyrmion above has $\hat{n} (r=0) = (0,0,1)$ and
$\hat{n}(|r| \rightarrow \infty) = (0,0,-1)$.} \label{skyr}
\end{figure}

\begin{figure}
\centerline{\includegraphics[width=3in]{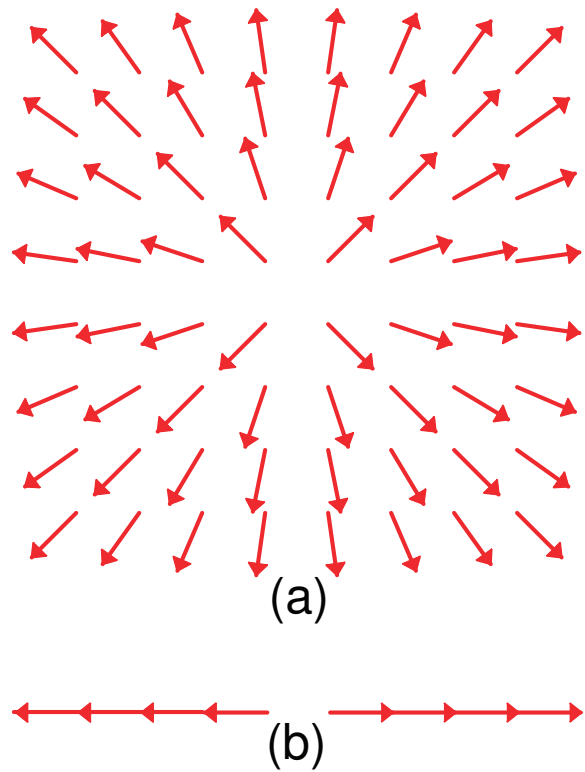}} \caption{A
monopole event, taken to occur at the origin of spacetime. An
equal-time slice of spacetime at the tunnelling time is
represented following the conventions of Fig~\ref{skyr}. So (a)
contains the vector $(n^x, n^y)$; the spin configuration is
radially symmetric, and consequently a similar picture is obtained
along any other plane passing through the origin. Similarly, (b)
is the representation of $(n^x, n^z)$ along the $x$ axis, and a
similar picture is obtained along any line in spacetime passing
through the origin. The monopole above has $\hat{n}_r = r/|r|$.}
\label{hedge}
\end{figure}

It is important to note that such monopoles have a natural
topological interpretation in terms of the conformations of the
N\'eel ordered state.  In particular, low but non-zero energy
configurations of the antiferromagnet are described by states with
slowly-varying N\'eel vector (at least at spatial infinity) of
constant amplitude,
\begin{equation}
\vec{N}_r=|\vec{N}|\hat{n}_r. \label{eq:Nn}
\end{equation}
Such classical configurations with finite energy admit topological
defects known as skyrmions (see Fig~\ref{skyr}).
The total {\em skyrmion number} associated with a configuration
defines an integer topological quantum number $Q$:
\begin{equation}
  \label{poynt} Q = \frac{1}{4\pi} \int d^2 r \, \hat n \cdot
  \partial_x \hat{n} \times \partial_y \hat{n}.
\end{equation}
Remarkably (see Sec.~\ref{sec:models}\ and
Ref.~\onlinecite{dadda}), the skyrmion density is simply related
to the magnetic flux of the gauge field $a_\mu$,
\begin{equation}
\label{eq:fluxeqskyrdens}
2\pi Q = \int d^2x (\partial_x a_y - \partial_y a_x).
\end{equation}
Thus the monopole instantons that change the gauge flux by $\pm
2\pi$ describe events in which the skyrmion number changes by $\pm
1$.  Thus the flux creation operator $v_{r\tau}^\dagger$ can also
be interpreted as a skyrmion creation operator.  The skyrmion
number changing events may be represented graphically as
``hedgehog'' configurations of the N\'eel vector in space-time
(See Fig.~\ref{hedge}).
The irrelevance of the monopole fugacities at the N\'eel-VBS
critical point is thus equivalent to the irrelevance of hedgehog
fugacities in a semiclassical description. Further, the
recognition that such monopole events can be characterized as
changing (as a function of time) the skyrmion number $Q$ enables
another interpretation of their irrelevance. In particular, in the
critical fixed point theory in Eq.~(\ref{sz}) -- {\em i.e.\/} at
low energies near the QCP -- the skyrmion number $Q$ is strictly
conserved. The emergence of this conserved topological quantum
number is the most fundamental meaning of the irrelevance of the
instantons.

We will also use this emergent topological conservation law as a
definition of a ``deconfined'' QCP.  Indeed, typically the gauge
theories that arise in various slave particle descriptions of
quantum magnets are {\em compact}. Specializing to a U(1) gauge
theory, the compactness means that instanton or monopole events in
which the magnetic flux changes by $2\pi$ are allowed
configurations of the gauge field in space-time. The proliferation
of these instanton events leads to confinement of the slave
particles in the gauge theory. In contrast, in a non-compact
theory -- which emerges at low energies when monopoles are
irrelevant -- the total magnetic flux is strictly conserved. This
is a topological conservation law and may be understood as a
global U(1) symmetry in an appropriate dual description.  Indeed,
we will explicitly construct such a dual theory for the case of
easy-plane anisotropy (and in some other related models). Quite
generally, then, the emergence of a non-compact U(1) gauge theory
at the critical point between the N\'eel and VBS phases signifies
an extra emergent (dual) global U(1) symmetry for the critical
theory that is not present in the microscopic Hamiltonian. This
provides a rather precise characterization of a `deconfined'
critical point.

An important property of the deconfined fixed points discussed in
this paper is the appearance of two {\em distinct} diverging
length (or equivalently two time) scales close to the transition -
one of which rises as a power of the other. This is directly due
to the dangerous irrelevance of monopoles. For the N\'eel-VBS
transition on approaching from the VBS side there is of course a
diverging spin correlation length $\xi$. However just beyond this
length scale the system has not yet chosen to pin itself into any
particular VBS ordered state. Rather it may be characterized as
fluctuating between different VBS configurations. It settles down
to a particular ordered state at a larger length scale $\xi_{\rm
VBS}$. This new length scale may also be characterized as the
thickness of a domain wall in the VBS order. The universal
crossovers associated with the critical fixed point describe the
behavior on passing through the length scale $\xi$. These are
described by the critical theory in Eq.~(\ref{sz}). As explained
above, this critical theory is monopole-free. The second crossover
associated with the length scale $\xi_{\rm VBS}$ describes how the
system evolves from the paramagnetic phase associated with the
monopole-free theory Eq.~(\ref{sz}) to the true VBS phase that
obtains when monopoles eventually proliferate. Further details of
the physics at the scales $\xi$ and $\xi_{\rm VBS}$ appear in
Section~\ref{sec:phys-prop-near}, where we also show in
Eq.~(\ref{eq:xispvsxi}) that $\xi_{\rm VBS}$ diverges as a power
of $\xi$ which is greater than unity.

Over the last several years we have become familiar with the
notion of fractionalization of quantum numbers in stable phases in
condensed matter. In contrast, the fractionalization phenomena
obtained in this paper are specific to the critical point
separating two conventional phases.  These `fractional' particles
-- the spinons -- are not present (i.e confined or condensed) at
low energies on either side of the transition but appear naturally
at the transition point. Likewise the emergent gauge field that
mediates interactions between the fractional particles is also
specific to the critical point. On approaching the critical point,
the confinement length scale diverges. Thus `deconfinement'
appears right at the transition.

We will also briefly discuss the phase transitions between
different quantum paramagnetic ground states. In particular we
will argue that the existing theory for the transition between a
VBS state and a fractionalized spin liquid implies that the
corresponding critical point is also described by a deconfined
U(1) gauge theory in precisely the same manner as above.
Furthermore, Refs.~\onlinecite{vbs,sondhivbs} argue that (at least
under certain conditions) there are direct transitions between two
different VBS phases that are also described by deconfined
critical points with a U(1) gauge structure.

There are several general lessons to be learnt from the results in
this paper.  First, we see that two dimensional spin-$1/2$ quantum
magnetism is full of examples of `deconfined' quantum critical
points which contradict the LGW paradigm for critical phenomena.
This suggests that in more complex quantum systems ({\em e.g.}
with fermions or disorder) novel critical phenomena may well be
quite commonplace. Such deconfinement may be at the root of
interesting non-Fermi liquid critical phenomena observed in the
heavy fermion materials and possibly in the cuprates as well.
Second, our results resolve some long-standing controversies in
the field of two dimensional quantum magnetism and have direct
implications for experiments and numerical work in the field.

Third, our results shed some light on questions of confinement in
gauge theories in two spatial dimensions.  It was shown by
Polyakov several years ago \cite{polyakov} that in two spatial
dimensions for pure gauge theories (i.e without any matter fields)
instantons generically always proliferate and drive the
theory into a confined phase. The behavior in the presence of
dynamic matter fields (particularly with fermionic matter) is much
less understood and is a subject of some controversy
\cite{iolar,wenqo,sudbo,igor,igorss}. The results in this paper show that with
bosonic matter there are at least isolated critical points
\cite{igorss} at which deconfinement is obtained (and the
instantons disappear at long scales). While typically reaching
criticality in a bosonic system requires some fine-tuning,
fermionic systems can have stable critical phases. This supports
the speculation that stable deconfined phases exist in two
dimensional compact U(1) gauge theories coupled to fermionic
matter \cite{ssfoot}. If true this would have interesting
implications for the theory of spin liquid phases of quantum spin
systems. These points are discussed further in
Section~\ref{sec:SJN} and Appendix~\ref{app:monoscr}.

Apart from these general notions, there are also a number of
specific physical ramifications of the proposed critical theory
for the N\'eel-VBS transition. One immediate consequence is that
the anomalous dimension of the magnon operator is much larger than
is usual at $D = 2+1$ dimensional fixed points. Thus the magnon
spectral function will be extremely broad right at the critical
point. Many other implications are explored in some detail later
in this paper and summarized in the overview in
Section~\ref{sec:overview}.

\section{Overview}
\label{sec:overview}

In this section we provide an overview of the main ideas in this
paper.

\subsection{History and Precedents}
\label{sec:history}

We begin by recalling some important prior results in the theory
of quantum magnetism on the two dimensional square lattice.  In
the N\'eel phase or close to it, the long distance low energy
fluctuations (of the orientation) of the N\'eel order parameter
are captured by the quantum O(3) Non-Linear Sigma Model
(NL$\sigma$M) with the Euclidean action (we have promoted the
lattice co-ordinate $r=(x,y)$ to a continuum spatial co-ordinate,
and $\tau$ is imaginary time):
\begin{eqnarray}
\mathcal{S}_n &=& \mathcal{S}_0 + \mathcal{S}_B \nonumber \\
 \mathcal{S}_0 &=& \frac{1}{2g} \int d\tau \int d^2 r
\left[
  \left(\frac{\partial \hat n}{\partial \tau}\right)^2 +
  c^2 \left(\nabla_r
    \hat n \right)^2 \right]
  \nonumber \\
  \mathcal{S}_B &=& i S \sum_r \epsilon_r \mathcal{A}_r
    \label{eq:nlsm}
\end{eqnarray}
Here $\hat n_r \propto \epsilon_r \vec S_r$ is a unit three
component vector that represents the N\'eel order parameter (the
factor $\epsilon_r$ is defined in Eq. (\ref{defeps})). The term
$\mathcal{S}_B$ contains crucial quantum-mechanical Berry phase
effects, and is sensitive to the precise quantized value, $S$ of
the microscopic spin on each lattice site: $\mathcal{A}_r$ is the
area enclosed by the curve mapped out by the time evolution of
$\hat{n}_r (\tau)$ on the unit sphere. These Berry phases play an
unimportant role in the low energy properties of the N\'{e}el
phase \cite{chn}, but are crucial in correctly describing the
quantum paramagnetic phase \cite{ReSaSuN,ReSa}. We will expand
these earlier results to show here that they also modify the
quantum critical point between these phases, so that the critical
exponents are distinct from the theory without $\mathcal{S}_{B}$
studied earlier \cite{chn,csy}.

To understand the summation over $r$ in $\mathcal{S}_B$, recall
that, as described in Section~\ref{sec:intro}, in two spatial
dimensions, smooth configurations of the N\'{e}el vector admit
skyrmion topological defects characterized by the integer
topological charge $Q$.  The Berry phase $\mathcal{S}_B$
vanishes\cite{hopf,subirbook} for all {\em smooth} configurations
even if they contain skyrmions. For such smooth configurations,
the total skyrmion number $Q$ is conserved. Crucially, however,
the skyrmion number changing monopole events are {\em not}
everywhere smooth.  It was shown by Haldane\cite{hald88} that the
summation over $r$ in $\mathcal{S}_B$ is non-vanishing in the
presence of such monopole events. Precise
calculation\cite{hald88,ReSa} gives a Berry phase associated with
each such skyrmion changing process which, for $2S=1\;({\rm mod}\,
4)$, oscillates rapidly on four sublattices of the dual lattice
(see Appendix~\ref{app:SJB}). This leads to destructive
interference which effectively suppresses all monopole events
unless they are quadrupled\cite{hald88,ReSaSuN,ReSa} ({\em i.e}
they change skyrmion number by four).

The NL$\sigma$M field theory augmented by these Berry phase terms
is, in principle, powerful enough to correctly describe the
quantum paramagnet.  Summing over the various monopole tunnelling
events shows that in the paramagnetic phase the presence of the
Berry phases leads to VBS order\cite{ReSaSuN,ReSa}.  This crucial
result from prior work identifies the VBS phase as resulting from
a proliferation of monopoles in the presence of non-trivial Berry
phases.  The non-trivial identification of the VBS order parameter
expressed from bond energies in Eq.~(\ref{eq:VBSorder}) as the
skyrmion creation operator is remarkable. When this operator
acquires an expectation value VBS order results.  In this manner
$\mathcal{S}_n$ contains within it the ingredients describing both
the ordered phases of $H$.

An important conclusion which follows from this body of work, via
the above identification, is that a direct second order transition
from the (collinear) N\'eel phase to a translation symmetric
`spin-liquid' is likely to be absent in two spatial dimensions. This still
leaves several possibilities for the evolution of the ground state
from the N\'eel to the VBS phases. For instance, there could be
two transitions with an intermediate phase that breaks both N\'eel
and lattice symmetries (an intermediate phase that breaks neither
symmetry is excluded), or simply a first order transition.

The possibility of a direct second order transition between N\'eel
and VBS phases is hinted at by several results in the existing
literature. First, note that in the N\'eel phase monopole
tunneling events are absent at long length and time scales. In the
quantum paramagnet these monopole tunneling events have
proliferated. The Haldane phases then lead to VBS order.  The
existence of a monopole condensate is clearly incompatible with
long range N\'eel order. Thus to the extent that the broken
lattice symmetry of the VBS state is a {\em consequence} of the
proliferation of monopoles it competes with the N\'eel state. A
direct transition from N\'eel to VBS then becomes conceivable
\cite{ReSaSuN,ReSa,SJ}.

A second hint comes from examining large-$N$ studies of SU$(N)$
quantum spin models \cite{ReSaSuN,ReSa}. In the limit $N = \infty$
(and in a bosonic representation of the spins) there is a second
order transition between the N\'eel phase and a quantum
paramagnet. In this limit the paramagnet breaks no symmetries
(including lattice symmetries). Furthermore, it supports gapped
spin-$1/2$ excitations (known as spinons). However, both features
are known to be artifacts of the limit $N = \infty$. Upon
including finite $N$ corrections, broken lattice symmetry appears
(leading to a VBS phase). The spinons also feel a gauge force that
leads to their confinement and disappear from the spectrum. What
about the vicinity of the transition? To answer this, it is
instructive to examine the various length scales in the problem in
the paramagnetic state. First, there is the spin correlation
length that diverges on approaching the transition.  Note that
deep in the paramagnetic phase this length stays constant when $N
\rightarrow \infty$. Then, there is the length scale at which VBS
order appears. As there is no VBS order at $N = \infty$ this
length scale must {\em diverge} as $N \rightarrow \infty$ -- hence
it must be much bigger than the spin correlation length in the
large-$N$ limit. Finally, there is a third (somewhat loosely
defined) length scale that may be thought of as the length scale
associated with spinon confinement. Clearly this scale also
diverges as $N \rightarrow \infty$ and is much bigger than the
spin correlation length. Indeed calculations of the VBS and
confinement length scales in the large-$N$ limit show that they
are $\mathcal{O}(\xi^{cN})$, where $\xi$ is the spin correlation
length \cite{MS} and $c$ is a constant.

This suggests the possibility of a direct transition between
N\'eel and VBS states in the large-$N$ limit where the monopoles
(and hence their Berry phases) are irrelevant at the critical
fixed point, but are important in producing VBS order and
confinement in the paramagnetic state. In critical phenomena
parlance, the monopoles are {\em dangerously} irrelevant at the
critical fixed point.

A picture similar to this was in fact proposed several years ago
by Chubukov {\em et al.} \cite{csy}. However, it was not
appreciated that the quadrupling of the monopoles, induced by the
Berry phases, renders {\em both} the monopoles and their Berry phases
irrelevant at the critical point (the distinction between the
relevance of single versus quadrupled monopoles is absent in the
large $N$ limit \cite{noteN}). In particular, it was assumed that
the confinement length scale will stay finite at the transition,
which was then modelled (for physical $SU(2)$ spins) by the LGW
theory obtained simply by neglecting $\mathcal{S}_B$ in
Eq.~(\ref{eq:nlsm}): this is the O(3)-invariant Wilson-Fisher
fixed point \cite{wilson}. In light of the discussion above, it is
clear that as the confinement goes hand in hand with the VBS order
both confinement and VBS length scales diverge at the transition.
Thus we might expect `deconfinement' to appear at the transition.

A  weakness in the arguments of Chubukov {\em et al.} \cite{csy}
was pointed out by Sachdev and Park\cite{sp}. The latter authors
argued that there was a finite density of monopoles in space-time
right at the critical point of the O(3) LGW model, and the Berry
phases then implied the presence of finite VBS order at any such
critical point. Based on this they suggested that a possible
evolution between the N\'eel and VBS phases was through a region
of coexistence of both broken symmetries. However, they left open
the possibility of a direct second order transition between the
N\'eel and VBS phases, but argued that any such transition could
not be described by the O(3) LGW model.

Our discussion here makes it clear it is necessary that the
corresponding fixed point have no monopoles at long scales. The
natural candidate is then precisely the fixed point governing the
transition in the model with monopoles forbidden. The arguments of
Motrunich and Vishwanath \cite{mv}, and our present analysis, on
such models show that the appropriate critical theory is that of
$\mathcal{L}_z$ in Eq.~(\ref{sz}). It must be kept in mind that
this critical theory is entirely distinct, with all critical
exponents different, from the O(3) LGW model obtained by dropping
$\mathcal{S}_B$ from Eq.~(\ref{eq:nlsm}). The first indication
that such a distinct continuous transition could exist in the
monopole suppressed O(3) NL$\sigma$M was from the work of Kamal
and Murthy \cite{KM}. Recently, the transition in this model with
monopole suppression was studied in Ref.~\onlinecite{mv}, where a
new approach that sidestepped the potential problems of
Ref.~\onlinecite{KM} was used. A continuous, non-Heisenberg
transitions with properties consistent with those of
Ref.~\onlinecite{KM} was found. Moreover, an independent numerical
simulation of a CP$^1$ model with a noncompact gauge field
was performed (essentially Eq.~\ref{sz}) which also yielded a
continuous transition and exponents consistent with the
simulations of the monopole suppressed O(3) NL$\sigma$M. This
provided a nontrivial check of both the essential correctness of
the numerical calculations and direct support for the
identification of Eq.~(\ref{sz}) as the critical theory for the
monopole suppressed O(3) NL$\sigma$M transition. The easy plane
deformation of these models was also studied in
Ref.~\onlinecite{mv}, where again a continuous transition was
obtained. This transition was argued to possess the remarkable
property of being {\em self-dual}.

The possibility of deconfinement of spinons at the critical point
between N\'eel and VBS phases is also hinted at by a different
consideration that is again motivated by the large-$N$
calculations. The excitations of both the N\'eel and VBS phases
are conventional ({\em i.e} do not contain any fractionalized
spinons). In a Schwinger boson description in terms of spin-$1/2$
spinons this is achieved through confinement. However the detailed
mechanism of such spinon confinement is different in the two
phases. In the N\'eel phase (described as a spinon condensate)
confinement is achieved through the usual Higgs mechanism. On the
other hand, in the VBS phase confinement is achieved through
proliferation of instantons. This difference in the confinement
physics then makes it conceivable that neither mechanism is
actually operational at the critical point and deconfinement
obtains.

\subsection{Numerics}

There have been a large number of numerical studies of the
destruction of N\'eel order in the $S=1/2$ square lattice
antiferromagnet \cite{misguich}. While there is evidence for the
existence of VBS order in the paramagnetic phase,
\cite{gsh,ziman,kotov,van,harada} the nature of the transition
between the N\'eel and paramagnetic states has been difficult to
address. A major obstacle is the well-known `sign' problem, which
prevents large-scale Monte Carlo simulations. Until recently, all
large scale studies of the destruction of N\'eel order have been
on models with an even number of $S=1/2$ spins per unit cell, with
a paramagnetic phase which does not break any lattice symmetries.
\cite{troyer,matsumoto}

The first large-scale study of the destruction of N\'eel order in
a $S=1/2$ square lattice antiferromagnet, in a Hamiltonian which
preserves a single $S=1/2$ spin per unit cell and the full
symmetry of the square lattice, was that of Sandvik {\em et al.}
\cite{sandvik}. This was on a model with a strong easy-plane
anisotropy. Such easy plane models have been studied analytically
previously \cite{Crtny,sp} and will be pursued further in the
present paper. Ref.~\onlinecite{sandvik} found convincing evidence
for VBS order in the paramagnetic phase. Furthermore, the VBS and
N\'eel order appear to vanish at points close to each other,
suggesting a direct second order transition in the class discussed
in the present paper.

We also note the wavefunction Monte Carlo work of Capriotti {\em
et al.} \cite{sorella} on the SU(2) $S=1/2$ antiferromagnet on the
square lattice with first and second neighbor exchange. They found
a `resonating valence bond' wavefunction characteristic of a spin
liquid state. Our results here suggest that they were perhaps
observing the deconfined state characteristic of the critical
point, and that they had not yet reached the crossover to VBS
order at the longest scales.

\subsection{Plan of Attack}

In this paper, the proposal of a deconfined continuous N\'eel-VBS
transition (as well as a VBS-spin liquid transition) is
substantiated by a variety of arguments.  First, in
Section~\ref{sec:models} we consider a concrete lattice $CP^{N-1}$
model which, for $N=2$, embodies the physics of the N\'eel state,
the monopoles and their Haldane Berry's phases (focusing on
$S=1/2$), and the VBS state. This model, introduced by Sachdev and
Jalabert \cite{SJ}, and referred to here as the SJ model, provides
a convenient starting point for theoretical analysis of the
$SU(2)$ invariant critical region. We address the nature of the
physically interesting $N=2$ case by showing that, in the two
limits $N=1$ (Section~\ref{sec:SJ1}) and $N=\infty$
(Section~\ref{sec:SJN})), this model indeed sustains a deconfined
critical point in the precise sense defined above. For $N=1$, this
can be directly shown using lattice duality transformations, which
demonstrate an exact equivalence of the SJ model to a $D=3$
classical XY model with a fourfold symmetry breaking term which corresponds
physically
to strength four monopoles. Such four-fold anisotropy is known to be irrelevant
at the $D=3$ XY transition
\cite{vicari}, establishing the deconfinement of this case. For
$N=\infty$, the scaling dimension of the 4-skyrmion creation
operator was computed previously by Murthy and Sachdev\cite{MS},
and is such that monopoles are again irrelevant. Hence we expect
by continuity that monopoles are irrelevant for {\em all} $N$,
including the interesting case $N=2$.

Second, in Section~\ref{sec:ep}, we consider specifically $N=2$,
in the presence of additional (strong) easy-plane anisotropy.  In
this case, the SJ model may be rewritten as a pair of O(2) rotors
(the phases of $z_\alpha$) interacting with a compact U(1) gauge
field.  The latter may be analyzed directly using duality
techniques (Appendix~\ref{app:SJ2}).  We obtain in this way an
explicit dual representation in terms of complex ``vortex''
annihilation operators $\psi_\alpha$ ($\alpha=1,2$) and a dual
{\em non-compact} gauge field $A_\mu$, whose flux represents the
(exactly) conserved uniform spin density $S^z$.  One may
understand the relation to the CP$^1$ variables by recognizing
that $\psi_1^\dagger$ creates a $+2\pi$ vortex in $z_2$, while
$\psi_2^\dagger$ creates a $-2\pi$ vortex in $z_1$, both of which
create physical $2\pi$ vorticity in $n^-=n^x-in^y = z_1^*
z_2^{\vphantom{*}}$. The dual theory, $\mathcal{L}_{\rm dual}$ for
$\psi_{\alpha}$ and $A_{\mu}$ is presented in Eq.~(\ref{crtny}).

The dual representation $\mathcal{L}_{\rm dual}$ has an appealing
semiclassical interpretation, described in detail in
Sec.~\ref{sec:semi-class-analys}.  Briefly, the two types of
vortices correspond to ``merons'' (half-skyrmions), in which the
N\'eel vector points either up or down inside the vortex core. The
skyrmion number changing monopole events thereby correspond
precisely to an event in which a vortex core tunnels from the up
to down staggered magnetization or vice-versa.

The advantage of this representation is that the (quadrupled)
monopole fugacity appears explicitly as a local operator in terms
of the vortex fields.  Remarkably, if this fugacity, $\lambda$, is
set to zero (as appropriate at the QCP provided it is, as we
argue, irrelevant), the dual action in Eq.~(\ref{crtny}) has
precisely the same form as the original one, Eq.~(\ref{sz}). More
precisely, an exact equivalence can be demonstrated between
lattice regularizations of the original and dual theory in the
absence of monopoles \cite{mv}.  Thus, as found in
Ref.~\onlinecite{mv}, the proposed critical theory in the
easy-plane case has an unusual self-duality property.

The irrelevance of monopoles can then be argued in several ways.
First, using the self-duality, each power of the skyrmion creation
operator has the same correlations at the QCP as the corresponding
power of the XY staggered raising operator $n^+$.  At the
deconfined critical point, fluctuations of $n^+$ are expected to
be stronger than they are at a conventional (confined) XY critical
point. The corresponding quadrupled operator is already irrelevant
in the latter case (as mentioned above), so we expect the
four-skyrmion fugacity to be only more irrelevant around the
deconfined critical theory.  This expectation is supported by an
explicit calculation in a large-$N$ (different from the $N$ in the
SJ model) generalization of the dual critical theory in
Section~\ref{sec:phase-ep}. Further arguments are given in
Appendix~\ref{app:est}.

In Section~\ref{sec:superfl-insul-trans} we demonstrate for the
easy plane case a direct derivation of the dual critical theory
from a microscopic bosonic representation of the underlying XY
model, without utilizing either the NL$\sigma$M or SJ models.

In Section~\ref{sec:vbssl}, we show that analogous deconfinement
obtains for a VBS to spin-liquid transition.  The latter has
already been discussed by several authors \cite{JaSa,sf}, and
shown to be equivalent to the transition in a fully frustrated
quantum Ising model, which has a simple XY critical fixed point
unaffected at low energies by an irrelevant 8-fold symmetry
breaking term.  We show that this description is in fact dual to a
deconfined gauge theory in the same sense as above, and that the
(dangerously) irrelevant 8-fold symmetry breaking term can
likewise be interpreted as an irrelevant monopole fugacity.

\subsection{Organization of paper}

We will begin in Section~\ref{sec:models} with a general
discussion of the important symmetries of the Hamiltonian, and
their action on a variety of order parameters and operators. This
section will also introduce the SJ model. The solution of the SJ
model in a variety of tractable limits appears in
Sections~\ref{sec:SJ} and~\ref{sec:epcp1}. Section~\ref{sec:ep}
also contains a general, semiclassical description of the physics
in the easy plane limit. The nature of the second-order critical
point between the N\'eel and VBS states is discussed in
Section~\ref{sec:phase}. A variety of predictions for the critical
properties on the N\'eel-VBS transitions with $SU(2)$ and XY
symmetries follow from our analysis. These are elaborated in
Sec.~\ref{sec:phys-prop-near}: readers not interested in the
detailed theoretical analysis may skip ahead to this section
without significant loss of continuity. Section~\ref{sec:vbssl}
describes the deconfined critical point between the VBS and
spin-liquid phases, as noted above. Finally,
Section~\ref{sec:anext} contains a variety of extensions of the
results in this paper. Section~\ref{sec:superfl-insul-trans} shows
that the easy-plane N\'{e}el to VBS transition can be
reinterpreted as a superfluid-insulator transition in an
interacting boson system; the insulator in this case contains a
density wave in the amplitude of the bosons to reside on `bond'
states. This approach also provides an alternative derivation of
the dual model Eq.~(\ref{crtny}). Section~\ref{sec:higherspin}
briefly discusses the extension of our results to antiferromagnets
with $S>1/2$, while section~\ref{sec:honeycomb} considers the case
of the honeycomb lattice.  Section~\ref{sec:ising-anisotropy}
discusses possible extension of our results to systems with Ising
anisotropy.

\section{Representations and Symmetries}
\label{sec:models}

In this section, we describe the representation of magnets with
local tendencies to N\'eel order in the NL$\sigma$M and CP$^1$ (SJ
model) representations.  We describe the action of the physical
symmetries on the corresponding $\hat{n}$ and $z_\alpha$ fields in
each case

We start from the action $\mathcal{S}_n$ in Eq.~(\ref{eq:nlsm}).
The all-important Berry phase term in $\mathcal{S}_B$ is defined
on the underlying square lattice, and it is clear that lattice
scale cancellations are important for the physics we are
interested in. It is therefore useful to return to a lattice formulation
to obtain
\begin{eqnarray}
\mathcal{S}_n & = & \mathcal{S}_{0} + \mathcal{S}_B \nonumber \\
\mathcal{S}_{0} & = & \int d\tau \left(\sum_r
  \frac{1}{2g}\left(\frac{d\hat{n}_r}{d\tau}\right)^2  -
  J\sum_{<rr'>}\hat{n}_r \cdot \hat{n}_{r'} \right) \nonumber \\
\mathcal{S}_B & = & i S\sum_r  \epsilon_r \int d\tau  \vec
A[\hat{n}]\cdot\frac{d\hat{n}_r}{d\tau}. \label{eq:nlsm2}
\end{eqnarray}
We have now rewritten the areas $\mathcal{A}_r$ in terms of
$\vec{A}$, which represents the vector potential of a magnetic
monopole with flux $4 \pi$ placed at the center of $\hat{n}$ space
at each lattice site. This lattice model is a faithful representation of the
original quantum antiferromagnet so long as $g$ is large.
The continuum limit of $\mathcal{S}_0$ in this model is clearly just what
appears in Eqn. \ref{eq:nlsm}.
The representation of the Berry phase used here leads
directly to Eq.~(\ref{eq:sbp}) in Appendix~\ref{app:SJB}.

The Berry phases are crucial for a correct description of
quantum paramagnetic phases.  As described in the previous
sections, it was shown by Haldane\cite{hald88} that the Berry
phases are non-vanishing only in the presence of monopole events.
The calculations in Refs.~\onlinecite{hald88,ReSa}\ give the total
phase (for spin-$1/2$ magnets that we consider -- for a derivation
see here, Appendix~\ref{app:SJB})
\begin{equation}
\prod_n \exp \left(i \frac{\pi}{2} \zeta_n \Delta Q_n \right).
\label{eq:haldane}
\end{equation}
Here the monopole is associated with a plaquette of the original
lattice (or equivalently with a site of the dual square lattice),
which is labelled by the index $n$. The product is over all
locations of monopoles, and $\Delta Q_n = \pm 1$ is the change in
skyrmion number associated with the monopole. Note that the
periodic boundary condition along the time direction requires that
the net change in skyrmion number is zero so that $\sum_n \Delta
Q_n = 0$. The fixed integer field $\zeta_n$ is $0,1,2,3$ depending
on whether the dual lattice coordinate is (even,even), (even,odd),
(odd,even) or (odd,odd), so that the phase factor associated with
each monopole is $1,i,-1,-i$ on these sublattices (see
Fig~\ref{offset} in Section~\ref{sec:SJ1}).

The oscillating nature of the Berry phase factors on adjacent
plaquette leads to destructive interference between different
tunnelling paths for single monopoles. Indeed this interference
effectively kills all monopole events unless they are quadrupled
({\em i.e} change skyrmion number by four).  Hence only such
quadrupled monopole events need be including in the quantum
statistical mechanical partition sum.

We have already indicated the remarkable identification of the VBS
order parameter defined in Eq.~(\ref{eq:VBSorder}) with the
skyrmion annihilation operator, $\psi_{\rm VBS}\sim v$, as shown
in Ref.~\onlinecite{ReSaSuN}. This provide the crucial confluence
of the {\em loss} of antiferromagnetic order (and consequent
proliferation of monopole events) with the {\em onset} of VBS
order, counter to conventional LGW wisdom. Because of the
importance of this result, we give a simplified derivation of this
relation here.

It is important to recognize that the VBS order parameter in
Eq.~(\ref{eq:VBSorder}) is entirely defined by its transformation
under the symmetries of the Hamiltonian.  Any other field with the
same symmetry properties as $\psi_{\rm VBS}$ will, on general
scaling and renormalization group grounds, be proportional to
$\psi_{\rm VBS}$ in the critical region.  Thus to prove the
identification of $\psi_{\rm VBS}$ with the skyrmion creation
operator $v$, it is sufficient to show that the latter transforms
identically to $\psi_{\rm VBS}$ under all symmetry operations.

As a topological index, the skyrmion number is unchanged under
smooth global $SU(2)$ spin rotations, hence the skyrmion number
changing operator is also an $SU(2)$ scalar.  Likewise, $\psi_{\rm
VBS}$, being defined through scalar bond operators, is $SU(2)$
invariant.  Let us consider the effect of lattice symmetry
transformations on $v$.  In the functional integral this operator
is defined by insertion of a space-time monopole.  It is easy to
see that under $\pi/2$ rotations in the counter-clockwise
direction about a direct lattice site (which we denote
$R_{\pi/2}$), the Berry phase associated with the skyrmion
creation event changes by $e^{i\pi S}$. Thus if we denote by
$v^{\dagger}$ the skyrmion creation operator and specialize to $S
= 1/2$, we have
\begin{equation}
R_{\pi/2}: v^{\dagger} \rightarrow iv^{\dagger}.
\end{equation}
The skyrmion creation operator is actually defined on a plaquette - for
the time being, we will label the plaquette by the lattice site at the
top right corner.

Lattice translation operations $T_{x,y}$ corresponding to
translations by one unit along $x,y$ directions of the microscopic
spin model have somewhat more subtle effects. First, in the rotor
representation of $\hat{n}_r$, these translations are represented
as
\begin{eqnarray}
T_x: \hat{n}_{r} & \rightarrow & - \hat{n}_{ r + \hat{x}} \\
T_y:\hat{n}_{ r} & \rightarrow & - \hat{n}_{ r + \hat{y}}
\end{eqnarray}
The change in sign of $\hat{n}$ is due to the staggering implicit
in its definition. Now note that the skyrmion number $Q$ is {\em
odd} under $\hat{n} \rightarrow -\hat{n}$. Consequently $T_{x,y}$
convert a skyrmion creation operator to an antiskyrmion creation
operator at the translated plaquette. Furthermore due to the
difference in the Berry phase factors for monopoles centered on
adjacent plaquettes, there is a phase factor that is introduced by
the translation.  Simple calculation gives the following
transformation properties for $v^{\dagger}$, specializing again to
$S = 1/2$:
\begin{eqnarray}
T_x: v^{\dagger}_{ r} & \rightarrow & -i v_{ r + \hat{x}} \\
T_y: v^{\dagger}_{ r} & \rightarrow & +i v_{ r+ \hat{y}}.
\end{eqnarray}

It is now clear that a paramagnetic state with a uniform
expectation value of $v^{\dagger}$ breaks these lattice
symmetries. For instance, if $\langle v^{\dagger} \rangle =
\langle v \rangle \neq 0$, then $R_{\pi/2}, T_x, T_y$ are all
broken. This suggests a plaquette ordered state such as that shown
in the lower right of Fig.~\ref{pdiag}. A straightforward
comparison shows that, up to an innocuous constant pre-factor, the
lattice transformation properties of $v$ are identical to those of
$\psi_{\rm VBS}$ determined from a more mundane analysis of
Eq.~(\ref{eq:VBSorder}).  In particular,
\begin{equation}
  \label{eq:vbseqskyr}
  v \sim e^{-i\pi/4} \psi_{\rm VBS},
\end{equation}
properly reproduces all the transformation properties of the VBS order
parameter.  Thus we may indeed identify the skyrmion creation operator with
the order parameter for the VBS order.

We have already introduced in the introduction the CP$^1$
``spinon'' fields to represent the N\'eel order parameter.  These
may be introduced on the lattice,
\begin{equation}
\label{cp1}
\hat{n}_r = z^{\dagger}_r \vec \sigma z^{\vphantom\dagger}_r.
\end{equation}
To maintain the unit magnitude of $\hat{n}_r$, the constraint
$|z_1|^2+|z_2|^2=1$ should be imposed upon the spinor $z = z(r , \tau)
= (z_1, z_2)$.

One can show\cite{dadda} that the partition function of the
continuum NL$\sigma$M (with the action in Eq.~(\ref{eq:nlsm}
neglecting Berry's phase terms) is exactly reproduced by the
continuum CP$^1$ model with the action
\begin{equation}
\mathcal{S}_{cp} = \int d\tau  d^2 r |\left(\partial_{\mu} -
  ia_{\mu}\right)z|^2.
\end{equation}
Here $a_\mu$ enters mathematically as a Hubbard-Stratonovich field,
and by considering its quadratic Euler-Lagrange equation, one can
deduce the relation of the skyrmion number $Q$ to the gauge flux of
$a_\mu$ given in Eq.~(\ref{eq:fluxeqskyrdens}).

As discussed above, to incorporate VBS phases it is important to
correctly account for the Haldane Berry phases associated with
these instantons.  An appropriate model has been constructed by
Sachdev and Jalabert\cite{SJ}.  The Euclidean action of the
Sachdev-Jalabert (SJ) model is
\begin{eqnarray}
\mathcal{S}_{\rm SJ} &=& \mathcal{S}_z + \mathcal{S}_a +
\mathcal{S}_B \nonumber \\ \mathcal{S}_z & = & -t\sum_i
z^{*}_{i\alpha} e^{ia_{\mu}}z_{i+\hat{\mu}, \alpha}
+ {\rm c.c.}, \label{eq:SJ1} \\
\mathcal{S}_a & = & \frac{K}{2} \sum\left(\epsilon_{\mu\nu\lambda}
  \Delta_{\nu}a_{\lambda} -
  2\pi q_{\mu}\right)^2 \label{eq:SJ2} \\
\mathcal{S}_B & = & i\frac{\pi}{2}\sum_n \zeta_n \Delta_{\mu}
q_{\mu}.\label{eq:SJ3}
\end{eqnarray}
Here we have put the complex spinon fields $z_{i\alpha}$ on the
sites, $i$, of a cubic space-time lattice in dimensions $D = 2+ 1$
(now, $n$ denotes the sites of the dual cubic lattice), and they
satisfy a unit length constraints $\sum_{\alpha} |z_{i\alpha}|^2
=1$ on each lattice site. The $a_{\mu}$ represent the compact U(1)
gauge field, and are defined on the links of the space-time
lattice. Note that the $z_{i\alpha}$ are minimally coupled to the
gauge field. The term $S_a$ represents the gauge field kinetic
energy. The quantity $q_{\mu}$ is an integer gauge flux that is
defined on the links of the dual cubic lattice. Its divergence
which enters the term $\mathcal{S}_B$ represents the number of
monopoles on the sites of the dual lattice. Consequently,
Eq.~(\ref{eq:SJ3}) is identical to the contribution in
Eq.~(\ref{eq:haldane}),  and $\mathcal{S}_B$ provides the Haldane
Berry phase factors that make the action appropriate for
describing spin-$1/2$ antiferromagnets on the square lattice. The
N\'eel ordered phase is a `Higgs' phase where the $z_i$ have
condensed, while the VBS phase is a `confined' phase where the
Berry phases have led to broken lattice symmetry.

The action $\mathcal{S}_{\rm SJ}$ is clearly closely related to
the lattice action $\mathcal{S}_n$ in Eq.~(\ref{eq:nlsm2}), after
replacing $\hat{n}$ by $z$ via Eq.~(\ref{eq:cp1}). However, the
corresponding $\mathcal{S}_B$ terms in Eqs.~(\ref{eq:nlsm2}) and
(\ref{eq:SJ3}) do appear rather different -- they are related by
the Berry phase summation carried out by Haldane \cite{hald88}.
Here, we establish the connection between these two forms of Berry
phases in Appendix~\ref{app:SJB}; further details on the
derivation of Eq.~(\ref{eq:SJ3}) from the microscopic
antiferromagnet appear in Refs. \onlinecite{SJ,sp,srev}.

If, as we will argue in the following, monopole events can indeed
be neglected at low energies near the QCP, we can set $q_\mu=0$.
Taking then a na\"ive continuum limit of
Eq.~(\ref{eq:SJ1}-\ref{eq:SJ3}) gives precisely the proposed field
theory of Eq.~(\ref{sz}).  We will, however, work directly with
the lattice SJ model including monopoles in several of the
sections to follow.

As in any critical phenomenon, symmetry plays a key role in the
discussion of the N\'eel-VBS transition.  We therefore list here
the various physical ({\em i.e.\/} non-gauge) symmetries of the
problem and their action upon the N\'eel and spinon fields.  The
only continuous physical symmetry is spin-rotational invariance,
either $SU(2)$ or U(1) in the case of easy-plane anisotropy. Under
such rotations, the N\'eel vector $\hat{n}$ and spinon field
$z_\alpha$ transform as global vectors and spinors, respectively.
The remaining unitary symmetries are discrete operations of the
space group of the square lattice, and can be composed from
$\pi/2$ rotations, translations, reflections, and inversions.  As
above, we denote $\pi/2$ clockwise rotations (around a direct
lattice site) by $R_{\pi/2}$ and unit translations in $x$ and $y$
by $T_x,T_y$ respectively.  Reflections $x\rightarrow -x$ or
$y\rightarrow -y$ around a lattice plane ({\em i.e.\/} leaving a
row or column invariant) are denoted ${\cal R}_x,{\cal
  R}_y$, and inversions about a site by $I$.  Finally, there is a
non-unitary time-reversal operation ${\cal T}$, which as usual
takes microscopic spins $\vec{S}_r \rightarrow -\vec{S}_r$.  The
transformation properties of the N\'eel, spinor, and gauge fields
are given in Table~\ref{tab:sym}.
\begin{table}[htbp]
  \centering
  \begin{ruledtabular}
  \begin{tabular}{l|c|c|c|c}
    Operation & Coordinates & N\'eel & spinor & gauge \\
    \hline
    $R_{\pi/2}$ & $x_i \rightarrow \epsilon_{ij} x_j$ & invariant &
    invariant & $a_i \rightarrow \epsilon_{ij} a_j$ \\
    $T_{x_i}$ & $x_i\rightarrow x_i+1$ & $\hat{n}
    \rightarrow -\hat{n}$ & $z_\alpha^{\vphantom\dagger} \rightarrow
    i\sigma^y_{\alpha\beta}z_\beta^\dagger$ & $a_\mu\rightarrow -a_\mu$ \\
    ${\cal R}_{x_i}$ & $x_i\rightarrow -x_i$ & invariant & invariant &
    $a_i\rightarrow -a_i$ \\
    $I$ & $r \rightarrow -r$ & invariant & invariant & $a_{x/y}
    \rightarrow -a_{x/y}$ \\
    ${\cal T}$ & $t \rightarrow -t$ & $\hat{n}\rightarrow -\hat{n}$ &
    $z_\alpha^{\vphantom\dagger} \rightarrow
    i\sigma^y_{\alpha\beta}z_\beta^\dagger$ &  $a_{x/y}\rightarrow -a_{x/y}$
  \end{tabular}
  \end{ruledtabular}
  \caption{Transformations of the N\'eel and spinor fields on the
square lattice under the
    discrete symmetry generators. Here $i=1,2=x,y$ is a spatial index,
  and $\epsilon_{ij}=i\sigma^y_{ij}$ is the fully antisymmetric
  rank two tensor.  Co\"ordinate transformations of the arguments of
  the fields have been suppressed.}
  \label{tab:sym}
\end{table}

Using these symmetry properties, one can determine the operators
of the field theory corresponding to physically interesting
microscopic quantities in a spin model.  Some of these are
tabulated in the second column of Table~\ref{tab:dual} of
Section~\ref{sec:epcp1} and describe general situations which allow for easy
plane anisotropy on the underlying magnet.
(The $SU(2)$ symmetric situation may be obtained as a special case).
These include the easy-plane and
hard-axis components of the N\'eel order parameter, $N^\pm = N^x
\pm i N^y$, $N^z$ and the VBS order parameter.  Also included are
the easy-plane and hard-axis components of the uniform
magnetization $\vec{M}$, all three being conserved for $SU(2)$
symmetry but only the latter being conserved with easy-plane
anisotropy.  Two currents are also of interest: the spatial
current $j_i^z$ of the conserved Ising (hard-axis) magnetization,
and the vorticity three-current $j_\mu^v$ - the latter being meaningful in the
presence of easy plane anisotropy.
Finally, we may
consider the CP$^1$ gauge three-current $j^{\scriptscriptstyle
  G}_\mu = \epsilon_{\mu\nu\lambda} \partial_\nu a_\lambda$, which in
the continuum theory is identified with the topological current
$j^{\scriptscriptstyle G}_\mu \sim
\frac{1}{4}\epsilon_{\mu\nu\lambda} \hat n \cdot \partial_\mu
\hat{n} \times \partial_\nu \hat{n}$.  In a microscopic model,
using the transformation properties of these operators, one can
construct a (rather complex) superposition of three-spin operators
with these same transformation properties. Consider for instance,
the time component $j^{\scriptscriptstyle G}_0 =
\epsilon_{ij}\partial_i a_j$. On a square plaquette with central
coordinate ${\sf r}$, number the sites starting at the uppper-left
corner of the plaquette and moving clockwise as $1,2,3,4$.  Then
one has
\begin{eqnarray}
  \label{eq:gaugeflux}
  j^{\scriptscriptstyle G}_0({\sf r}) & \sim & (-1)^{\sf r} \big[
  \vec{S}_1\cdot\vec{S}_2\times\vec{S}_3
    - \vec{S}_2\cdot\vec{S}_3\times\vec{S}_4  \nonumber \\
    & & +
    \vec{S}_3\cdot\vec{S}_4\times\vec{S}_1  -
    \vec{S}_4\cdot\vec{S}_1\times\vec{S}_2 \big],
\end{eqnarray}
where the $(-1)^{\sf r}$ takes opposite signs on the two sublattices of
the dual lattice.

\section{SJ Models}
\label{sec:SJ}

One useful generalization of the SJ model is to allow
$\alpha=1\ldots N$ above, so that $z$ is an $N$-component complex
vector of unit magnitude. The $SU(2)$ spin model corresponds to $N
= 2$. It will be possible to analyse the limits $N = 1$ and $N =
\infty$. As argued in Section~\ref{sec:stag}, the $N = 1$ model
actually may be realized in a spin-$1/2$ model in a staggered
Zeeman field. The large-$N$ limit describes ordering transitions
of certain SU$(N)$ quantum antiferromagnets and is less directly
physical. Its main utility is its tractability.  Similar behavior
in both extreme limits -- in particular the irrelevance of
monopoles in both cases -- suggests the same is true for the
models with intermediate $N$.

\subsection{SJ model at $N = 1$}
\label{sec:SJ1}

Consider first $N = 1$ where $z_i \equiv e^{i\phi_i}$ is simply a
complex number of unit magnitude. Then
\begin{equation}
\mathcal{S}_z = - 2t \sum_{\ell} \cos \left({\boldsymbol{\Delta}}
\phi - {\bf a} \right) \label{eq:SJ4},
\end{equation}
where sum is over the links $\ell$ of the cubic lattice. We
indicate spacetime 3-vectors here in bold face, and the discrete
lattice gradient by ${\boldsymbol\Delta}$. As discussed by SJ,
this $N = 1$ model displays a transition between a Higgs and a
translation broken phase. The latter has a four-fold degenerate
ground state due to lattice symmetry breaking. Simple symmetry
arguments suggest a transition modelled by a $Z_4$ clock model -
as the four fold anisotropy is irrelevant at the $D = 3$ XY fixed
point,\cite{vicari} this is in the $3D$ XY universality class. SJ
also provided numerical evidence supporting this expectation. As
shown below, all of this is readily established by a duality
transformation of the $N = 1$ model.

\begin{figure}[t]
\centerline{\includegraphics[width=2in]{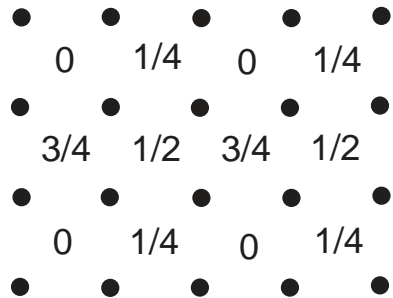}}
\caption{Specification of the fixed field $\vartheta = -\zeta/4$.
The filled circles are the sites of the direct lattice, and
$\vartheta$ resides on the sites of the dual
lattice.}\label{offset}
\end{figure}

To dualize the $N = 1$ SJ action we use a Villain representation
of the $\mathcal{S}_z$ term in Eq.~(\ref{eq:SJ4}):
\begin{equation}
\mathcal{S}_z \rightarrow \sum_{\ell} \left[
\frac{1}{2\tilde{t}}|{\bf j}|^2 - i{\bf j}\cdot
\left({\boldsymbol{\Delta}} \phi - {\bf a} \right) \right],
\end{equation}
The integer valued field ${\bf j}$ represents the current of the
$z$ field.  We also decouple the $S_a$ term in Eqn. \ref{eq:SJ2} by a
Hubbard-Stratanovich field ${\bf b}$ to write
\begin{equation}
\mathcal{S}_a \rightarrow \sum \left[\frac{1}{2K}|{\bf b}|^2 + i
{\bf b}\cdot ({\boldsymbol\Delta} \times {\bf a} - 2\pi {\bf q})
\right]
\end{equation}
Here, and below, the leading sum in the action extends over all
sites/links/plaquettes over the cubic lattice, as needed.
Performing the sum over the integer field ${\bf q}$, we get
\begin{equation}
{\bf b} - {\boldsymbol \Delta} \vartheta = {\bf B} \label{eq:SJ10}
\end{equation}
with $\vartheta_r=-\zeta_r/4$ (see Fig~\ref{offset}) and ${\bf B}$
an integer.
If we now integrate over $\phi$, we get the current conservation
condition
\begin{equation}
{\boldsymbol\Delta}\cdot{\bf j} = 0.
\end{equation}
This may be solved by writing
\begin{equation}
{\bf j} = {\boldsymbol\Delta} \times{\bf  A}
\end{equation}
with ${\bf A}$ an integer. Integrating over the gauge field ${\bf
a}$, we obtain
\begin{equation}
{\boldsymbol\Delta} \times{\bf B} = {\bf j}. \label{eq:SJ11}
\end{equation}
This may be solved by writing
\begin{equation}
{\bf  B} = {\bf A} + {\boldsymbol\Delta} \chi \label{eq:SJ12}
\end{equation}
with $\chi$ an integer. The action then reads
\begin{equation}
\mathcal{S} = \sum \left[
\frac{1}{2\tilde{t}}\left({\boldsymbol\Delta} \times {\bf A}
\right)^2 + \frac{1}{2K}\left({\boldsymbol\Delta}(\chi +
  \vartheta) + {\bf A} \right)^2 \right].
\label{eq:discretedual}
\end{equation}
The hard integer constraints on ${\bf A}, \chi$ may be softened by adding
terms
\begin{equation}
-t\cos(2\pi{\bf A})- \sum_n \lambda_n \cos(2 \pi n\chi)
\label{eq:soften1}
\end{equation}
We may now shift $\chi \rightarrow \tilde{\chi} = \chi +
\vartheta$, ${\bf
  A} \rightarrow {\bf A'} = {\bf A} + {\boldsymbol\Delta} \tilde{\chi}$.
The ${\bf A'}$ field is massive and may be integrated out. The remaining
action for the $\tilde{\chi}$ reads
\begin{equation}
\mathcal{S} = \sum \left[ -t \cos(2\pi {\boldsymbol\Delta}
\tilde{\chi}) - \sum_{n}\lambda_n \cos[2\pi n(\tilde{\chi} -
\vartheta)] \right]
\end{equation}
This describes an XY model with various $n$-fold anisotropy terms
of strengths $\lambda_n$ The shift by $\vartheta$ leads to rapid
spatial oscillations of these anisotropy terms unless $n = 0\,
({\rm mod} \,4)$. Near the critical point in the continuum limit,
the leading non-vanishing anisotropy term is at $n = 4$. The
critical properties are therefore that of an XY model with
four-fold anisotropy $\lambda_4$. The latter has a scaling
dimension $\Delta_4 > 3$, which renders it irrelevant at the $D=3$
XY critical point.\cite{vicari}

An overly cautious reader may object that uncontrolled approximations
have been made in softening the integer constraints on the ${\bf
  A},\chi$ fields.  However, all manipulations up to
Eq.~(\ref{eq:discretedual}) are exact, and from this point an
exact world-line representation may be obtained by implementing
the integer constraints using the Poisson resummation formula. The
latter representation clearly describes charged relativistic
particles for which charge non-conservation events oscillate spatially
unless the charge is changed in multiples of four.  On
universality grounds, one expects this model to be in the same
universality class as an XY model with 4-fold anisotropy. Though
we will not pursue it, a similar exact duality can be performed on
the $N=2$ SJ model in a world-line representation, and may be used
to somewhat more rigorously argue for self-duality of the critical
theory in this case.

The results above can be interpreted physically as follows.  Let
us first consider the vortices in the $z$ condensate. These will
carry gauge flux that is quantized in units of $2\pi$. Such a
$2\pi$ flux can end at a space-time monopole. Hence monopoles act
as sources of the vortices of the $z$ field. The Berry phases
imply that these monopole events are quadrupled so that only
processes where four vortices disappear (or are created) together
are important in the continuum limit. Now if we forbid monopoles
by hand, then the usual duality arguments map the model to a
global XY model in terms of the vortex fields. The dual global
U(1) symmetry of this XY model is precisely associated with
conservation of vorticity. Including monopoles (which act as
sources for 4 vortices) introduces a four-fold anisotropy on this
global XY model. Such an anisotropy is irrelevant at the $3D$ XY
critical fixed point. Thus monopole events are again irrelevant
and (in the original representation) a theory where the $z$ boson
is coupled to a non-compact U(1) gauge field describes the
transition.

\subsection{SJ model at large $N$}
\label{sec:SJN}

Now let us consider $N$ large. In the limit $N \rightarrow \infty$
the gauge field is non-fluctuating and can be taken as a classical
`background' in which the $z$ particles move. The minimum energy
saddle point corresponds to $a_{\mu} = 0$ (up to gauge rotation).
The $z$ bosons are gapped and free in the paramagnetic state,
while they are condensed in the ordered state. Now consider the
nature of both states, and the transition, upon including
fluctuations in a $1/N$ expansion. It is useful to discuss the
effects of instantons separately from other fluctuations. Ignoring
instantons, the $1/N$ expansion proceeds along standard lines. In
the ordered state, the gauge fields acquire a mass by the usual
Anderson-Higgs mechanism. The gauge flux is quantized in units of
$2\pi$ - the associated point defects are the large-$N$ avatars of
the skyrmion described previously directly at $N = 2$. However, on
the paramagnetic side the gauge fields are gapless and describe a
`photon' which disperses linearly at low energies. The transition
is described by a field theory of $z$ bosons coupled minimally to
a non-compact U(1) gauge field. This transition is second order
with critical exponents that evolve continuously from their values
at $N = \infty$. In particular, consider the gauge invariant
physical spin operator (which is the appropriate generalization to
large-$N$ of the familiar N\'eel order parameter at $N = 2$). This
is bilinear in the $z$ fields. At $N = \infty$, the spin operator
therefore has a large anomoulous dimension $\eta = 1$. This will
acquire (calculable) corrections\cite{irkhin,starykh} of
$\mathcal{O} (1/N)$ upon considering finite but large $N$. Hence
$\eta$ will be large for large but finite $N$.

Now consider including instantons. It is important to realize that
the entire gauge action is of order $N$ in this theory.
Consequently,\cite{MS} the `bare' instanton core action, obtained
by integrating out the $z$ fields in the presence of a background
instanton configuration of the gauge fields in space-time, is of
order $N$. Thus the bare instanton fugacity is small
(exponentially small in $N$). In the ordered state, the inclusion
of instanton events means that point defects with quantized $2\pi$
flux are no longer stable. The physics in the paramagnetic state
is more interesting. Here the instantons proliferate and lead to
confinement of the gapped $z$ bosons. Furthermore, the gapless
photon (present in the non-compact model) is rendered unstable.
The Haldane Berry phases associated with the instantons lead to
lattice symmetry breaking. As explained in
Section~\ref{sec:models}, this follows from the observation that
the instanton operators transform non-trivially under lattice
symmetries. Hence if they acquire an expectation value, lattice
symmetry is broken.

Now let us consider the effect of instantons at the transition.
>From the discussion in preceding sections, it is clear that the
crucial question is whether the four-monopole event is
relevant/irrelevant at the fixed point of this non-compact model.
The scaling dimension of the $p$-monopole operator in this model
was computed by Murthy and Sachdev\cite{MS}.  For $p =4$, their
results give a scaling dimension $\propto N$. Hence the instantons
are strongly irrelevant \cite{noteN} for large $N$.

This then implies that the critical point of the non-compact
theory is stable to inclusion of instanton events, even though the
states on both sides of the critical point are qualitatively
changed. In particular, consider approaching the transition from
the paramagnetic side. The proliferation of instantons in the
paramagnetic state had two effects - to confine the spinons and to
produce VBS order. The irrelevance of the instantons at the
critical fixed point implies that both the VBS order and the
spinon confinement disappear at the transition. We note that as
the bare instanton fugacity is exponentially small in $N$, this
perturbative analysis of their relevance/irrelevance is sufficient
to determine the nature of the transition. In particular the
alternate possibility that there is a coexistence region with
width shrinking to zero as $N \rightarrow \infty$ appears unlikely
at large-$N$.

It is also useful to interpret the results above in the context of
other recent discussions of instantons in the
literature\cite{sudbo,igor,igorss,ssfoot}. The strategy of these,
and other works, is to integrate out the $z$ bosons, and to work
with an effective action for the gauge field. This action will be
of order $N$. Consequently, it seems reasonable to assume that the
gauge field dynamics is described to leading order in $1/N$ by a
Gaussian action (this is equivalent to the RPA approximation), and
to address issues of instanton physics within this Gaussian gauge
action. Such an approach will correctly describe the qualitative
physics of the paramagnetic state. For the critical point itself,
the form of the Gaussian gauge action is determined by scaling to
be \cite{sudbo,igor} (see also Eq.~(\ref{eq:amuprop}))
\begin{equation}
\label{eq:RPA} \mathcal{S}_G = \int \frac{d^3 K}{(2\pi)^3}
N\sigma_0 |K| \left| {\bf a}_T({\bf K})\right|^2
\end{equation}
where ${\bf K}$ is the Euclidean $3$-momentum, ${\bf a}_T$ refers
to the transverse part of the gauge field, and $\sigma_0$ is a
universal constant associated with the universal critical
conductivity of the $z$ bosons at the transition at $N = \infty$.
Note that this action is more singular than the usual `Maxwell'
action - this originates in the integration over the massless
critical modes of the $z_\alpha$ fields. The action for a single
instanton can be calculated within this Gaussian approximation,
and is of order $N \ln L$ where $L$ is the system size. This
suggests that instanton-anti-instanton pairs interact
logarithmically with each other. It also suggests that the effect
of instantons could be captured by studying a classical three
dimensional Coulomb gas of instantons with pairwise logarithmic
interactions. If this gas is in a plasma phase, free instantons
have proliferated. On the other hand, one might also conceive a
different phase where instanton-anti-instanton pairs are strongly
bound to each other. For the classical $3D$ Coulomb gas,
examination of this issue\cite{igor,igorss} has led to the
conclusion that the logarithmic interaction is screened at long
length scales into a short-ranged interaction, by bound instanton
pairs at shorter scales (however, it was noted\cite{igor,igorss}
that the screening could fail at fine-tuned critical points). This
screening then forces proliferation of free instantons, so that
the Coulomb gas is in a plasma phase. How are we to reconcile this
apparently general conclusion with our claim that the instantons
are suppressed at the N\'eel-VBS critical point?

This conundrum is resolved as follows. The Gaussian `RPA' action
does not properly account for the effects of highly non-linear
perurbations such as instantons. This is already clear from the
results of Ref.~\onlinecite{MS}. Within the Gaussian theory, the
action of a strength $p$ instanton scales with $p$ as $p^2$. This
would imply that the scaling dimension of the $p$ instanton
operator scale as $p^2$ - this disagrees with results of
Ref.~\onlinecite{MS}, which obtained a highly non-trivial
dependence on $p$. In other words, even in the large-$N$ limit,
the Gaussian action is not sufficient to correctly calculate the
scaling dimension of the instantons: the non-linear terms in the
gauge action all contribute in determining the instanton action
\cite{Murthy} even at $N=\infty$. More significantly, we can
likewise conclude that the RPA treatment of instanton interactions
by a simple pairwise interaction is inadequate. The true instanton
gas (even in the large-$N$ limit) has a rather specific structure
of higher order interactions, some of whose features are
universally determined by the fact that they arose from
integrating out particular gapless critical modes. If we attempt
to compute the screening of instanton interactions by integrating
out bound instanton-anti-instanton pairs, effects which
renormalize the screening length are intricately intertwined with
those that shift the position of the critical point between the
magnetic and paramagnetic phases. Indeed, fine-tuning to be at the
critical point between the magnetic and paramagnetic phases is all
that is needed to also suppress the instanton plasma phase, and
the `na\"ive' conclusion that the instantons are irrelevant at
this critical point \cite{MS} is correct.
Appendix~\ref{app:monoscr} considers a specific toy model for
which these arguments can be demonstrated explicitly.

As we noted at the end of Section~\ref{sec:intro}, the above
reasoning may also apply to fermionic models which have a line of
critical points\cite{igor,ssfoot}: in this case, the suppression
of instantons may occur along the entire line, and not just at an
isolated point.

\section{Spin models with easy plane anisotropy}
\label{sec:ep}

An alternate and particularly fruitful deformation of the model is
provided by the situation where there is some easy-plane anisotropy on
the underlying $SU(2)$ spin model.  Such an anisotropy tends to orient
the spins preferentially perpendicular to the $z$-axis in spin space.
Indeed precisely such an easy plane spin-$1/2$ model with both two
particle and four-particle ring exchanges has recently been studied
numerically\cite{sandvik}.  A direct transition between N\'eel and valence
bond solid phases was found.

Consider first the fate of the global symmetries in the presence
of easy plane anisotropy. A U(1) subgroup of symmetry of spin
rotations about the $z$-axis of spin still survives.  In addition
there are a number of discrete symmetries.  Either under a unit
translation or time reversal (see Table~\ref{tab:sym}), the N\'eel
vector changes sign
\begin{equation}
  \hat{n}_r \rightarrow -\hat{n}_r. \label{eq:StominusS}
\end{equation}
This may be combined with a U(1) spin rotation in the XY-plane
which restores the sign of $n^x,n^y$ to simply change the sign of
$n^z$ alone. Thus $n^z \rightarrow -n^z$ is a discrete symmetry in
the easy-plane case.

Easy plane anisotropy is readily incorporated into the non-linear
sigma model description in Eqs.~(\ref{eq:nlsm}) or
(\ref{eq:nlsm2}) as a term
\begin{equation}
  \mathcal{S}_{\rm ep} = - \int d\tau d^2 x ~w (n^z)^2
\end{equation}
with $w < 0$ (this is clearly related to Eq.~(\ref{eq:epterm})).
The global U(1) symmetry simply corresponds to a uniform rotation
of all the $\hat n$ vectors about the $z$-axis.

\subsection{Semi-classical analysis}
\label{sec:semi-class-analys}

Let us first think classically about this easy plane model. By
classical we mean to focus on time independent configurations of
the $\hat n$-field and to ignore the Berry phase effects. The
classical ground state simply consists of letting $\hat n$ be
independent of position and lie entirely in the spin XY plane.
Topological defects in this ground
state will play an important role.  With the easy plane
anisotropy, these are simply vortices in the field $n^+ = n_x +
in_y$. More precisely, on going around a large loop containing a
vortex the phase of $n^+$ winds around by $2\pi m$ with $m$ an
integer.

\begin{figure}
\centerline{\includegraphics[width=2.7in]{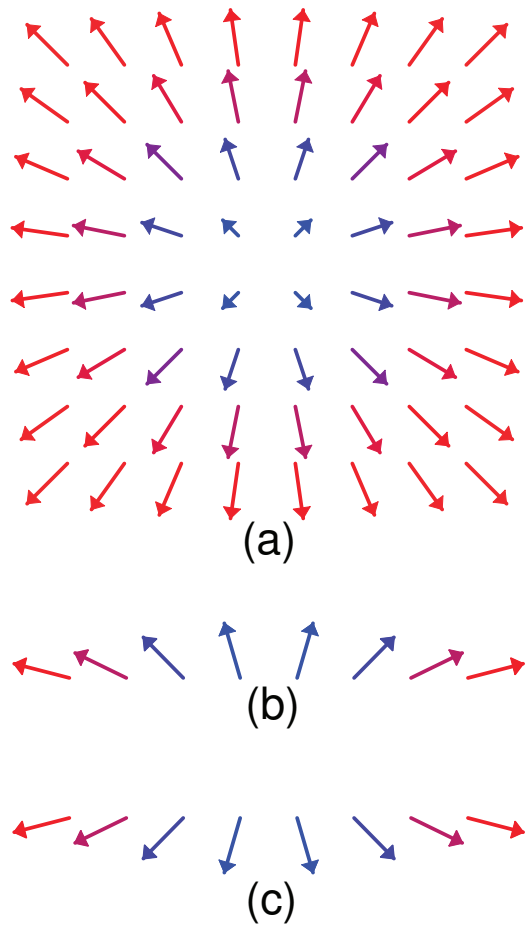}} \caption{The
`meron' vortices in the easy plane case. There are two such
vortices, $\psi_{1,2}$, and $\psi_1$ is represented in (a) and
(b), while $\psi_2$ is represented by (a) and (c), following the
conventions of Fig~\protect\ref{skyr}. The $\psi_1$ meron above
has $\hat{n} (r=0) = (0,0,1)$ and $\hat{n} ( |r| \rightarrow
\infty ) = (x,y,0)/|r|$; the $\psi_2$ meron has $\hat{n} (r=0) =
(0,0,-1)$ and the same limit as $|r| \rightarrow \infty$. Each
meron above is `half' the skyrmion in Fig~\protect\ref{skyr}: this
is evident from a comparison of (b) and (c) above with
Fig~\protect\ref{skyr}b. Similarly, one can observe that a
composite of $\psi_1$ and $\psi_2^\ast$ makes one skyrmion.}
\label{mer}
\end{figure}

What is the nature of the core of these vortices? In the core the
XY order will be suppressed and the $\hat n$ vector will point
along the $\pm \hat z$ direction. In terms of the microscopic spin
model, this corresponds to a non-zero staggered magnetization of
the $z$ component of the spin in the core region.  Thus at the
classical level there are two kinds of vortices depending on the
direction of the $\hat n$ vector at the core (see Fig.~\ref{mer}).
Note that either kind of vortex {\em breaks} the Ising-like $n^z
\rightarrow -n^z$ symmetry at the core.

Clearly this breaking of the Ising symmetry is an artifact of the
classical limit - once quantum effects are included, the two broken
symmetry cores will be able to tunnel into each other and there will be
no true broken Ising symmetry in the core.  This tunneling is often
called an `instanton' process that connects two classically degenerate
states.

Interestingly such an instanton event is physically the easy-plane
avatar of the space-time monopole described above for the fully
isotropic model. This may be seen pictorially. Pictorially each
classical vortex really represents half of a skyrmion
configuration. Such half-skyrmions are known as merons. As shown
in Fig.~\ref{mer}, the tunneling process between the two merons is
equivalent to creating a full skyrmion. This is precisely the
monopole event.

Now the results of Haldane imply once again that (in the continuum)
monopole events are quadrupled. Thus the only tunneling processes that
survive in the continuum limit are those in which four merons with core
spins along one direction come together and collectively flip the
orientation of their core spins to produce four merons of the opposite
kind.

\subsection{Easy plane in the CP$^1$ representation}
\label{sec:epcp1}

It is extremely useful to also consider easy plane anisotropy in
the framework of the CP$^1$ representation. In this
representation, the easy plane anisotropy was already presented in
Eq.~(\ref{eq:epterm}). Let us first translate the above classical
thinking into the CP$^1$ representation.  Suppose for this purpose
$w$ is negative but small, so that states with $|z_1|=|z_2|$ are
favored but not rigidly enforced. Clearly, the preferred uniform
classical ordered states satisfy
\begin{equation}
|\langle z_1\rangle| = |\langle z_2\rangle|  \neq 0,
\end{equation}
so that $n^+ = z^*_1 z^{\vphantom{*}}_2$ is ordered {\em and}
there is no average value of $n^z = |z_1|^2 - |z_2|^2$.  Now
consider vortex configurations.  Clearly a full $2\pi$ vortex in
$n^+$ can be achieved by either having a $2\pi$ vortex in $z_1$
and not in $z_2$ or a $2\pi$ anti-vortex in $z_2$ and no vortex in
$z_1$.  Far from the vortex core both fields will have equal
amplitude, but in the first choice the amplitude of the $z_1$
condensate will be suppressed at the core but $\langle z_2\rangle$
will be unaffected. Consequently $n^z = |z_1|^2 - |z_2|^2$ will be
non-zero and negative in the core.  The other choice also leads to
non-zero $n^z$ which will now be positive. Thus we may identify
the two kinds of meron vortices with $2\pi$ vortices in the spinon
fields $z_{1,2}$ respectively.

To explore this analytically, we consider the behavior now {\em
deep} in the easy plane limit, in which $n^z =
|z_{i1}|^2-|z_{i2}|^2 \approx 0$. Together with the CP$^1$
constraint $|z_{i1}|^2+|z_{i2}|^2=1$, this implies fixed magnitude
for each component of $z_i$, so we may write
\begin{equation}
  z_{i\alpha} \sim \frac{1}{\sqrt{2}}e^{i\phi_{i\alpha}},
\end{equation}
where $\phi_{i\alpha} \in [0, 2\pi)$ is the phase of the spinon
field. The ``kinetic'' term of the SJ model action in
Eq.~(\ref{eq:SJ1}) is then modified to
\begin{equation}
\mathcal{S}_z  =  -t\sum_{\ell,\alpha}\cos
\left({\boldsymbol\Delta} \phi_{\alpha} - {\bf a} \right) ,
\label{eq:SJ21}
\end{equation}
with the other terms ($\mathcal{S}_a,\mathcal{S}_B$) given as
before in Eqs.~(\ref{eq:SJ2}) and (\ref{eq:SJ3}).

It is very useful both for further insight and for concrete
calculations to explore a `dual' representation which focuses on
the meron vortex fields.  Although the form of the dual action is
dictated completely by the general considerations of the previous
subsection, we provide an explicit derivation in
Appendix~\ref{app:SJ2} by proceeding as in Section~\ref{sec:SJ1}
with the duality transformation.  We obtain the dual action
$\mathcal{S}_{\rm dual} = \int d^2 r d \tau \mathcal{L}_{\rm
dual}$ with
\begin{eqnarray}
\label{crtny} \mathcal{L}_{\rm dual}& = & \sum_{a =
1,2}|\left(\partial_{\mu} - iA_{\mu}\right)\psi_{a}|^2 + s_d
|\psi|^2 + u_d\left(|\psi|^2 \right)^2 \nonumber \\
&& \hspace{-0.6in} + w_d |\psi_1|^2 |\psi_2|^2 +
\kappa_d\left(\epsilon_{\mu\nu\kappa}\partial_{\nu}A_{\kappa}\right)^2
 - \lambda {\rm Re}[\left(\psi^*_1 \psi^{\vphantom{*}}_2 \right)^4].
\end{eqnarray}
We have used the short-hand notation $|\psi|^2 \equiv |\psi_1|^2 +
|\psi_2|^2$. Here $\psi_{1,2}$ denote quantum fields that destroy
meron vortices whose core points in the up direction for $\psi_1$
and down for $\psi_2$.

We now show how the dual action of Eq.~(\ref{crtny}) can be
understood entirely on general grounds. As usual in dual theories,
the net vorticity is conserved, corresponding to the overall U(1)
symmetry of Eq.~(\ref{crtny}). This symmetry is gauged by the
non-compact vector potential $A_{\mu}$, as usual in dual
descriptions of two dimensional bosonic systems. Physically, the
gauge field is required to embody spin ($S^z$) conservation of the
original model, $j_\mu=\epsilon_{\mu\nu\lambda}\partial_\nu
A_\lambda/\pi$ being the $3$-current of $S^z$.  Hence the dual
magnetic and electric fields correspond to the spin density and
spin current, respectively. Minimal coupling of the vortex fields
to $A_\mu$ also gives them proper logarithmic interactions and
magnus force dynamics.

Clearly under the discrete Ising-like $n^z\rightarrow -n^z$ symmetry, the two
vortices get interchanged, {\em i.e} $\psi_1 \rightarrow \psi_2$ and
vice-versa.  The dual action must therefore be invariant under
interchange of $1$ and $2$ labels.

Finally, if monopole events were to be completely ignored ({\em
i.e} disallowed by hand) the total skyrmion number must be
conserved.  As is apparent from Fig~\ref{mer}, a composite of a
$\psi_1$ vortex and a $\psi^*_2$ antivortex is precisely a
skyrmion configuration of the $\hat n$ field. Thus we may view
skyrmion number conservation as the conservation of the {\em
difference} of the total number of either species of vortices.
This implies the global U(1) symmetry
\begin{eqnarray}
\label{dglobu1}
\psi_1 & \rightarrow & \psi_1 \exp{(i\varrho)} \nonumber \\
\psi_2 & \rightarrow & \psi_2 \exp{(-i\varrho)}
\end{eqnarray}
where $\varrho$ is a constant.

As discussed at length above, monopole events destroy the
conservation of skyrmion number, and hence this dual global U(1)
symmetry. However as the monopoles are effectively quadrupled due
to the Berry phase terms, skyrmion number is still conserved (mod
$4$). Thus the dual global U(1) symmetry must be broken down to
$Z_4$.

The dual Lagrangian $\mathcal{L}_{\rm dual}$ in Eq. (\ref{crtny})
is the simplest one that is consistent with all these
requirements. In particular, we note that at $\lambda=0$ the dual
global U(1) transformation in Eq.~(\ref{dglobu1}) leaves the
Lagrangian invariant.  The $\lambda$ term breaks this down to
$Z_4$ as required. Thus we may identify $\lambda$ as the fugacity
of the (quadrupled) monopole tunneling events, $\lambda \sim
\lambda_4$ in Eq.~(\ref{eq:smono}).

Actually this action was first derived by completely different
means in Refs.~\onlinecite{Crtny} and~\onlinecite{sp}. The
discussion above is however more directly physical, and gives an
interpretation of the $\lambda$ term and of the other symmetries
of this dual action.

An important coupling constant in the above dual action is $w_d$,
which appears as a sort of ``anisotropy''.  The exact lattice
duality in the appendix in fact leads to a ``hard spin'' (rotor)
model in which $|\psi_1|=|\psi_2|=1$.  The above continuum theory
is arrived by ``softening'' this constraint.  However, it is clear
that the appropriate sign of $w_d$ to connect microscopically to
the original SJ model is $w_d<0$ (and large).  While the above
symmetry arguments do not specify this sign, the model with
$w_d>0$ presumably corresponds to rather different physics, and
has no clear connection to the original SJ model. See the
discussion in Sec.~\ref{sec:ising-anisotropy} for a possible physical
application of this case.

Note that apart from the $\lambda$ term,
Eq.~(\ref{crtny}) has exactly the same structure as the continuum
CP$^1$ theory $\mathcal{L}_z$ in Eq.~(\ref{sz}) in the presence of the easy
plane anisotropy Eqn. ~(\ref{eq:epterm}). As we will
discuss below, the $\lambda$ term (which represents the quadrupled
monopole tunneling events) are irrelevant at the QCP:
consequently, the QCP has a self-dual structure.

We should further note that there is no connection between $w_d$
in the vortex action and the analogous parameter $w$ in the
continuum limit of the SJ model.  The latter is clearly simply
related to the physical spin anisotropy, corresponding for $w<0$
to easy-plane and for $w>0$ to easy-axis anisotropy.  The point
$w=0$ describes the SU(2) invariant magnetic QCP.  This {\em does
not} correspond to $w_d=0$ in the dual theory.  Indeed, there may
be no dual theory whatsoever for any but the easy plane case
(though see Sec.~\ref{sec:ising-anisotropy}).

\begin{table}[htbp]
  \centering
  \begin{ruledtabular}
  \begin{tabular}{l|c|c}
   Field & CP$^1$ & dual \\
    \hline
    $N^+$ & $z^\dagger \sigma^+ z^{\vphantom\dagger}$ & $e^{i\int
      E_j  {\cal A}_j }$ \\
    $\psi_{\rm VBS}$ & $e^{i\int
      e_j  {\cal A}_j }$ & $\psi^\dagger \sigma^+
    \psi^{\vphantom\dagger}$ \\
    $N^z$ & $z^\dagger \sigma^z z^{\vphantom\dagger}$ & $\psi^\dagger \sigma^z
    \psi^{\vphantom\dagger}$ \\
    $M^+$ & $i z^\dagger \sigma^+ (\tensor\partial_0-i a_0)
    z^{\vphantom\dagger} $ & $e^{i\int
      E_j  {\cal A}_j }\psi^\dagger \sigma^z
    \psi^{\vphantom\dagger}$ \\
    $M^z$ & $i z^\dagger \sigma^z (\tensor\partial_0-i a_0)
    z^{\vphantom\dagger} $ & $\epsilon_{ij} \partial_i A_j /\pi$ \\
    ${j}_i^z$ & $i z^\dagger \sigma^z (\tensor\partial_i-i a_i)
    z^{\vphantom\dagger} $ & $\epsilon_{ij}
    (\partial_0 {A}_j-\partial_j A_0)/\pi$ \\
    $j_\mu^{v}$ & vorticity &
    $i\psi^\dagger (\tensor\partial_\mu-i
    A_\mu)\psi^{\vphantom\dagger}$ \\
    $j^{\scriptscriptstyle G}_\mu$ & $\epsilon_{\mu\nu\lambda}
    \partial_\mu {a}_\lambda/\pi$ &
    $i\psi^\dagger \sigma^z (\tensor\partial_\mu-i
    A_\mu)\psi^{\vphantom\dagger}$ \\
  \end{tabular}
  \end{ruledtabular}
  \caption{  \label{tab:dual} Operators in the easy plane CP$^1$ (column 2) and
    dual (column 3) representations corresponding to physical operators
    (column 1), in the notations of Sec.~\ref{sec:models}.   Here
    we have introduced the classical gauge field configuration for a
    unit point flux, with $\epsilon_{ij}\partial_i {\cal A}_j(x) =
    2\pi \delta^2(x)$. The symbol $e_j$ represents the $j$- componenet of the
electric field operator that corresponds to the
gauge field in the $CP^1$ representation.  We have also used the symbol
    $\tensor{\partial}_\mu$, defined by $f \tensor{\partial}_\mu g =
    \frac{1}{2}[f \partial_\mu g - (\partial_\mu f) g]$.  The symbol
    $j_i^z$ is the current of conserved
    magnetization, while $j_\mu^v$ and $j^{\scriptscriptstyle G}_\mu$
    are the three-currents of vorticity and gauge flux, respectively.}
\end{table}
A list of the representation of physical operators of interest in
the original and dual representations is given in
Table~\ref{tab:dual}.  In the dual vortex theory, the XY ordered
phase is simply characterized as a dual `paramagnet' where both
the $\psi_{1,2}$ fields are gapped. On the other hand, true spin
paramagnetic phases correspond to condensates of the fields
$\psi_{1,2}$, which break the dual gauge symmetry. In particular
if both $\psi_1$ and $\psi_2$ condense with equal amplitude
$\langle\psi_1\rangle = \langle\psi_2\rangle \neq 0$, then a
paramagnetic phase where the global Ising symmetry is preserved
results. Note the strong similarity between the description of the
phases in this dual theory with that in terms of the spinon fields
of the CP$^1$ representation if we interchange the role of the XY
ordered and paramagnetic phases. This is a symptom of an exact
duality between the two descriptions that obtains close to the
transition.  At this point, the two descriptions do not appear
wholly identical, due to the $Z_4$ symmetry-breaking term
$\lambda$ not present in the CP$^1$ theory, and the compactness of
the CP$^1$ gauge field not present in the dual one. As argued
above, the two differences represent one and the same physics,
since the vortex tunneling events generated by $\lambda$ represent
exactly the non-oscillating four-monopole events allowed in the SJ
model.  In the next section, we will argue that these events are
irrelevant in the scaling limit near the QCP, making the duality
between the two descriptions complete.

As indicated in Table~\ref{tab:dual}, the combination $\psi_{\rm
VBS}\sim \psi^*_1\psi_2^{\vphantom{*}}$ serves as the order
parameter for the translational symmetry broken VBS ground state.
This may be seen from the analysis of Refs.~\onlinecite{Crtny,sp}.
Alternately this may be seen by the identification described in
Sec.~\ref{sec:models} of the skyrmion creation operator with the
order parameter for translation symmetry breaking.  Such a
condensate of $\psi_{1,2}$ breaks the global $Z_4$ symmetry of the
action in Eq.~(\ref{crtny}).  The preferred phase of $\psi_{\rm
VBS}$ depends on the sign of $\lambda$, the two inequivalent sets
of preferred directions corresponding to columnar and plaquette
patterns of translational symmetry breaking.

\section{Phase transitions}
\label{sec:phase}

\subsection{Easy plane limit}
\label{sec:phase-ep}

Consider the dual vortex action in Eq.~(\ref{crtny}). In mean
field theory the transition happens when the parameter $s_d$
becomes smaller than zero and can clearly be second order.
Fluctuation effects will modify the mean field behavior in
important ways. Consider first the properties of the transition
when $\lambda = 0$, {\em i.e} in the absence of instanton events.
The resulting model has recently been studied in
Ref.~\onlinecite{mv}. Remarkably, as argued there, the model has
the property of being self-dual - the ordered and paramagnetic
phases get interchanged under the duality transformation.  To
understand this first note that in the $\lambda = 0$ limit, the
dual action Eq.~(\ref{crtny}) has precisely the same structure as
an easy-plane CP$^1$ model with a {\em non-compact} U(1) gauge
field (as in Eq.~(\ref{sz})). As this same limit actually
corresponds to disallowing all monopole events, in the spinon
description we must work with a {\em non-compact} gauge field.
Then the exact same field theory obtains both in terms of the
spinon fields $z$ (in Eq.~(\ref{sz})) and in terms of the meron
vortices $\psi$ in the easy plane limit (in Eq.~(\ref{crtny})) and
ignoring instantons.

It was established in Ref.~\onlinecite{mv} via numerical Monte
Carlo simulations that a continuous ordering transition exists in
this model with the non compact gauge field. The fixed point
controlling this transition in this limit is therefore described
by a self-dual field theory. Note that in either representation
the natural fields of the theory are not those associated with the
`physical' boson operator (either $n^+$ or the skyrmion creation
operator). Rather the theory is expressed most simply in terms of
`fractionalized' fields - namely the spinons or the meron
vortices. In particular, the physical $n^+$ field is a composite
of two spinon fields and likewise the skyrmion field is a
composite of the two meron fields.

Let us now imagine including instanton events. This is most easily
accessed in the vortex representation where it simply amounts to
letting $\lambda \neq 0$.  This is the main advantage of the dual
representation -- the non-trivial non-local effect of instantons
is represented as a simple local perturbation in the dual theory.
We may now address the question of relevance/irrelevance of
instantons at the $\lambda = 0$ fixed point. This is determined by
the scaling dimension $\Delta$ of the $\left(\psi^*_1
\psi_2^{\vphantom{*}}\right)^4=\psi_{\rm VBS}^{(4)}$ operator, in
principle determined from the two-point correlation function of
this operator in the (non-trivial) theory with $\lambda=0$:
\begin{equation}
  \label{eq:deltadef}
  \left\langle \psi_{\rm VBS}^{(4)}(x)
    \psi_{\rm VBS}^{(4)*}(x') \right\rangle_{\lambda=0}
  \sim \frac{1}{|x-x'|^{2\Delta}},
\end{equation}
where $x,x'$ are space-time coordinates.  Hence $\Delta$ is
determined by the correlations of the fourth power of the physical
VBS order parameter, and one requires $\Delta>D=3$ for
irrelevance.  Being self-dual, the same anomalous dimension should
be ascribed to the correlations of the physical boson (XY order
parameter).  The $\lambda = 0$ critical fixed point describes an
XY ordering transition where the physical boson field is a
composite of the fundamental fields of the theory.  We therefore
expect that correlators of the physical boson (and its various
powers) will decay with an anomalous dimension that is {\em
  larger} than the corresponding one for the ordinary XY transition in
$D = 2 + 1$ dimensions.  Now for the usual XY fixed point
four-fold symmetry breaking perturbations are known to be {\em
irrelevant}, {\em i.e.\/} have a scaling dimension $\Delta_4>3$.
This then implies that a small $\lambda$ will be irrelevant by
power counting at the $\lambda = 0$ fixed point of the present
model as well (see also Appendix \ref{app:est}).

This latter expectation can be checked in an appropriate large $N$
generalization.  In particular, consider the non-compact gauge
theory with Lagrangian
\begin{eqnarray}
  \label{eq:largeNxy}
  \mathcal{L} &= &  \sum_{i=1}^{2N} [|(\partial_\mu -i
    A_\mu)\psi_i|^2 +r|\psi_i|^2 + u |\psi_i|^4] \nonumber \\
  &+ &  \kappa N (\epsilon_{\mu\nu\lambda} \partial_\nu A_\lambda)^2
  -\lambda\sum_{i=1}^{N}[ (\psi_i^*\psi_{i+N}^{\vphantom{*}})^4 +
  {\rm c.c.}] ,
\end{eqnarray}
where the $U(1)\times U(1)$ symmetry (for $\lambda=0$) of the dual
action has been elevated to a $U(1)^{2N}$ invariance (under
independent phase rotations of each $\psi_i$ field).  Of this,
only the single U(1) subgroup of identical rotations of all $2N$
fields has been gauged with the non-compact gauge field $A_\mu$.
The $\lambda$ term breaks the $U(1)^{2N}$ symmetry to $U(1)^N$, of
which $N-1$ are global, and the single gauge U(1) is preserved. In
addition, there is a residual global $Z_4^N$ symmetry under
$\psi_j \rightarrow e^{i n_j
  \pi/4} \psi_j$, $\psi_{j+N} \rightarrow e^{-i n_j \pi/4} \psi_{j+N}$,
with $n_j \in \{0,1,2,3\}$, for $j=1\ldots N$.

In the $N=\infty$ limit, the theory may be analyzed by saddle point
methods.  In particular, consider for simplicity the partition function
with $\lambda=0$, which may be formally written
\begin{eqnarray}
  \mathcal{Z} &=& \int [dA]\, \exp\Bigl\{ -N[2 \mathcal{S}^{\rm eff}_{XY}(A_\mu)
\nonumber \\
  &~&~~~~~~~~~~+ \int \!
    d^2 r d \tau \, \kappa (\epsilon_{\mu\nu\lambda} \partial_\nu
    A_\lambda)^2 ]\Bigr\},  \label{eq:pf}
\end{eqnarray}
where
\begin{eqnarray}
  \mathcal{S}_{XY}^{\rm eff}  &=& -\ln \int [d\psi] \exp\Biggl\{ -\int\! d^2
r d\tau \Bigl[ \nonumber \\
    &~&~|(\partial_\mu -i
    A_\mu)\psi|^2 +r|\psi|^2 + u |\psi|^4 \Bigr] \Biggr\}   \label{eq:seffxy}
\end{eqnarray}
is the effective action of the $D=3$ XY model as a functional of
$A_\mu$.  Formally, at $N=\infty$, from Eq.~\ref{eq:pf}, a saddle
point approximation in $A_\mu$ is justified, with the saddle point
value being zero, $A_\mu^*=0$.  The relevance of $\lambda$ is then
determined by the two-point function of $\psi^{(4)}_{VBS;N}
=\sum_{i=1}^N (\psi_i^*\psi_{i+N}^{\vphantom{*}})^4$ in the saddle
point theory with $A_\mu=0$.  Since in this theory the $\psi_i$
are decoupled XY fields (fluctuating according to the non-trivial
$3D$ XY fixed point), one has then
\begin{eqnarray}
  \label{eq:deltadef1}
&&   \left\langle \psi_{{\rm VBS};N}^{(4)}(x)
    \psi_{{\rm VBS};N}^{(4)*}(x') \right\rangle_{\lambda=0}
  \\
  && \sim N \left|\left\langle (\psi^{*}(x))^4 (\psi(x'))^4
    \right\rangle_{3DXY}\right|^2\sim
  \frac{N}{|x-x'|^{4\Delta_4}}. \nonumber
\end{eqnarray}
The final expression obtains since the expectation value in the
second line is none other but the two-point function of the
four-fold symmetry breaking field at the $D=3$ XY fixed point.
Hence, one has $\Delta=2\Delta_4$, implying $\Delta>6$ since
$\Delta_4>3$.  Thus in this limit the ``monopole'' (symmetry
breaking) terms are strongly irrelevant.  We note in passing that
the irrelevance of $\lambda$ can also be established by working
for $\lambda\neq 0$, and taking the large $N$ limit.  The saddle
point remains at $A_\mu=0$, and irrelevance follows simply from
the irrelevance of ``bi-quartic'' coupling between two $3D$ XY
models by their four-fold symmetry breaking fields.

We conclude that a direct second order transition with irrelevant
instanton tunneling events is possible in this easy plane case.  Note
the crucial role played by the Berry phase term for the instantons in
reaching this conclusion. Indeed it was the Berry phases that forced
quadrupling of instantons thereby increasing their scaling dimension and
making it possible for them to be irrelevant.

While the $\lambda$ term may be irrelevant at the critical fixed point
it is clearly very important in deciding the fate of either phase. In
particular in the paramagnetic phase it picks out the particular pattern
of translation symmetry breaking (columnar versus plaquette) and forces
linear confinement of spinons.  In critical phenomena parlance, it may
be described as a {\em dangerously irrelevant} perturbation.



\subsection{Isotropic magnets}

In the context of the SJ models, the results of previous sections
show that the $N = 1$, $N = \infty$, and easy plane $N = 2$ models
all provide the same picture. A direct second order transition
between the $z$ condensed and VBS phases is possible with a
`deconfined' critical point. Right at this point, monopole
tunneling events become irrelevant and spinon degrees of freedom
emerge as the natural fields of the critical theory. This provides
strong evidence that the same thing happens for the $SU(2)$
symmetric model ({\em i.e} at $N = 2$).

What then is the proposed description of the critical point in the
$SU(2)$ symmetric model? This is simply the CP$^1$ model with a
non-compact gauge field and no Berry phase terms in
Eq.~(\ref{sz}). Equivalently it may be thought of as the critical
point of the $D = 3$ classical O(3) model when monopoles have been
forbidden by hand. This transition was first studied by Kamal and
Murthy\cite{KM} and more recently by Motrunich and
Vishwanath\cite{mv}, where it was established that a continuous
transition indeed exists that is different from the Heisenberg
transition. The non-compact $CP^1$ theory Eqn.~(\ref{sz}) was also directly
studied via
numerical Monte Carlo methods and found to possess a continuous
transition with the same universal properties as the monopole
suppressed O(3) NL$\sigma$M.  Numerical results for exponents
associated with several observables are available.  Further
evidence for the continuous nature of the transition in the CP$^1$
model coupled to a noncompact gauge field is obtained by
considering the larger class of models with CP$^{N-1}$ fields
coupled to a noncompact gauge theory. It is well known that the
$N=1$ model has a continuous transition \cite{dasgupta} which is
dual to the XY transition, and a continuous transition is also
expected for large values of $N$. Thus, the model of interest
$N=2$ is sandwiched between these two extremes where a continuous
transition is well known to obtain.

\section{Physical properties near the `deconfined' critical point}
\label{sec:phys-prop-near}

We now discuss the consequences of the theory for the physical
properties near the direct N\'eel-VBS transition. We will first discuss
those properties that follow generally from the (dangerous) irrelevance
of monopoles. Later we will specialize to the easy plane limit where the
self-duality enables more progress.

It is useful to think first about the various length scales in the
problem in the VBS phase.

First there is the spin-spin correlation length $\xi$ which will
diverge at the transition. Second, there is a length scale
$\xi_{\rm VBS}$ associated with the `thickness' of the domain
walls of the (discrete) VBS order. The latter is clearly
determined by the strength of the quadrupled monopole operator,
$\lambda \equiv \lambda_4$, in Eq.~(\ref{eq:smono}); in the easy
plane case, $\lambda$ appears as the co-efficient of a local term
in the dual action Eq.~(\ref{crtny}). These two length scales will diverge
differently - the domain wall thickness will diverge faster than
the spin-spin correlation length.  One can determine the scaling
of $\xi_{\rm VBS}$ with $\xi$ by a matching argument. On scaling
grounds, one expects
\begin{equation}
  \label{eq:xiscaling}
  \xi_{\rm VBS} \sim \xi f(\lambda \xi^{3-\Delta}),
\end{equation}
where $f$ is a scaling function, and $3-\Delta$ is the RG
eigenvalue of $\lambda$ assuming the scaling dimension of the four
monopole operator is $\Delta>3$, and $d=2,z=1$.  Beyond the scale
of the correlation length $\xi$, one can regard the VBS phase as
XY ordered in $\psi_{\rm VBS}$, though with very weak four-fold
anisotropy since $\lambda$ is irrelevant at the QCP.  Hence the
low-energy variations of the phase $\theta$ of $\psi_{\rm VBS}
\sim |\psi_{\rm VBS}|e^{i\theta}$ are described as a
pseudo-Goldstone mode, with energy
\begin{equation}
  \label{eq:pg}
  E(\theta) = \int \! d^2x\, \left[\frac{\tilde{K}}{2}|\nabla \theta|^2 -
  \tilde\lambda \cos 4\theta \right],
\end{equation}
where $\tilde{K}$ and $\tilde\lambda \propto \lambda$ are
renormalized parameters on the scale of $\xi$.  A twist of
$\theta$ (of e.g. $\pi/2$) is carried hence by a domain wall
which, by dimensional analysis, has width $\xi_{\rm VBS} \sim
\sqrt{\tilde{K}/\tilde{\lambda}}$.  Knowing then that $\xi_{\rm
VBS} \sim \lambda^{-1/2}$, one requires that $f(x)\sim x^{-1/2}$
in Eq.~(\ref{eq:xiscaling}), which implies \cite{csy}
\begin{equation}
  \label{eq:xispvsxi}
  \xi_{\rm VBS}\sim
  \xi^{(\Delta-1)/2}.
\end{equation}
Since $(\Delta-1)/2>1$, $\xi_{\rm VBS}$ indeed grows more rapidly
than $\xi$ as the QCP is approached.

Thus there are two independent diverging length scales. Either of
these length scales may be given several different
interpretations.  For instance, the spin correlation length $\xi$
may also be interpreted as the length scale at which correlations
of the dual global order parameter crossover from that of the
critical fixed point to that of the (unstable) fixed point which
breaks the dual global continuous symmetry. Similarly, the domain
wall thickness $\xi_{\rm VBS}$ of the VBS order is also the length
scale at which the photon that couples to the spinons acquires a
mass due to instanton effects.  This is also the length beyond
which the logarithmic Coulomb potential between spinons crosses
over to a linear confining one.  This is distinct from the
`confinement' length scale describing the spatial size of the
resulting two-spinon bound states (triplons).  This length scale
is actually a non-trivial combination of the two other diverging
length scales. It however diverges faster than the spin correlation length
$\xi$.

Note that the critical theory is isotropic in space-time and
therefore has dynamic scaling exponent $z = 1$. The values of
other critical exponents may be obtained from the numerical work
of Ref.~\onlinecite{mv}. In the $O(3)$ symmetric case the
correlation length exponent $\nu \approx 1$ (for $\xi$), and the
N\'eel order parameter exponent $\beta \approx 0.80$. Perhaps most
remarkably the anamolous dimension of the N\'eel order parameter
field $\eta$ is large ($\approx 0.6$). This should be contrasted
with the extremely small value for $\eta$ at the usual
Wilson-Fisher $O(3)$ fixed point in $D = 3$ dimensions (and indeed
for most other familiar three dimensional critical points). The
large value of $\eta$ can be rationalized by the thinking that the
N\'eel order parameter field decays into spinons right at the
critical point. Indeed as argued in previous sections it is the
spinons which appear as the more natural degrees of freedom at the
deconfined critical point. We note however that the spinons are
not to be considered `free particles' - they are critical and
furthermore interact through the coupling to the non-compact gauge
field.

Consider the effect of twisting the boundary conditions on the VBS
order -- for instance, for columnar dimerization prefer even
columns at one boundary and odd columns at the opposite boundary.
Let us suppose the twist is applied between the top and bottom
ends ($y=0,W$) of an $L\times W$ sample.  On general grounds, the
energy cost at long scales will be $E \sim \sigma L^{d-1}=\sigma
L$, where $\sigma$ is a ``surface tension'' or domain wall energy
per unit length.  This surface tension is set, however, by the
irrelevant monopole term and vanishes in a manner set by the
divergence of the domain wall thickness.  In particular, the
surface tension scaling obtains only for twists sustained over a
distance $W\gtrsim \xi_{\rm VBS}$.  For twists of the VBS order
over a shorter distance $W$ such that $\xi_{\rm VBS} \gg W \gg
\xi$, the energy cost for this twist is greatly reduced to $E\sim
K L/(2W)$, where $K$ is the ``stiffness'' associated with the
continuous dual global symmetry (and we are at length scales where
the system has not realized this symmetry is actually discrete).
The two energy costs for the twist become comparable for $W\sim
\xi_{\rm VBS}$, so that one expects $\sigma \sim K/\xi_{\rm VBS}$.
This stiffness itself vanishes upon approaching the quantum
critical point in a manner set by the divergence of the spin
correlation length.  Furthermore, the corresponding exponent is
the same as for the spin stiffness on the other side of the
transition. Specifically, the VBS stiffness $K \sim \xi^{2-d-z}$
where $\xi$ is the dual correlation length, $d = 2$ is the spatial
dimension, and $z = 1$ is the dynamic critical exponent.  Thus $K
\propto 1/\xi$.

Note that this is {\em not} a test of self-duality but rather a
test of the irrelevance of monopoles: the scaling of the VBS
stiffness is a consequence of dual current conservation which
obtains if monopoles are irrelevant. Thus the same behavior is
also expected for the isotropic model.

In practice, a measurement of the domain wall energy in the
columnar state is likely best obtained by simply comparing
energies of systems of size $L\times W$ (in the $x$ and $y$
directions, respectively) with periodic boundary conditions in
both directions ({\em i.e.\/}  on the torus) and varying $W$.  In
particular, let us consider $W > L$, with $L$ odd.  In this case,
the columns will prefer to align along the short direction ({\em
i.e.\/} columns parallel to the $x$ axis, breaking translational
symmetry along $y$) in order to avoid introducing a domain wall
(which would be required in the other orientation due to the odd
$L$) with energy cost $\sigma W$.  If $W$ is odd, there will still
be a domain wall required, but it is shorter and less costly, with
energy $\sigma L$.  Hence one expects the ground state energy of
the system to be $E_{LW} \approx \epsilon W L$ for $W$ even, and
$E_{LW} \approx \epsilon W L + \sigma L$ for $W$ odd.  Here
$\epsilon$ is the ground state energy density.  Hence the surface
tension can be obtained by
\begin{equation}
  \label{eq:oddeven}
  E_{L,W+1}-2E_{L,W}+E_{L,W-1} \sim 2\sigma L(-1)^W ,\qquad L \; {\rm odd}.
\end{equation}
This behavior will obtain provided $W >L \gg \xi_{\rm VBS}$.  When
the system is smaller than the domain wall thickness, however,
this energy is determined instead by the stiffness, {\em i.e.\/}
\begin{equation}
  \label{eq:oddeven1}
  E_{L,W+1}-2E_{L,W}+E_{L,W-1} \sim K\frac{L}{W} (-1)^W ,\qquad L \; {\rm odd},
\end{equation}
for  $\xi \ll L<W \ll \xi_{\rm VBS}$.

It is also clear that at the critical point, both columnar dimer
and plaquette order parameters will have power law correlators
with the same exponent. This is independent of which one of these
two phases we eventually end up in.  This is because both order
parameters are contained in the dual global boson creation
operator. They correspond to the phase of this boson locking in
different directions.  Which one of these phases is selected is
determined by the sign of the anisotropy term, but as this is
renormalizing to zero at the critical fixed point, there will be
no distinction between the two order parameter correlations at the
critical point.

This suggests the following interesting numerical check. Consider,
for instance, the situation where the disordered phase has
columnar VBS order.  Now consider measuring the stiffness to {\em
plaquette} order in this columnar phase.  Since this order is not
spontaneous in the VBS phase, measuring this stiffness cannot be
accomplished as above by simply comparing systems with odd and
even lengths. Instead, one should imagine introducing e.g. two
rows separated by half $W$ on which the magnetic couplings have
been increased or decreased in a pattern mimicking the strong
bonds of the plaquette state.  The ground state energy of this
system should be compared with that obtained by shifting one of
these rows by one lattice spacing, and the difference of these two
energies interpreted as the energy cost for a twist of the
plaquette order parameter. Deep in the phase, this energy cost
will be exponentially small, $\Delta E \sim L e^{-W/\xi_{\rm
VBS}}$.  However, in the regime $\xi_{\rm VBS} \gg W \gg \xi$ the
cost for the twist of the plaquette order parameter will be
determined by the stiffness, {\em i.e.\/} $\Delta E \sim K_{pl}
L/(2W)$. There will thus be a dramatic change from exponential to
power-law behavior in this quantity on approaching the critical
point.

The coefficient of proportionality $K_{pl}$ will equal the
corresponding stiffness $K$ for columnar order and scale
identically to the physical spin stiffness on the other side of
the transition. Once again, this is not a test of self-duality but
follows from the irrelevance of monopoles and will hold in the
isotropic case as well.

Specializing to the easy plane case, the self-duality of the
critical fixed point implies some further interesting properties.
First, it is clear that the dual global boson will have the same
power law decay as the physical spin-correlator.  The former is
identified with the VBS order parameter while the latter
corresponds to the staggered XY correlators in the microscopic
spin model. We thus have the remarkable result that the columnar
dimer, plaquette, and staggered XY magnetization all decay with
the same power law right at the critical point. Furthermore, the
$\beta$ exponent for the particular VBS order that actually
develops will be the same as the $\beta$ for the spin order. This
is because the anisotropy only serves to lock the phase of the
dual order parameter. The amplitude is already non-zero in the
scaling limit near the critical point.

\subsection{Ordered state}
\label{sec:ordered}

As elaborated in the previous subsection, there are two diverging
length (or equivalently time) scales upon approaching the
transition from the VBS side. How does this manifest itself in the
N\'eel ordered side? To understand this first note that in the
ordered phase close to the transition there will be `soft' modes
that correspond to the incipient VBS order on the other side of
the transition. Indeed the frequency of these modes will go to
zero at the critical point. For concreteness consider the case
where the VBS order that develops is columnar. Then as is natural
there will be a soft mode corresponding to columnar order with
frequency vanishing on approaching the critical point. Remarkably
there will be other soft modes that correspond to plaquette
ordering whose frequency also vanish on approaching the
transition. Once again this is despite there being no such order
in the VBS phase. This result is already implied by the discussion
in the preceding subsection. Indeed both the plaquette and
columnar order parameters have power law correlations at the
critical point regardless of which one of the two orders actually
develops in the VBS phase. Thus it is natural that the frequencies
of both modes go to zero on approaching the critical point from
the ordered side. Furthermore, both soft mode frequencies vanish
in exactly the same universal way on approaching the transition.
Formally the columnar and plaquette order parameters are
distinguished only by the orientation of the phase of the complex
VBS order parameter. In the absence of monopole tunneling events
(which tend to pin this phase to certain values), these two
distinct order parameter fields will both behave in a common
manner determined by the complex VBS order parameter. Thus in the
scaling limit near the critical point both the plaquette and
columnar order parameters will display the same universal
behavior. On general scaling grounds we expect the VBS soft mode
frequency $\omega_{\rm VBS} \sim \rho_s$ where $\rho_s$ is the
ground state spin stiffness of the N\'eel ordered state.

Despite the common universal behavior of the vanishing frequency
of the two distinct VBS soft modes, there will be a small
splitting between the two frequencies that is due to the
irrelevant (quadrupled) monopole tunneling events. Indeed the
information about which of the two VBS orders eventually develops
in the paramagnetic phase is contained in this small splitting. If
columnar order develops, then the corresponding soft mode will
have slightly lower frequency in the ordered phase as compared to
the plaquette soft mode. Note that this splitting will go to zero
as the critical point is approached as the monopole fugacity
renormalizes to zero. This will however vanish in a very different
way from the overall VBS energy scale $\omega_{\rm VBS}$. Indeed
this splitting defines a new energy scale that vanishes faster
than $\omega_{\rm VBS}$. Thus we see that two different energy
scales also characterize the physics of the ordered phase.

In the easy plane case, one intriguing aspect of our theory is the
physics of the vortex cores in the XY ordered phase close to the
transition.  As discussed extensively, there are two kinds of
classical meron vortices which tunnel into each other in the
quantum theory. However, the irrelevance of these instanton
tunneling events near the transition implies that the Ising order
in the core will survive for a very long time.

We have, up to this point, not considered any effects which
explicitly break the lattice translational symmetry.  Hence the
discussion should be read as appropriate for extended, plane wave
states of vortices.  Crucially, these states may be classified by
their (quasi-)wavevectors.  Consider the continuum theory in which
the spatially-oscillating single instanton events are neglected,
and only the (irrelevant) quadrupled instanton fugacity $\lambda$
is included. In the dual formulation, the two Ising vortex states
appear as relativistic charged particles. They carry a conserved
U(1) non-compact gauge charge (physically their vorticity), {\em
i.e.\/} the number of these vorticies $N_1+N_2$ is conserved.
There is a discrete $Z_4$ global symmetry, which implies that
$N_1-N_2$ is conserved modulo 8. The latter is a consequence of
the continuum limit which removes the single instanton events, and
is promoted to a continuous U(1) symmetry (with $N_1-N_2$ fully
conserved) if the quadrupled instantons are also neglected.
Physically, then, excitations of the vortex vacuum, {\em i.e.\/}
the XY ordered state, can be labelled by these quantum numbers.
And one should expect there to be ``quasiparticles'' (really
``quasivortices'' or ``quasimerons'') carrying an elementary unit
($\pm 1$) non-compact gauge charge (physically $\pm 2\pi$
vorticity) and an elementary $Z_4$ charge $N_1-N_2 =\pm 1$.  These
vortices (as befits an XY ordered phase), with a `core energy' scaling as
$1/\xi$. There will
be other gapped excitations (also with a gap of
$\mathcal{O}(1/\xi)$) with zero non-compact gauge charge (zero
vorticity) and $N_1-N_2 = \pm 2$, which can be viewed as the VBS soft
modes, or alternatively as ``vortex excitons''.

Let us now consider the effects of the neglected oscillating
single monopole term.  Na\"ively, this violates the $Z_4$
conservation law, and can mix the quasimeron states with
$N_1-N_2=\pm 1$.  However, due to the 4-sublattice oscillation,
the process in which one meron state is converted to another is
accompanied by the addition of a large momentum. In the absence of
a sink for this momentum, therefore, {\em even the
  single instanton term cannot mix the two meron quasiparticle states}.
At non-zero temperature, thermally excitated excitations with a
large gap ($\gg 1/\xi$) can provide this momentum, but are present
only with an exponentially small probability due to their gap.
Hence violations of the conservation of the ``Ising'' charge of a
single quasiparticle are exponentially weak at low temperature.
Of course, {\em four} quasiparticles with ``up'' Ising cores can
scatter off one another to produce four quasiparticles with
``down'' Ising cores via the non-oscillating $\lambda$ term.  The
amplitude for even this process is, however, suppressed by the
irrelevance of $\lambda$ if one is near the QCP.

Thus we arrive at the remarkable conclusion that the
``elementary'' gapped vortex excitations of the XY ordered phase
carry a sharp extra Ising quantum number.  Ramifications will be
explored in Sec.~\ref{sec:VL}.

A number of other predictions may also be made on the effects of various
perturbations that can be added to the Hamiltonian near the zero
temperature critical point.

\subsection{Uniform Zeeman field}
\label{sec:zeeman}

First let us think about the excitation structure in the
paramagnetic side. Deep in the phase, the lowest spinful
excitations will be $S = 1$ which will be gapful. On approaching
the transition, due to the diverging `confinement' length one
might na\"ively think that this will break up into spinons.
However even with a non-compact gauge field due to the log
attraction coming from the photon there will be logarithmic
confinement and the lowest energy spin carrying excitations will
continue to have $S = 1$  This ``magnon'' is a gauge neutral bound
state of two spinons. Now imagine sitting in the paramagnetic side
close to the transition, and turn on a Zeeman field along the
$z-$axis in spin space. Once the Zeeman energy exceeds the magnon
gap, the chemical potential for such magnons becomes positive and
they should condense to modify the ground state.  This leads to XY
antiferromagnetic order.  In the non-compact approximation,
because this condensate is gauge neutral, it does not create a gap
for the photon via the Higgs mechanism.  In reality ({\em i.e.\/}
beyond the non-compact approximation) what this means is that
there will be coexistence between VBS and XY order.  As the
spinons are not condensed in this phase, there is no disruption of
the VBS order (this can be seen {\em e.g.} from the fact that the
spinons appear as dual ``vortices'' in the VBS order parameter,
while the magnons carry zero dual vorticity).

These considerations hold only provided another transition does
not pre-empt magnon condensation as the Zeeman field is increased.
This will happen, {\em e.g.} if the magnons experience attractive
interactions with one another.  Indeed, in a Coulomb interacting
system, it is natural to expect that the magnons, which are the
analog of excitons, will have attractive interactions with one
another at long distances, due to the analog of ``van der Waals''
forces between their fluctuating dipole moments.  This attraction,
however, competes with other local interactions due to the complex
critical physics on scales $\lesssim \xi$, so the outcome is not
clear.  Therefore, we do not see a clear argument against a
continuous magnon condensation transition into a coexistence
phase.  Likewise, of course, we cannot rule out a direct first
order transition.  In any case it therefore seems as though a
direct second order transition between the magnetically ordered
and VBS phases is unlikely at non-zero Zeeman field.

If the coexistence phase exists, it is interesting to contemplate
the transition between the coexistence phase and the pure
magnetically ordered one (with canted antiferromagnetic order). In
the system with XY symmetry, a transition with exactly these
symmetries has been studied in Ref.~\onlinecite{BFrey}.   In this
paper, it was shown that, despite coexisting superfluid order and
the consequent gapless goldstone mode, this transition is in the
universality class of a $D=3$ XY model, the $Z_4$ symmetry
breaking perturbation {\em and} the coupling to the goldstone mode
being irrelevant.  Thinking in terms of the dual formulation
suggests this analysis should apply here.  In particular, both
vortex fields $\psi_{1,2}$ remain gapped across the transition,
since both phases are XY ordered.  Only the composite order
parameter $\psi_{\rm VBS}=\psi_1^\dagger
\psi_2^{\vphantom\dagger}$ is involved in the criticality, and it
does not couple minimally to the non-compact gauge field.
Integrating out the massive vortex fields while keeping a
composite $\psi_{\rm VBS}$ field and the gauge fluctuations
describing the Goldstone mode, one arrives at a model equivalent
to Ref.~\onlinecite{BFrey}. Note that this result implies that this critical
point is also deconfined in precisely the same sense
as the others discussed in this paper.

\subsection{Staggered Zeeman field}
\label{sec:stag}

Consider the effect of a staggered Zeeman field on the original
spin model. First assume easy plane anisotropy in the plane
orthogonal to the applied field.
 The staggered field will always induce some staggered
magnetization but we can ask about XY or VBS order superimposed on
this. In the CP$^1$ description, a staggered Zeeman field
corresponds to a uniform `magnetic' field that couples to the $z$
component of $\hat{n}$. It is quite clear that there will now be a
split transition between the VBS and XY ordered phases with an
intermediate phase with neither XY nor VBS order (but of course
with a staggered magnetization).

Consider the approach from the VBS phase. In the presence of the
staggered field, one of the two CP$^1$ fields will condense first.
This transition is described by the $N = 1$ SJ model (and is
inverted $D=3$ XY). The resulting phase is the advertised phase
with neither XY or SP order. Actually it is more useful to think
of this critical point in the gauge language as a `deconfined'
critical point than just as inverted XY. This is so particularly
if one asks about the magnon spectral function at this transition.
This will be determined by the spinon dynamics which in turn are
coupled to a non-compact U(1) gauge field. Thus we might again
expect anamolously broad spectral functions even though both
phases are confined. (Note that total $S^z$ is still conserved).

Eventually, as one tunes towards the XY ordered phase, the other
CP$^1$ field will also condense leading to XY order. This
transition is actually exactly dual to the other one discussed
above.  This is because the staggered Zeeman field couples to the
same operator in both the CP$^1$ and dual representations. For
small staggered Zeeman fields, the phase boundaries must come in
with the same exponents, etc.

Finally, in the O(3) model, the second transition will not happen
as the N\'eel vector will immediately line up with the staggered
field. However, the first transition will continue to be described
by the $N=1$ SJ model. Details of the slopes of various phase
boundaries etc. may be found Ref.~\onlinecite{mv} - in the
terminology of that reference, a uniform field corresponds to the
staggered Zeeman field discussed here.

\subsection{Finite temperature transitions}
\label{sec:temp}

Finite temperature properties near the transition may also be
discussed. Here it is clearly necessary to distinguish between the
easy plane and isotropic cases.  In the latter, the N\'eel order
does not survive for any non-zero $T$ while in the former case
there is power law order at low $T$ which eventually disappears
through a Berezinski-Kosterlitz-Thouless (BKT) transition.  In
both cases however the discrete broken lattice symmetry of the VBS
phase will survive upto a non-zero finite $T$.  The associated
finite-$T$ transition will be in the universality class of the
$Z_4$ clock model in $d = 2$.  This transition is known to be
described by a {\em line} of fixed points with continuously
variable ({\em i.e.\/} nonuniversal) critical exponents.  The line
of fixed points results from the exact marginality of the fourfold
symmetry breaking term $\psi_{\rm VBS}^4$, {\em i.e.\/} our
$\lambda$ coefficient.  As $\lambda$ approaches zero, the nature
of the criticality approaches that of a simple $D=2$ classical XY
model, i.e it becomes BKT-like.  Thus $\eta\rightarrow 1/4$ and
$\nu \sim 1/|\lambda|$ diverges in this limit.  Since at the zero
temperature QCP instantons are irrelevant, we may conclude that
the fixed point value of $\lambda$ at the {\em
  classical} VBS-paramagnet transition (which is generally finite and
non-zero at $T_c$) approaches zero as $T_c \rightarrow 0$.  Hence
the non-universal critical behavior of the VBS-paramagnet
transition becomes arbitrarily close to BKT behavior as this
transition line is followed into the $T=0$ QCP.  This conclusion
is independent of the XY or O(3) symmetry of the magnetic
ordering.

In the XY case, the self-duality of the easy plane fixed point
implies further that the phase boundaries associated with the
finite-$T$ transition from both the N\'eel and VBS phases have the
same shape at low $T$.  Note that both are BKT-like for $T_c$
asymptotically close to zero ({\em i.e.\/} near the QCP),
consistent with duality.  Indeed, one expects not only the phase
boundary but also all critical correlations to match in this
limit, including amplitude ratios.

\section{Deconfined quantum criticality at the VBS to spin liquid transition}
\label{sec:vbssl}

In this section we argue that the transition between a valence
bond solid and a fractionalized spin liquid is also an example of
a deconfined quantum critical point in a precise sense.

In two spatial dimensions, a fractionalized spin liquid is
expected to be described as the deconfined phase of a $Z_2$ gauge
theory with a gapped $Z_2$ vortex - the vison. This $Z_2$ gauge
field is minimally coupled to spin-$1/2$ spinon excitations. We
only consider the case where the spinons are gapped. A precise
theoretical characterization is given by the notion of topological
order\cite{Wen,topth}.

Consider the evolution of the ground state of a spin-$1/2$ system
(or equivalently for bosons at half-filling) between such a
fractionalized spin liqiud and a VBS on, say, a square lattice.
Despite the lack of any obvious local order parameter for the spin
liquid there is a close similarity with the N\'eel-VBS transition.
Indeed both the spin liquid and VBS are characterized by two
distinct types of order (the former by topological order and the
latter by broken lattice symmetry). Na\"ive thinking might then
suggest that a direct second order transition is not possible.
Rather one might have expected two transitions with an
intermediate ``coexistence region'' which breaks lattice symmetry
but is also topologically ordered (a VBS$^*$ phase, in the
notation of Ref.~\onlinecite{sf}). Once again this na\"ive
expectation is incorrect and a direct second order transition is
indeed possible. Furthermore the critical theory may be regarded
as a {\em non-compact} U(1) gauge theory with an extra emergent
dual global U(1) symmetry.

It is convenient to begin with a theoretical formulation that is
powerful enough to describe both phases and all of their distinct
excitations. Such a formulation is provided in the work of
Refs.~\onlinecite{ReSaSpN,ijmp,Wen}. As before, the underlying spin
model is first reformulated as a theory of spin-$1/2$ spinon
fields that are minimally coupled to a compact U(1) gauge field
with Berry phases. The VBS corresponds to a confined paramagnet
where the spinons have disappeared from the spectrum. The spin
liquid obtains when a singlet pair of spinons - which carries
gauge charge-2 - condenses {\em i.e} enters a Higgs phase. Let us
represent this Higgs field by the operator $Q=e^{i \varphi}$. We
imagine  integrating out
the individual spinon fields (this is permissible
because all spin carrying excitations
 are gapped in both the VBS and spin liquid phases),
and obtain the following theory for the transition \cite{JaSa,sf}
\begin{eqnarray}
\mathcal{S} &=& \mathcal{S}_\varphi + \mathcal{S}_a +
\mathcal{S}_B \nonumber \\
\mathcal{S}_\varphi &=& - 2t \sum_{\ell} \cos
\left({\boldsymbol{\Delta}} \varphi - 2 {\bf a} \right)
\label{eq:SJ5},
\end{eqnarray}
where $\mathcal{S}_a$ and $\mathcal{S}_B$ are defined in
Eqs.~(\ref{eq:SJ2}) and (\ref{eq:SJ3}). Note that
Eq.~(\ref{eq:SJ5}) is just the $N=1$ SJ model studied in
Section~\ref{sec:SJ1}, but with the crucial difference that
$\varphi$ carries charge 2 (compare with Eq.~(\ref{eq:SJ4})). The
duality transformations of Section~\ref{sec:SJ1} are easily
applied to Eq.~(\ref{eq:SJ5}), and we obtain an XY model with
8-fold anisotropy which is irrelevant at the transition (this
contrasts with the 4-fold anisotropy obtained in
Section~\ref{sec:SJ1}).

In the spirit of previous Sections (particularly
Section~\ref{sec:SJ1}), these results may be understood physically
as follows. In the fractionalized phase the condensation of the
charge-$2$ scalar leads to vortex excitations (the visons) which
carry $\pi$ gauge flux. In the fractionalized phase instanton
effects kill visons in pairs -  indeed this is precisely what
leads to their `Ising' nature. The transition to the confined VBS
phase occurs when the visons condense. But near the transition,
and in the continuum limit, we expect once again that all
instanton events are quadrupled. Thus the $\pi$ flux vortices can
only disappear 8 at a time. This gives the XY model with 8-fold
anisotropy.

We thus see that the 8-fold anisotropy in the dual XY model should
be interpreted as instanton tunneling events in the original
compact gauge theory. Consequently as before we conclude that
instantons are irrelevant at the critical fixed point so that a
gapless non-compact U(1) gauge theory obtains. (We remind the
reader that the global XY model is the dual of the condensing
charge-2 scalar coupled to a non-compact gauge field).

Note once again the crucial role played by the Berry phases which
are responsible for leading to an 8-fold anisotropy (as opposed to
2-fold as would obtain in their absence).

Note also that spinons are well-defined in the fractionalized
phase but are confined in the VBS phase.  What is the fate of the
gapped spinons right at the transition between these two phases?
The arguments above show that at the critical point the spinons
are minimally coupled to a {\em non-compact} U(1) gauge field
descending from $a_{\mu}$ (which in turn is also coupled to the
critical spinon pair field). The strong scaling properties of this
critical point (which is dual to the $D=3$ XY model) implies that
the gauge-field has the following two-point correlator at
criticality
\begin{equation}
\langle a_\mu (p) a_\nu (-p) \rangle \sim \frac{1}{p} \left(
\delta_{\mu\nu} - \frac{p_\mu p_\nu}{p^2} \right),
\label{eq:amuprop}
\end{equation}
where $p_\mu$ is the spacetime 3-momentum. Note that this
propagator does not have the Maxwell $1/p^2$ scaling, but a $1/p$
dependence fixed by the scaling dimension $\mbox{dim}[a_\mu ] =
1$. This implies a $1/r$ interaction between static massive
spinons at criticality.

It is sometimes stated that the transition between the VBS and
spin liquid phases is described by a $Z_2$ gauge theory. The
results here however show that the transition is in fact described
as a deconfined U(1) gauge theory in a very precise sense. It is
the spin liquid phase itself (as opposed to the transition) that
is described as a (deconfined) $Z_2$ gauge theory.

\subsection{Spin liquids that break lattice symmetry}
\label{sec:sllat}

An important subtlety has been glossed over in the analysis so far
in this section. Spin liquid states with no broken lattice
symmetries are certainly possible \cite{Wen,sst,sf,moessner}, and
for these the above analysis applies. However, in bosonic mean
field theories of SU(2) spin liquid states on a variety of
lattices \cite{ijmp,ross,chung,ybkim,kimchung}, the spin liquid
state is commonly found to break a global lattice rotation
symmetry \cite{higgs} -- such a state has `bond-nematic' order.
The spin liquid is associated with short-range, incommensurate
spin correlations at a wavevector $K$, and the choice of $K$ often
breaks a lattice symmetry {\em e.g.\/} a spin liquid state at $K =
(k,k)$ is distinct and inequivalent to a state at $K=(k,-k)$. Such
states appear naturally at the boundary of a VBS state\cite{ijmp},
and for these the theory above has to be reconsidered. Before
doing this, we note one important case for which this is {\em
not\/} necessary: the Cs$_2$CuCl$_4$ lattice\cite{ross,ybkim},
which interpolates between the square and triangular lattices.
Within a large $N$ bosonic mean field theory treatment, the ground
state in the square lattice limit is a VBS, while in the
triangular limit it is spin liquid which breaks no lattice
symmetries: the transition between these states is described by
the theory in Eq.~(\ref{eq:SJ5}).

Turning to a spin liquid that does break lattice symmetries,
consider {\em e.g.\/} the transition on the square
lattice\cite{ijmp} from the VBS (Fig~\ref{pdiag}) with
short-ranged spin correlations peaked at the wavevector
$(\pi,\pi)$, to a bond-nematic spin liquid at wavevector $K =
(k,k)$ or $K = (k,-k)$. The choice of either of the latter states
breaks a symmetry of reflection about the principal square lattice
axes. In mean field theory,\cite{ijmp} this transition is
characterized by the condensation of two Higgs fields, which we
denote as $Q_x = e^{i \varphi_x}$ and $Q_y = e^{i \varphi_y}$.
These fields are odd under the lattice reflections $\mathcal{R}_x$
and $\mathcal{R}_y$ in Table~\ref{tab:sym}
respectively,\cite{higgs}, and this prohibits terms which are
linear in either Higgs field in the effective action. Using these
symmetries, and the requirements of gauge invariance, we
generalize Eq.~(\ref{eq:SJ5}) to
\begin{eqnarray}
\mathcal{S} &=& \mathcal{S}_\varphi + \mathcal{S}_a +
\mathcal{S}_B \nonumber \\
\mathcal{S}_\varphi &=& - \sum_{\ell} \Bigl[ 2t \cos
\left({\boldsymbol{\Delta}} \varphi_x - 2 {\bf a} \right)
\nonumber \\&~&\!\!\!\!\!\!\!\!\!\!\!\!\!\!\!+ 2t \cos
\left({\boldsymbol{\Delta}} \varphi_y - 2 {\bf a}
\right)+2t^{\prime} \cos(2 \varphi_x - 2 \varphi_y) \Bigr]
\label{eq:SJ51},
\end{eqnarray}
where, as before, $\mathcal{S}_a$ and $\mathcal{S}_B$ are defined
in Eqs.~(\ref{eq:SJ2}) and (\ref{eq:SJ3}). Note the crucial factor
of 2 in the argument of the third cosine in Eq.~(\ref{eq:SJ51}):
this is required by the inversion constraints above. Apart from
the usual compact U(1) gauge invariance, Eq.~(\ref{eq:SJ51}) is
also invariant under the global $Z_2$ transformation
\begin{eqnarray}
\varphi_x &\rightarrow& \varphi_x + \pi/2 \nonumber \\
\varphi_y &\rightarrow& \varphi_y - \pi/2 \label{eq:phiz2}
\end{eqnarray}
which realizes the lattice reflection symmetry. (Note that the
square of the transformation in Eq.~(\ref{eq:phiz2}) is equivalent
to the identity modulo a compact U(1) gauge transformation.)
Consequently, there are now two inequivalent Higgs phases, with
$\langle \varphi_x - \varphi_y \rangle = 0$ (or $\langle Q_x
\rangle = \langle Q_y \rangle$) and $\langle \varphi_x - \varphi_y
\rangle = \pi$ (or $\langle Q_x \rangle = -\langle Q_y \rangle$),
and these correspond\cite{ijmp} to the two possible spin liquid
phases at $K=(k,k)$ and $K=(k,-k)$. The theory Eq.~(\ref{eq:SJ51})
can be analyzed by the same duality transformation applied to
Eq.~(\ref{eq:SJ5}), but the critical properties have not been
determined.

\section{Analogies and extensions}
\label{sec:anext}

\subsection{Superfluid-insulator transition of correlated bosons}
\label{sec:superfl-insul-trans}

The models and the phenomena discussed in this paper can be
fruitfully discussed from a different point of view. Consider a
system of bosons with short-ranged repulsive interactions on a
square lattice such that there is half a boson per site on
average. It has long been appreciated that such a bosonic system
is closely related to quantum spin models with easy plane (or easy
axis) anisotropy.  Indeed, there is an exact equivalence in the
hard-core limit in which at most one boson occupies each lattice
site. Specifically, one may consider a model of bosons (described
as O(2) quantum rotors) on a square lattice:
\begin{equation}
\label{o2rot} H = U\sum_r \left(n_r - \frac{1}{2} \right)^2 -
t\sum_{\langle rr'\rangle} \cos(\phi_r - \phi_r')+ ....
\end{equation}
Here $\phi_r \in [0,2\pi)$ represents the boson phase, $n_r$ is
the conjugate boson number and is an integer $\in [-\infty,
\infty]$. The ellipses represent other short ranged terms that can
be tuned to drive transitions from a superfluid to (for instance)
the bond stripe insulator.  To relate the above boson Hamiltonian to the
antiferromagnetic systems considered in the bulk of the paper, we note
that one may define
\begin{eqnarray}
  \label{eq:bosontospin}
  S_r^\pm & = & \epsilon_r e^{\mp i\phi_r}, \\
  S_r^z & = & n_r - \frac{1}{2}.
\end{eqnarray}
For large $U$ this gives a faithful representation of an easy-plane
spin-$1/2$ antiferromagnet, and the universal physics is expected to be
unchanged at smaller $U$.

Clearly a superfluid phase of the bosons is possible (and
corresponds to the XY ordered phase in the magnet analogy).
Various kinds of Mott insulating ground states are also possible.
(These correspond to quantum paramagnets in the magnetic case). A
simple Mott state corresponds to the bosons forming a checkerboard
ordered pattern in which the sites of one sublattice are
preferentially occupied. This will be stabilized by large nearest
neighbor repulsion and corresponds to the Ising ordered
antiferromagnet. In the boson language the columnar VBS state may
be understood as a `bond-centered' stripe (or a bond density wave)
-- a state in which each boson is shared in a bond between two
nearest neighbor sites such that these favored bonds have lined up
in columns. The considerations of Section~\ref{sec:ep} more or
less apply directly to the transition between the superfluid and
the bond-centered stripe insulator (or the analogous plaquette
ordered insulator). In particular the critical theory is
`deconfined' and is expressed in terms of two fields each with
boson charge-$1/2$ that are minimally coupled to a non-compact
U(1) gauge field. However the discussion in Section~\ref{sec:ep}
was intended for weak easy plane anisotropy on an isotropic spin
model. It is somewhat more satisfying to derive the crucial field
theory Eq.~(\ref{crtny}) directly for the bosonic system. We point
out that the approach of Ref.~\onlinecite{Crtny} provides such a
direct derivation of the required dual action.  However the close
connection with fractionalized charge degrees of freedom is
somewhat obscured by that approach.  We therefore sketch in
Appendix~\ref{sec:direct-deriv-dual}  a derivation proceeding in a
manner more similar to the considerations of the previous
sections, in particular going directly from the boson model of
Eq.~(\ref{o2rot}) to the dual meron action obtained earlier.

In the context of boson models (in view of potential applications
e.g. to atomic bosons in optical lattices or to electronic systems where
the bosons are Cooper pairs), some physical properties arise which are
less natural in the context of quantum antiferromagnets discussed
earlier.  In particular, it is interesting to consider the effects of an
applied orbital magnetic field coupling.  This can bring out the unusual
physics of Ising ordering in the vortex cores discussed earlier in
Sec.~\ref{sec:ordered}.

\subsubsection{Orbital magnetic field}
\label{sec:VL}

Let us consider the structure in an applied {\em orbital} magnetic
field $B$.  The QCP at zero field describes a transition between a
superfluid phase and a bond-centered striped phase.

Suppose the system is on the superfluid (XY ordered) side of the QCP,
and a small magnetic field is applied (we use the internal field $B$).
This field produces vortices, separated by an average distance
$\ell=\sqrt{\phi_0/B}$, where $\phi_0=hc/q$ is the flux quantum, and
$q$ is the boson charge.  In the weak field limit, where the length
$\ell$ is large, one expects these vortices to form an Abrikosov
lattice, since the long-range logarithmic interactions between
vortices dominate their kinetic energy.  Now suppose one is near the
QCP, so that the correlation length $\xi$ is large. To a first
approximation, one can neglect instanton events, and treat the Ising
quantum number of the vortices as conserved.  Then each vortex in the
Abrikosov lattice has a definite Ising ``charge", and hence the system
as a whole some sort of Ising magnetic order.  It is straightforward
to see that the basic interactions between these Ising ``spins'' are
antiferromagnetic, and that these interactions decay rapidly when the
two vortices in question are separated by a distance much larger than
$\xi$.  These interactions arise because the two types of vortices
carry opposite gauge flux $\int\! d^2 x \, \epsilon_{ij} \partial_i
a_j = \pm \pi$. This gauge flux is confined to a region of the size of
the (gauge) ``penetration depth''. Since the gauge field fluctuations
are part of the quantum critical theory, this penetration depth is of
the $\mathcal{O} (\xi)$.  Due to the Maxwell term in the action, two
nearby vortices have lowest energy with opposite gauge fluxes (and
hence smaller total gauge flux), provided the two fluxes overlap. A
mean field analysis following Abrikosov leads to the same
conclusion as discussed below. In particular, consider the Lagrangian ${\cal
L}(z_\alpha)$ (in Eqn.~\ref{sz}) for the $z_\alpha$ fields.  We are
interested in $w<0$ (in Eqn.~\ref{eq:epterm}),
and it is convenient to consider the limit $w=-2u + \delta w$, with
$0<\delta w \ll 2u$.  For $\delta w=0$, the mean field theory (which
neglects fluctuations of $\vec{a}$) comprises simply of two decoupled
copies of Abrikosov's lattices for $z_1$ and $z_2$.  Thus the solution consists
of a triangular vortex lattice in each $z_\alpha$, with lattice
spacing $\sqrt{2}\ell$ (since each $z_\alpha$ has charge $q/2$).
These two lattices are completely decoupled in this approximation.
With $\delta w>0$, the energy is minimized when the integral of
$|z_1|^2|z_2|^2$ is smallest.  This is accomplished by placing the
$z_1$ and $z_2$ vortices as far apart as possible, so that
$|z_1|^2|z_2|^2$ is reduced over the maximum spatial area.  The
solution is to choose the two triangular $z_\alpha$ vortex lattices as
the two sublattices of a honeycomb lattice.  This corresponds to an
antiferromagnetic orientation of the Ising vortex cores on the
honeycomb.

To establish the stability of this order, we must reconsider the
effect of instanton events in this phase.  The important events
are {\em single} instantons, which act like a transverse field on
the Ising quantum number.  While these average away in the
continuum theory, the finite lattice spacing $\sim\ell$ provides
an upper length cutoff for the oscillations of the single
instanton fugacity, which can therefore have an effect.  Near the
QCP, it is possible for both $\ell$ and $\xi$ to be large, but to
have $\ell$ not much greater than $\xi$.  In this limit, which we
consider, the overlap of the vortex cores is strong, hence the
Ising antiferromagnetic ``exchange energy'' between neighboring
vortices is large, {\em i.e.\/} of order $1/\xi$ by scaling.  The
effective transverse field on the vortices is more difficult to
estimate.  In a mean field treatment, one simply averages the
oscillating instanton fugacity over the two dimensional $\sim
\xi^2$ using some smooth envelope function.  This gives a
transverse field $\sim \lambda_0/\xi^2$.  Fluctuation effects may
be expected to further decrease this field.  Hence, the transverse
field is much weaker for large $\xi$ than the antiferromagnetic
coupling between cores.

Thus we arrive at the remarkable conclusion that the vortex state
near the QCP exhibits antiferromagnetic Ising LRO of the staggered
Ising magnetization of the vortex cores.  Note that this analysis
applies when the magnetic length $\ell\sim \xi$.  For very small
fields, or further from the QCP, $\ell \gg \xi$, and the
antiferromagnetic interactions between cores ($\sim
e^{-\ell/\xi}$) decay exponentially, while the transverse field is
likely of power law form.  Hence for very small fields it seems
probable that the Ising cores become disordered.  In this case,
the physical manifestation of the long-lived Ising staggered
magnetization is the presence of a low energy ``antibonding''
excitation of each vortex.

Clearly, upon increasing the quantum fluctuations, this vortex
lattice must disappear, since the VBS state on the other side of
the critical point is an orbital paramagnet, {\em i.e.\/} no
change in symmetry occurs on applying a weak field to it.  Hence
there is at least one phase boundary separating the Ising ordered
vortex lattice from the VBS phase that persists at $B>0$.  We will
not address this ``vortex lattice melting'' physics here, except
to say that first order, continuous, and multi-stage transitions
(with intermediate partially ordered phases) are all possible in
principle (and difficult to distinguish between on purely
theoretical grounds).

\subsection{Higher spin}
\label{sec:higherspin}

In the bulk of this paper we have focused on the spin-$1/2$ square
lattice antiferromagnetic model. Here we briefly discuss the fate
of higher spin models on the square lattice. It should be clear by
now that the answers depend crucially on the Haldane phases that
obtain for higher spin. In the isotropic model, if $2S = 1 ({\rm
mod}~4)$ then the monopoles are quadrupled. Thus for all such
values of the spin, a direct second order N\'eel-VBS transition
described by the same deconfined critical theory as for spin-$1/2$
obtains. If $2S = 0 ({\rm mod}~4)$, then there are {\em no}
oscillating phase factors for the monopoles. This has the
consequence that a translation symmetric quantum paramagnetic
state is now possible. The transition to this state from the
N\'eel state will be described by the usual LGW O(3) fixed point
({\em i.e.\/} with monopoles present). If $2S = 2 ({\rm mod}~4)$
then the appropriate Haldane phases lead to doubling of monopole
events. Now confined paramagnetic states necessarily break lattice
symmetries. Whether a direct second order N\'eel-VBS transition is
allowed or not depends on the scaling dimension of the 2-monopole
operator at the monopole-suppressed fixed point. If this is
irrelevant, then the same deconfined critical theory as for
spin-$1/2$ will obtain.

It is interesting to consider the spin-$1$ case in the presence of
some easy plane anisotropy. This may equivalently be viewed as a
model of  bosons at integer filling - unlike in the isotropic
limit,  a translation symmetric confined paramagnet is clearly
possible. A direct transition between the XY ordered phase and
such a paramagnet is clearly possible and will be in the usual $D
= 3$ XY universality class.

However, presumably the interesting question even in the easy
plane case is whether a direct second order transition is possible
between the XY ordered phase and a lattice symmetry {\em broken}
confined paramagnet with bond order.  To answer this question and
to obtain a description of such a paramagnet, it is convenient to
start from the isotropic limit and introduce weak easy plane
anisotropy. In the isotropic limit the confined paramagnetic
states will break lattice symmetries and this will be preserved
upon turning on some easy plane anisotropy. In a CP$^1$
description, there will now be monopole Berry phases that
oscillate on two sublattices of the dual lattice. We may now
dualize in the easy plane limit to the meron vortex degrees of
freedom. In this description it is clear that the translation
broken VBS state is again described by an equal amplitude
condensate of both vortex fields. Now the instanton term converts
two merons of one kind into two of the other kind {\em i.e.\/} the
co-efficient of the $\lambda$ term in Eq.~(\ref{crtny}) is
$\mbox{Re}[(\psi_1^{\ast} \psi_2)^2]$, with $\lambda \equiv
\lambda_2$. The relevance/irrelevance of this at the self-dual,
easy plane, non-compact CP$^1$ fixed point will determine whether
a direct second order transition obtains or not: this question
remains open at present.

Note that near the transition to the usual paramagnet with no
broken symmetries there will only be single species of vortex with
a featureless `paramagnetic' core. On the other hand near the
transition to the VBS phase there will once again be two species
of (nearly) stable vortices with cores that have very long-lived
Ising order (if as discussed above such a direct continuous
transition is possible).

\subsection{Honeycomb lattice}
\label{sec:honeycomb}

The considerations of this paper generalize readily to other two
dimensional bipartite lattices. For instance on the honeycomb
lattice the Haldane Berry phase calculation implies that all
monopole events are tripled (rather than quadrupled as on the
square lattice). This implies that the issue of whether or not a
direct second order transition described by a `deconfined'
critical point obtains between the N\'eel and VBS phases is
determined by the scaling dimension of the 3-monopole operator at
the monopole-suppressed fixed point. Unlike the square lattice it
is however less clear that the 3-monopole operator will be
irrelevant. For instance for the $N = 1$ SJ model, the Higgs-VBS
transition is determined by the $Z_3$ clock universality class
which is distinct from the XY universality class. In other words
the 3-fold anisotropy which represents instantons is {\em
relevant} at the deconfined fixed point at $N = 1$. At large-$N$
all monopoles continue to be irrelevant. The fate of the physical
models (with and without easy plane anisotropy) can only be
settled by direct numerical computation of the scaling dimension
of the $3$-monopole operator.

\subsection{Ising Anisotropy and Other Transitions}
\label{sec:ising-anisotropy}

In this paper, we have focused on the properties of spin-$1/2$
antiferromagnets with full SU(2) spin rotational symmetry, or its
easy-plane reduction to U(1).  These two cases are amenable to
analysis due, on the one hand, to natural continuations of the
SU(2) invariant CP$^1$ representation to CP$^{N-1}$, and through
standard XY duality. One may also ask whether similar deconfined
critical points might arise in systems with easy-axis (i.e. Ising)
anisotropy, which also retain the U(1) subgroup of SU(2).
Unfortunately, this limit is much less amenable to microscopic
duality transformations on the lattice level, and so it is
difficult to make firm statements.  While some of us suspect that
no deconfined critical behavior is likely in this case, it is
nevertheless of interest to present candidate field theories for
such deconfined transitions.

Very na\"ively, one may attempt to begin with the CP$^1$
representation of the quantum antiferromagnet, and simply change
the sign of the anisotropy term, taking $w>0$ in
Eq.~(\ref{eq:epterm}).  In a mean-field analysis of the continuum
field theory of Eq.~(\ref{sz}), including anisotropy of this sign
would indeed have the desired effect of yielding a transition
between an Ising ordered phase (for $s<0$) and a VBS phase (for
$s>0$ -- actually one na\"ively obtains the Coulomb phase of the
gauge theory, neglecting the dangerously irrelevant instantons).
There are, however, several caveats to this candidate theory that
must be mentioned. First, supposing the gauge field $a_\mu$
non-compact, fluctuation effects are known in some situations
({\em e.g.\/} the classical Abrikosov transition between the
normal state and vortex lattice at $H_{c2}$) to drive na\"ively
continuous transitions involving gauge fields first order.  While
we believe this does not occur in the cases of SU(2) invariance
and XY anisotropy, these conclusions are based on several exact
dualities and the numerical results of Ref.~\onlinecite{KM,mv}
directly on models in which instantons have been suppressed, and
direct simulations of CP$^1$ models coupled to non compact gauge
fields \cite{mv}. Second, even if the non-compact transition is
continuous, to constitute a stable deconfined QCP, it must be
stable to the (quadrupled) instanton events allowed by the
microscopic compact model.

At present we do not have supporting evidence in favor of either
of these two conditions.  It would be of some interest to develop
a semiclassical description of the above scenario to better
evaluate it in physically intuitive terms.  We note that XY
anisotropy, for very simple reasons, favors a ``deconfined''
critical scenario.  In particular, weak XY anisotropy converts the
topological defects (solitons) of the antiferromagnet from
skyrmions to merons, ``fractionalizing'' them already in the
antiferromagnetic phase.  Adding Ising anisotropy instead renders
the topological defects local ``droplets'', or domains of
antiphase ordering.  These can apparently be viewed as distorted
skyrmions, in which the smooth rotation from the anti-aligned core
to infinity is replaced by a domain wall of finite width.  Thus
there is no fractionalization of the topological defects in the
Ising antiferromagnet, although they do appear to carry the
integer skyrmion quantum number.

Nevertheless, the action in Eqs.~(\ref{sz}) and (\ref{eq:epterm})
appears to describe a putative deconfined Ising AF to VBS
transition. There is clearly no self-dual description of this QCP,
since neglecting instantons, the VBS phase is replaced by a
Coulomb phase with a gapless photon, while the Ising AF has no
gapless excitations. Formally, however, one may wonder what
physics might be represented by considering the mathematically
similar ``anisotropy'' in the dual meron theory, {\em i.e.\/}
taking $w_d >0$ in Eq.~(\ref{crtny}). Provided this transition
remains continuous in $2+1$ dimensions and the $\lambda$ term
remains irrelevant in this case, this would describe a {\em
different} quantum phase transition.  In particular, for $s_d>0$,
the ground state has no vortices and there is a Meissner response
(Maxwell term for $A_\mu$), hence it describes an XY superfluid.
For $s_d<0$, with $w>0$, one or the other (not both) types of
merons condense and the $A_\mu$ gauge field develops a Higgs mass.
Hence this describes a non-superfluid state.  From
Table~\ref{tab:dual}, one sees that the non-zero expectation value
of $|\psi_1|^2-|\psi_2|^2$ implies Ising AF order.  Thus this
critical point describes a putative direct continuous transition
between XY and Ising antiferromagnets.

Thus these two theories describe different potential routes of
``disordering'' a quantum Ising antiferromagnet with U(1)
spin-rotation symmetry, either to a VBS phase or an XY
antiferromagnetic phase.  If they are indeed continuous, with the
instanton fugacity and $\lambda$ term respectively irrelevant,
they are not self-dual but instead dual to one another.  It would
be interesting to determine with more certainty whether these
putative critical theories can survive fluctuation effects. We
note that the numerical simulations of Ref.~\onlinecite{sandvik}\
observed a first order transition between VBS and Ising
antiferromagnetic phases.  While this does not rule out the
possibility of a continuous transition in other microscopic
models, it is perhaps some evidence to the contrary.

\section{Experiments}
\label{sec:exp}

We now briefly discuss the implications of the phenomena discussed
in this paper for experiments on quantum magnets. What are some
good signatures of these phenomena in experiments? Imagine a quasi
two dimensional Mott insulator where each layer has a square
lattice of localized spin-$1/2$ moments. Ignoring all effects due
to coupling between the layers and to other degrees of freedom
(phonons,etc), a direct second order zero temperature transition
between the N\'eel ordered and translation broken VBS phases
should be possible (for instance by application of pressure) with
properties described as in previous sections. It is first
important to emphasize that the proposal of a deconfined critical
point is on firmest ground  for a system with a spin-$1/2$ moment
per unit cell. With higher spin or with more than one spin-$1/2$
moment per unit cell other (more ordinary) kinds of phase
transitions may well obtain. In practice (even with spin-$1/2$ per
unit cell)  the growing VBS fluctuations associated with the
lattice symmetry breaking will couple strongly to lattice
disortions particularly at low temperature . If the phonons can be
regarded as three dimensional (even though the magnetic
interactions may be well approximated as two dimensional), a small
region of coexistence will most likely be introduced at very low
temperature. This may be roughly understood as follows. The
elastic energy cost of a latttice disortion of magnitude $x$ that
couples to the VBS order parameter is of order $x^2$. However the
electronic energy gain is much bigger (as the susceptibility
associated with the VBS order parameter diverges at the
transition), going as $x^\kappa$ with $\kappa <2$. In the easy
plane case $\kappa \approx 1.35$ from the numerical
results\cite{mv}. Thus the phonons will then prempt a direct
N\'eel-VBS transition and introduce a small coexistence region. It
will thus be necessary to look at temperatures that are not too
low to meaningfully compare with experiments.

Barring these caveats the interesting critical phenomena discussed
in this paper should be visible in a number of different
experimental probes. Scaling forms for various experimentally
physical observables are readily written down. For instance, right
at the critical point the spin response function $\chi(k_i,
\omega)$ near the ordering wavevector ($Q_i = (\pi, \pi)$) will
take the form
\begin{equation}
\label{eq:spscform} \chi(k_i, \omega) \sim
\frac{1}{k^{2-\eta}}F\left(\frac{\omega}{ck},
\frac{\hbar\omega}{k_B T}\right)
\end{equation}
Here $k_i$ is assumed to measure the deviation of the physical
wavevector from $Q_i$, and $T$ is the temperature. The
corresponding spectral function can be directly measured in
neutron scattering experiments. At a fixed small wavevector $k_i$
and temperature $T$, this will show sharp spin wave peaks as a
function of frequency in the N\'eel state and similar sharp
`triplon' peaks near the spin gap in the VBS state. However right
at the critical point, there will be an anamolously broad power
law peak (due to the large $\eta$).

The large value of $\eta$ will also directly manifest itself in
NMR experiments. Indeed the nuclear spin lattice relaxation rate
is essentially given by
\begin{equation}
\label{eq:nmrt1}
\frac{1}{T_1} \sim T \int d^2k \lim_{\omega \rightarrow 0}\frac{\chi''(\vec k,
\omega)}{\omega}
\end{equation}
where $\chi''$ is the imaginary part of the spin response
function. It is now easy to see from scaling that $1/T_1 \sim
T^{\eta}$ at finite temperatures in the ``quantum critical''
region. Thus this experiment provides a direct measurement of
$\eta$. It is therefore an excellent way to experimentally
distinguish the predictions of the present paper from those of the
earlier accepted theory\cite{csy} of the N\'eel-VBS transition
which gives $\eta \approx 0$.

In this context it is interesting to reconsider experiments
measuring the spin- lattice relaxation rate of the $Cu$ ions in
the undoped and lightly doped cuprates\cite{Imai}. Remarkably at
high temperature the $1/T_1$ saturates to a temperature {\em and}
doping independent value. One suggested explanation\cite{csy} is
that at these high temperatures in the undoped sample the system
is in a quantum critical regime associated with a disordering
transition of the N\'eel order such that $\eta \approx 0$.
Furthermore the effects of doping has been suggested to only make
the system appear closer to the critical point (at least for the
high temperature spin physics). The results of the present paper
imply that if this interpretation of the experiments in terms of
proximity to quantum criticality is correct, then the
corresponding transition cannot be from the N\'eel to the VBS
state.

\section{Discussion}

This paper has described a variety of quantum critical points in
two dimensions which can be understood using the new paradigm of
`deconfined' quantum criticality \cite{shortpap}. The critical
point has an emergent topological conservation law, and the
critical theory is expressed most naturally in terms of
fractionalized degrees of freedom. The order parameters
characterizing the phases flanking the critical point emerge at
large length scales as composites of the fractionalized modes, or
their duals. These examples clearly violate the LGW paradigm, in
that the order parameters are not directly related to the critical
modes.

Our primary example was the N\'eel to VBS transition for the
$S=1/2$ square lattice antiferromagnet. We showed that a
deconfined critical point scenario emerged in a number of
tractable deformations of models appropriate to describe this
transition. Several results existing in the literaure (for
instance on large-$N$ models) were shown to support this proposal
when correctly interpreted. We briefly reiterate a few key
physical aspects of this theory. First, the critical theory
possesses an extra global topoogical conservation law (associated
with skyrmion number). It is most naturally expressed not in terms
of the natural order parameters of either phase but in terms of
new spin-$1/2$ spinon degrees of freedom that are specific to the
critical point. The emergence of these fractional spin fields at
the critical fixed point manifests itself quantitatively in the
large value of the anamoulous exponent $\eta$ at the transition.
The extra topological conservation law is obtained because
monopole events at which skyrmion number can change are irrelevant
and disappear at long scales at the critical fixed point. However
they are relevant in the paramagnetic phase and lead to the
appearnce of broken lattice symmetry. There are two diverging
length or time scales near the critical point. In the paramagnetic
phase the first (shorter) length is the spin correlation length.
There is a longer length scale at which the VBS order gets pinned.

We also considered a number of other critical points in this
paper. The transition between VBS and spin liquid states was
discussed in Section~\ref{sec:vbssl}, and described by critical
theory closely related to that for the SJ models. The
superfluid-insulator transition of bosons at half-filling on the
square lattice was considered in
Section~\ref{sec:superfl-insul-trans}: the insulator exhibits
bond-density-wave order and the theory for the critical point
(Eq.~(\ref{crtny})) had been obtained earlier\cite{Crtny}. Here we
provided a physical reinterpretation of this theory, and showed
how it could also be understood as a deconfined QCP.

Overall our work shows that such deconfined quantum criticality is
quite common in two dimensional systems with a spin-$1/2$ moment
per unit cell. This leads us to suspect that the scope for finding
deconfined QCP's in other correlated electron systems, including
those with fermionic excitations, is bright. These QCP's naturally
have large anomalous dimensions for observable order parameters
\cite{rbl}, and so hold the prospect of explaining a variety of
experimental puzzles.

\begin{acknowledgments}

This research was supported by the National Science Foundation
under grants DMR-0308945 (T.S.), DMR-9985255 (L.B.),
DMR-0098226 (S.S.)  and DMR-0210790, PHY-9907949 (M.P.A.F.). We
would also like to acknowledges funding from the NEC Corporation
(T.S.), the Packard Foundation (L.B.), the Alfred P.  Sloan
Foundation (T.S., L.B.), a Pappalardo Fellowship (A.V.) and an
award from The Research Corporation (T.S.).  We thank the Aspen
Center for Physics for hospitality.

\end{acknowledgments}

\appendix

\section{Berry phases in the SJ model}
\label{app:SJB}

The non-linear sigma model representation in Eq.~(\ref{eq:nlsm2})
associates the Berry phases with a summation over the individual
Berry phases of each spin. Each such contribution measures the
area enclosed by the path of the spin on the unit sphere, and this
is represented by $\mathcal{S}_B$ in Eq.~(\ref{eq:nlsm2}). Upon
transforming to the $z$ variables via Eq.~(\ref{eq:cp1}), there is
a simple way of measuring this area \cite{SJ,sp,srev}: it is the
Polyakov loop of the U(1) gauge field of the CP$^1$ model. This
connection suggests that the $\mathcal{S}_B$ in Eq.~(\ref{eq:SJ3})
should be replaced by
\begin{equation}
\mathcal{S}^{\prime}_B = i \sum_i \epsilon_i a_\tau,
\label{eq:sbp}
\end{equation}
where $\epsilon_i$ is the cubic lattice, $\tau$-independent
representation of the square lattice sublattice staggering factor
$\epsilon_r$. One can now consider a `modified' SJ model \cite{sp}
with action $\mathcal{S}^{\prime}_{\rm SJ} = \mathcal{S}_z +
\mathcal{S}_a + \mathcal{S}^{\prime}_B$ defined by
Eqs.~(\ref{eq:SJ1}), (\ref{eq:SJ2}) and (\ref{eq:sbp}). This
appendix will argue that the properties of $\mathcal{S}_{\rm SJ}$
and $\mathcal{S}^{\prime}_{\rm SJ}$ are very similar, and
universal features are identical.

\begin{figure}[t]
\centerline{\includegraphics[width=1.7in]{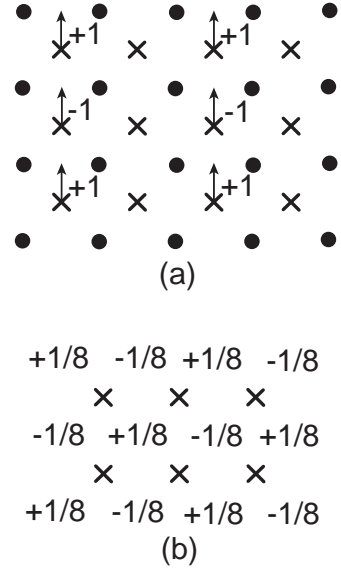}}
\caption{Specification of the non-zero values of the fixed fields
(a) ${\bf B}_0$ and (b) ${\boldsymbol\beta}$. The circles are the
sites of the direct lattice, $j$, while the crosses are the sites
of the dual lattice, $\bar{\jmath}$; the latter are also offset by
half a lattice spacing in the direction out of the paper (the $\mu
= \tau$ direction). The $B_{0\mu}$ are all zero for $\mu=\tau,x$,
while the only non-zero values of $B_{0y}$ are shown in (a). Only
the $\mu=\tau$ components of $\beta_{\mu}$ are non-zero, and these
are shown in (b). }\label{ssfig}
\end{figure}

First, we show that for $t=0$ (this is deep in the VBS phase), the
two theories are, in fact, exactly equivalent. In this limit, we
can proceed with a duality mapping as in Section~\ref{sec:SJ1},
and obtain a dual representation of $\mathcal{S}_{\rm SJ}$ which
is Eq.~(\ref{eq:discretedual}) with ${\bf A} = 0$:
\begin{equation}
\mathcal{S} = \frac{1}{2K} \left({\boldsymbol\Delta}(\chi +
  \vartheta) \right)^2 \label{eq:ss1}
\end{equation}
Proceeding with the analogous duality transformation to
$\mathcal{S}^{\prime}_{\rm SJ}$, we find instead
\begin{equation}
\mathcal{S}^{\prime} = \frac{1}{2K} \left({\boldsymbol\Delta}\chi
+ {\bf B}_0\right)^2 \label{eq:ss2}
\end{equation}
where ${\bf B}_0$ is a fixed integer-valued field on the links of
the dual lattice chosen so that
\begin{equation}
 {\boldsymbol\Delta} \times {\bf B}_0
= \epsilon \hat{\boldsymbol\tau} \label{eq:ss3}
\end{equation}
where $\hat{\boldsymbol\tau}$ is a unit vector in the $\tau$
direction. A convenient choice for ${\bf B}_0$ is shown in
Fig~\ref{ssfig}a.
Now note that we can write
\begin{equation}
{\bf B}_0 = {\boldsymbol\Delta} \vartheta + {\boldsymbol\Delta}
\times {\boldsymbol\beta} \label{eq:ss4}
\end{equation}
with $\vartheta_r$ defined below Eq.~(\ref{eq:SJ10}) and shown in
Fig~\ref{offset}, ${\boldsymbol\beta}$ is a fixed vector field on
the links of the dual lattice with only its temporal components
non-zero as shown in Fig~\ref{ssfig}b. It is now evident that
Eqs.~(\ref{eq:ss2}) and (\ref{eq:ss4}) are exactly equivalent to
Eq.~(\ref{eq:ss1}), as the couplings between ${\boldsymbol\beta}$
and $\chi$, $\vartheta$ vanish.

Moving to the general case with $t \neq 0$, let us examine the
fate of the modified SJ model for $N=1$ as in
Section~\ref{sec:SJ1}. In this case, Eq.~(\ref{eq:discretedual})
is replaced by
\begin{equation}
\mathcal{S}^{\prime} =
\frac{1}{2\tilde{t}}\left({\boldsymbol\Delta} \times {\bf A}
\right)^2 + \frac{1}{2K}\left({\boldsymbol\Delta}(\chi +
  \vartheta) + {\bf A} + {\boldsymbol\Delta}
\times {\boldsymbol\beta} \right)^2. \label{eq:ss5}
\end{equation}
It is now not difficult to show that the additional term
associated with ${\boldsymbol\beta}$ above makes little difference
to the universal properties of theory: integrating out the massive
${\bf A}$ modes is now a little more involved, but the final
theory for $\chi$ has the same structure as that in
Section~\ref{sec:SJ1}. Further details may be found in
Ref.~\onlinecite{srev}.

Similar comments apply to the modified SJ model at $N=2$, with
easy plane anisotropy, as discussed in Appendix~\ref{app:SJ2}.

\section{Duality transformation with easy plane anisotropy}
\label{app:SJ2}

The duality transformation for the SJ model at $N = 2$ in the easy
plane limit proceeds very similarly to that of the $N =1$ case
discussed in Section~\ref{sec:SJ1}. We begin by rewriting the
`boson hopping' term in Eq.~(\ref{eq:SJ21}) in a Villain
approximation:
\begin{equation}
\mathcal{S}_z \rightarrow \sum_{\ell,\alpha} \left[
\frac{1}{2\tilde{t}}|{\bf j}_\alpha |^2 - i{\bf j}_{\alpha} \cdot
\left({\boldsymbol{\Delta}} \phi_\alpha - {\bf a} \right) \right],
\end{equation}
where $\alpha = 1,2$ labels the two species of bosons and ${\bf
j}_{\alpha}$ are the corresponding integer valued currents defined
on the links of the square lattice. Proceeding as before in
Section~\ref{sec:SJ1}, integrating out the $\phi_{\alpha}$ fields
leads to the current conservation conditions
\begin{equation}
{\boldsymbol\Delta}\cdot{\bf j}_{\alpha} = 0.
\end{equation}
which can be solved by writing
\begin{equation}
{\bf j}_\alpha = {\boldsymbol\Delta} \times{\bf  A}_\alpha
\end{equation}
with ${\bf A}_\alpha$ integer fields. We treat the gauge field
kinetic energy term as in Section~\ref{sec:SJ1} by first
decoupling it with the ${\bf b}$ field, and then summing over the
integer ${\bf q}$ to obtain
\begin{equation}
{\bf b} - {\boldsymbol \Delta} \vartheta = {\bf B}
\end{equation}
with ${\bf B}$ an integer field. Integrating over the gauge field
${\bf a}$ now replaces Eq.~(\ref{eq:SJ11}) by
\begin{equation}
{\boldsymbol\Delta} \times{\bf B} = {\bf j}_1 + {\bf j}_2
\end{equation}
This may be solved by writing
\begin{equation}
{\bf B} = {\bf A}_1 + {\bf A}_2 + {\boldsymbol\Delta} \chi
\end{equation}
with $\chi$ an integer. The action then reads
\begin{eqnarray}
\mathcal{S} &=& \sum \left[
\frac{1}{2\tilde{t}}\left({\boldsymbol\Delta} \times {\bf A}_1
\right)^2 + \frac{1}{2\tilde{t}}\left({\boldsymbol\Delta} \times
{\bf A}_2 \right)^2  \right. \nonumber \\
&~&~~~~~~\left. + \frac{1}{2K}\left({\boldsymbol\Delta}(\chi +
  \vartheta) + {\bf A}_1 + {\bf A}_2 \right)^2 \right].
\label{eq:discretedual2}
\end{eqnarray}

As in Eq.~(\ref{eq:soften1}) we may soften the integer constraints
on ${\bf A}_\alpha , \chi$ by adding the terms
\begin{equation}
-t \cos(2\pi {\bf A}_1)-t \cos(2\pi {\bf A}_2) - \sum_n \lambda_n
\cos(2\pi n \chi)
\end{equation}
with $n$ running over all positive integers. Now we can shift
$\chi \rightarrow \tilde \chi = \chi + \vartheta$. Then we can put
$2\pi \chi = \theta_1 - \theta_2$, and integrate over both phase
fields, $\theta_\alpha$, leaving the partition function unchanged
up to an overall multiplicative constant. Upon shifting the two
fields ${\bf A}_1 \rightarrow {\bf A}_1 -
{\boldsymbol\Delta}\theta_1/2\pi$ and ${\bf A}_2 \rightarrow {\bf
A}_2 + {\boldsymbol\Delta}\theta_2/2\pi$, the last term in
Eq.~(\ref{eq:discretedual2}) takes the form $({\bf A}_1 + {\bf
A}_2)^2$. We can then define ${\bf A}_+ = {\bf A}_1 + {\bf A}_2$
and ${\bf A} = \pi ({\bf A}_1 - {\bf A}_2)$, and integrate out the
massive field ${\bf A}_+$.  Up to irrelevant terms we thereby
obtain for the full action:
\begin{eqnarray}
\mathcal{S} & = & \mathcal{S}_\lambda + \sum \Biggl[
\frac{4\pi^2}{\tilde{t}} \left({\boldsymbol\Delta} \times {\bf A}
\right)^2 - t \cos({\boldsymbol\Delta} \theta_1 - {\bf A})
\nonumber \\ &~&~~~~~~~~~~~~~~~~ -t \cos( {\boldsymbol\Delta}
\theta_2 - {\bf A}) \Biggr]
\nonumber \\
\mathcal{S}_{\lambda} & = &  - \sum \left[\sum_n \lambda_n
\cos(n(\theta_1 - \theta_2 - 2\pi\vartheta)) \right]
\end{eqnarray}
Once again as $e^{2i\pi \vartheta}$ oscillates on four
sublattices, for smooth variations of $\theta_{1,2}$, the lowest
value of $n$ that survives is $n = 4$. We may therefore replace
$\mathcal{S}_{\lambda}$ by
\begin{equation}
\mathcal{S}_{\lambda} = -\sum \left[ \lambda \cos(4(\theta_1 -
\theta_2)) \right]
\end{equation}
 with $\lambda \equiv \lambda_4$. The resulting action is then a
`hard-spin' version of the action $\mathcal{S}_z$ in
Eq.~(\ref{crtny}) of Section~\ref{sec:ep}, with the identification
of the vortex operators, $\psi_\alpha \sim e^{i \theta_\alpha}$ .

\section{Estimate of monopole scaling dimension}
\label{app:est}

We can ask about the answer for all these scaling dimensions that
would be obtained by first integrating out the matter fields,
truncating the resulting gauge action to quadratic order, and
using that theory to calculate the scaling dimension. This would
be an estimate, albeit uncontrolled; one might hope that it will
correctly capture the trends.

Anyway, it turns out that this can be done without serious
calculation. First, note that in this procedure the answer depends
only on the universal conductivity of the matter fields at the
transition ignoring all coupling to the gauge field. The higher
this universal conductivity the higher the instanton anomalous
dimension.

The simplest case is the $N = 1$ SJ model. Here the relevant
universal conductivity is that of a single boson species. We know
that at this transition the four instanton operator is irrelevant.
Now in the $N = 2$ cases with either easy plane or full SU(2)
symmetry, it is clear that the universal conductivity will only be
larger than at $N = 1$. Thus we should expect a higher anamolous
dimension. This would then predict irrelevance of the four
instanton term with or without easy plane anisotropy for the
physical case of $N = 2$ in agreement with other expectations.

\section{SJ models in one dimension}

There is a close and useful analogy between some of the phenomena
explored in this paper and corresponding ones in one spatial
dimension. Specifically consider a one dimensional spin-$1/2$
magnet in the presence of some easy plane anisotropy. The analog
of the N\'eel phase in $d = 1$ is a phase with power law
correlators for the staggered XY magnetization. This phase may be
described as a Luttinger liquid. There is a direct second order
transition between this phase and a VBS phase where there is
spontaneous dimerization of the spin chain.

A useful theoretical description of this transition is obtained by
focusing on vortices in the space-time configuration of the
staggered XY order parameter field. From a quantum mechanical
point of view such vortices correspond to phase-slip (or
instanton) events. It is well-known that a $2\pi$ phase slip event
carries a momentum $\pi$, and hence is not allowed as a term in
the Hamiltonian. (In an equivalent description of this Luttinger
liquid phase in terms of interacting spinless fermions, these
$2\pi$ phase slips correspond to interchange of left and right
movers). In the VBS phase these phase slip events have
proliferated. Indeed it is precisely the $\pi$ momentum that is
carried by the $2\pi$ phase slip that is responsible for the
broken translation symmetry of the VBS. A convenient order
parameter for the VBS phase is therefore provided by the $2\pi$
phase slip operator.

Though $2\pi$ phase slip terms are not allowed in the Hamiltonian
$4\pi$ phase slips (which carry zero crystal momentum) are clearly
allowed. The transition from the Luttinger liquid to the VBS is
driven by the proliferation of these $4\pi$ phase slips. These
`doubled' phase slips are irrelevant throughout the Luttinger
liquid phase and are marginally irrelevant at the critical fixed
point.

The analogy with the two dimensional situations considered in this
paper is now clear. The $2\pi$ phase slip is the direct analog of
the skyrmion tunneling ({\em i.e} single instanton) event. In both
$d =1,2$ the VBS phase is understood as a condensate of the
appropriate single instanton event. In $d = 1$ the transition is
driven by doubled instanton events (similar to the quadrupling of
instantons in two dimensions) which stay irrelevant at the
critical fixed point. At a formal level it is possible to
construct an appropriate `SJ' model that correctly describes the
transition even in $d = 1$.

A complete presentation closely related to the discussion in this
appendix appears in Ref.~\onlinecite{sp}.

\section{Direct derivation of dual meron action}
\label{sec:direct-deriv-dual}

We pass from Eq.~(\ref{o2rot}) to the analog of a CP$^1$
representation by letting
\begin{eqnarray}
e^{i\phi_r} & = & b^{\dagger}_{1r}b^{\vphantom\dagger}_{2r} =
e^{-i\left(\phi_{1r} - \phi_{2r}\right)}  \\
n_r & = & \frac{n_{1r} - n_{2r} + 1}{2}  \\
n_{1r} + n_{2r} & = & \epsilon_r .
\end{eqnarray}
Here $b_{1,2}= e^{i\phi_{1,2}}$ represent charge-$\pm 1/2$ bosonic
operators and $n_{1,2}$ are the corresponding boson numbers, and
$\epsilon_r$ was defined in Eq.~(\ref{defeps}). Note that
$b_{1,2}$ are not canonical Bose operators, and the relevant
commutation relations here are $[n_1, \phi_1]=-i$ and $[n_2,
\phi_2]=-i$. The eigenvalues of $n_{1,2}$ are integers which run
from $-\infty$ to $\infty$. As is usual, there is a gauge
redundancy associated with an arbitrary choice of the local phase
of the $b_{1,2}$ fields. The last equation is a constraint that
requires the total number of both species of bosons to be fixed at
$+1$ on the A sublattice and $-1$ on the B sublattice. The
left-hand side of this constraint equation is precisely the
generator of the local gauge transformation. We have chosen to
stagger this gauge charge on the two sublattices.

The Hamiltonian in Eq.~(\ref{o2rot}) is readily rewritten in terms
of these new variables:
\begin{eqnarray}
H & = & H_U + H_t \\
H_U & = & \frac{U}{2}\sum_r \left[\left(n_{1r} -
\epsilon_r n_0 \right)^2 + \left(n_{2r}- \epsilon_r n_0 \right)^2 \right] \\
H_t & = & -t\sum_{\langle rr'
\rangle}\left[\left(b^{\dagger}_{1r}b^{\vphantom\dagger}_{2r}\right)
\left(b^{\dagger}_{2r'}b^{\vphantom\dagger}_{1r'}\right)+ {\rm
H.c.} \right]
\end{eqnarray}
We have introduced a term proportional to $n_0$ which describes a
`chemical potential' for the total on-site gauge charge. As the
total gauge charge is fixed to $\pm 1$ on each site, this addition
is completely innocuous (for any value of $n_0$). Later we will
choose $n_0$ appropriately to ensure that the $n_{1,2}$ fields
have zero mean value.  While this step is not necessary, it is
convenient, and will be commented upon further at the appropriate
point.

Now we proceed to a path integral representation to write
\begin{eqnarray}
\mathcal{S} & = & \mathcal{S}_U + \mathcal{S}_{\tau}
+ \mathcal{S}_{a_0} + \mathcal{S}_t \nonumber \\
\mathcal{S}_U & = & \sum_r \int \! d\tau\, \frac{U}{2}\sum_r
\left[\left(n_{1r} - \epsilon_r n_0 \right)^2 + \left(n_{2r}-
\epsilon_r n_o \right)^2 \right] \nonumber \\
\mathcal{S}_{\tau} & = & \sum_r \int d\tau  \left[
in_{1r}\partial_\tau\phi_{1r} + in_{2r}
\partial_\tau\phi_{2r} \right]\nonumber  \\
\mathcal{S}_{a_0} & = & \sum_r \int d\tau
\left[ ia_0 \left(n_{1r} + n_{2r} - \epsilon_r \right) \right] \nonumber \\
\mathcal{S}_t & = & \int d\tau H_{t}
\end{eqnarray}
As usual, it is assumed that there is a fine discretization of the
imaginary time index $\tau$, and that the integer-valued boson
numbers $n_{1r}, n_{2r}$ live on the temporal links at each
spatial point. The `gauge' constraint is imposed by means of a
Lagrange multiplier field $a_0$ which will be interpreted as the
time-component of a gauge field. We now proceed as is usual in
slave particle theories of correlated systems. We decouple the
interactions in $H_t$ using a complex auxiliary field $\chi_{rr'}$
defined on each spatial link to write
\begin{eqnarray}
e^{-\mathcal{S}_t} &=&\prod_{\langle rr'\rangle} \int d\chi_{rr'}
\exp \Biggl[-
\frac{|\chi_{rr'}|^2}{t} \nonumber \\
& & +  \chi_{rr'}\sum_{\alpha=1}^2e^{i(\phi_{\alpha r} -
\phi_{\alpha r'})} + {\rm c.c.} \Biggr],
\end{eqnarray}
with $\chi_{rr'} = \chi^*_{r'r}$. The fluctuations in the
amplitude of the $\chi$ field are expected to be innocuous. Hence
we will write
\begin{equation}
\chi_{rr'} \approx \chi_0 e^{ia_{rr'}} ,
\end{equation}
with $\chi_0$ a constant that simply renormalizes the boson
hopping amplitude. As usual the $a_{rr'} = -a_{r'r}$ will be
interpreted as the spatial component of a gauge field.  The full
action now is invariant under the gauge transformation
\begin{eqnarray}
e^{i\phi_{\alpha r} (\tau)} & \rightarrow & e^{i\gamma_{r} (\tau)}
e^{i\phi_{\alpha r} (\tau) } \\
a_0 & \rightarrow & a_0 - \frac{d\gamma_{r} (\tau) }{d\tau} \\
a_{rr'} & \rightarrow & a_{rr'} - \gamma_{r} (\tau)  + \gamma_{r'}
(\tau),
\end{eqnarray}
where $\alpha=1,2$. To examine universal critical properties near
the phase transitions of interest it is legitimate to add various
possible local terms that are consistent with the global
symmetries and gauge structure of the action. It is particularly
useful to add a `kinetic energy' term for the gauge fields on all
plaquettes (spatial and space-time) in Villain form:
\begin{eqnarray}
\mathcal{L}_E & = &  \frac{u}{2}E_i^2 + iE_i \left(\partial_{\tau}a_i - \Delta_i
a_0 \right) \\
\mathcal{L}_B & = & \frac{u}{2} B^2 + iB\left(\epsilon_{ij}
\Delta_i a_j \right) .
\end{eqnarray}
Here $E_i$ is an integer-valued `electric field' defined on the
spatial links at each time-slice ($i,j$ extend over the spatial
co-ordinates $x,y$) and $B$ is the corresponding integer-valued
magnetic field on a spatial plaquette. We have introduced the
vector notation $ a_i = (a_x, a_y) = (a_{\vec r, \vec r+\hat x},
a_{\vec r, \vec r+ \hat y})$. For simplicity we have chosen the
same constant $u$ multiplying the $E_i^2$ and $B^2$ terms. The
original microscopic action is formally obtained in the large-$u$
limit ($u \rightarrow \infty$).

To proceed it is first useful to note that the background gauge
charge present in this formulation will lead to a background
electric field about which the true electric field will actually
fluctuate. It will be convenient to incorporate this effect by
finding a suitable mean field for the various fields in the
action. Consider a mean field (saddle point) of the action where
the non-zero expectation values are
\begin{eqnarray}
\langle n_1 \rangle =  \langle n_2 \rangle & = & \overline{n} \nonumber \\
\langle a_0 \rangle & = & \overline{a}_0 \nonumber \\
\langle E_i \rangle & = & E_{i0}.
\end{eqnarray}
The saddle point equations are obtained by varying the action with
respect to these fields:
\begin{eqnarray}
u E_{i0} & = & i\Delta_i \overline{a}_0  \nonumber \\
\Delta_i E_{i0} & = & -2\overline{n} + \epsilon_r \nonumber \\
U(\overline{n} - \epsilon_r n_0) &=& - i\overline{a}_0
\end{eqnarray}
We now use our freedom in choosing the constant $n_0$ to set
\begin{equation}
n_0 = \frac{i\overline{a}_0 \epsilon_r}{U}
\end{equation}
so that $\overline{n} = 0$. One may worry that this special choice of
$n_0$ might indicate some non-generic nature of the resulting theory.
We note, however, that qualitatively identical results are obtained
for any other choice of $n_0$ -- in the dual action with such a choice
the merons see some non-zero but spatially oscillating flux.  Having a
zero spatial average, this flux does not qualitatively effect the low
energy (extended) meron states.  The above choice simply renders the
low energy behavior more transparent.  We then have
\begin{equation}
\Delta_i E_{i0}  =   \epsilon_r
\end{equation}
Note also that $E_{i0}$ is the gradient of a potential determined
by $a_0$. These conditions determine $E_{i0}$ (the background
electric field) uniquely to have the value $1/4$ oriented from the
$A$ to $B$ sublattice.

We may now examine the full theory by first shifting $a_0 =
\overline{a}_0 + \delta a_0$. Straightforward manipulation shows
that the electric field now fluctuates about a background value
$E_{i0}$ so that the $E_i$ dependent terms in the action read
\begin{equation}
\mathcal{L}_E = \frac{u}{2}\left( E_i -  E_{i0} \right)^2 + i E_i
\left(\partial_{\tau} a_i - \Delta_i a_0 \right)
\end{equation}
All other terms remain unchanged (after the replacing $a_0$ with
$\delta a_0$). We may now dualize this action as in previous
sections to directly derive the dual meron action of
Section~\ref{sec:ep}.

\section{Breakdown of the `screening argument' in the monopole gas}
\label{app:monoscr}

In this Appendix we will consider a simple toy model of a compact
U(1) gauge theory without Berry phases which can be shown to
possess a deconfined critical point. This will enable us to
understand clearly the claim of Section~\ref{sec:SJN} that the
specific monopole gases that obtain at the critical points studied in this paper
evade the general monopole screening arguments
\cite{igor,igorss} for a three dimensional Coulomb gas with logarithmic
interactions.

We consider a model of charge $n$ bosons ($n \geq 4$) coupled to a
compact U(1) gauge field in $D = 2+1$ dimensions with Euclidean
action
\begin{equation}
\mathcal{S} = -J \sum\cos({\boldsymbol\Delta} \phi - n {\bf a}) -
K \sum_P \cos( {\boldsymbol\Delta} \times {\bf a} ).
\label{eq:znmodel}
\end{equation}
Here $e^{i\phi_i}$ represents a boson field on the sites $i$ of a
three dimensional cubic lattice. The sum in the first term is over
the links of the lattice, while that in the second term is over
the plaquettes $P$. The field $a_{ij} \equiv a_{ij} + 2\pi$ is a
compact U(1) gauge field, and the integer $n$ is the charge of the
boson. This model has two phases. For large $J$, there is a Higgs
phase where the boson field has `condensed'. Following the
arguments of Ref.~\onlinecite{FrSh}, the effective theory of this
phase is a $Z_n$ gauge theory in its deconfined phase in $2+1$
dimensions. The excitations in this phase are stable vortices that
carry flux $2\pi q/n$, for $q = 1,......n-1$, of the gauge field
$a$.  For small $J$, on the other hand, there is a different phase
which is associated with confinement of the U(1) gauge theory. In
particular, the $Z_n$ vortices that appear in the Higgs phase are
no longer present in the spectrum. As also argued in
Ref.~\onlinecite{FrSh}, the transition between these two phases is
actually described by that in a $Z_n$ gauge theory. This latter
theory is dual to the global $Z_n$ clock model - for $n \geq  4$
the clock anisotropy is irrelevant and the universality class is
$3D$ XY.

The $n=1$ case of Eq.~(\ref{eq:znmodel}) was considered in
Ref.~\onlinecite{sudbo}. For this case, the `clock' anisotropy is
strongly relevant (it rounds out the transition into a crossover),
and the physics is very different from $n \geq 4$.

Formally, the action above is readily dualized (as in many of the
other examples discussed at length in previous sections). The dual
action takes the form
\begin{equation}
\mathcal{S}_{\rm dual} = -t\sum \left[\cos(2\pi
{\boldsymbol\Delta} \chi) - \lambda \cos(2 \pi n\chi) \right]
\label{eq:sdualchi}
\end{equation}
and has the expected structure of a global XY model with $n$-fold
anisotropy.

It is useful, for our purposes, to have a physical interpretation
of these results. In the Higgs phase, the vorticity of the $\phi$
field is quantized in units of $2\pi/n$ (as is natural for a
charge $n$ condensate). The presence of instantons implies that
the total flux can change by $2\pi$, so that $n$ of these vortices
can appear or disappear together. Thus the vortices only carry a
$Z_n$ quantum number. The dual description focuses on these $Z_n$
vortices. Without instantons, the $2\pi/n$ vortex is globally
conserved and its physics is described by a global XY model (this
is just the duality in Ref.~\onlinecite{dasgupta}). The presence
of instantons leads to the $n$-fold anisotropy (the $\lambda$ term
in Eq.~(\ref{eq:sdualchi})) for this global XY model, leading to
the global $Z_n$ model. Thus the irrelevance of the $n$-fold
anisotropy, for $n \geq 4$, should be interpreted as the
irrelevance of instantons at the transition. Indeed the XY
universality class, is the exact dual of the condensation
transition of the charge $n$ boson coupled to a {\em non-compact}
U(1) gauge field.\cite{dasgupta}

Now let us analyze the transition in the `RPA' approach outlined
in Section \ref{sec:SJN}. The transition is associated with the
condensation of the $e^{i\phi}$ field. We therefore integrate out
this field in the presence of a non-trivial gauge potential, and
truncate the resulting gauge action to quadratic order (initially
ignoring instantons). The result is, as in Eq.~(\ref{eq:RPA}),
\begin{equation}
\mathcal{S}_{G} = \int \frac{d^3 K}{(2\pi)^3}  \sigma_0 n^2 |K|
|{\bf a}_T({\bf K})|^2 + ...
\end{equation}
Here $\sigma_0$ is a universal constant, and ${\bf a}_T$ is the
transverse component of the gauge field. The Maxwell term present
in the bare action is less important at long distances than the
term in the action displayed above, and we have dropped it. We now
examine the stability of $\mathcal{S}_{G}$ to instantons. First,
we dualize $\mathcal{S}_{G}$ to obtain\cite{sudbo,igor}
\begin{equation}
\mathcal{S}_{G,{\rm dual}} =  \int \frac{d^3K}{(2\pi)^3}
\frac{K^3}{\sigma_0} |\chi(K)|^2  - \sum \lambda \cos(2\pi n
\chi).
\end{equation}
The last term represents instanton events. Note the $K^3$ in the
first term. Now the $\lambda$ term will in general generate a
$K^2$ term in the Gaussian $\chi$ action which will then
eventually make instantons relevant. A fine-tuning is
required\cite{igor,igorss} to prevent the generation of the $K^2$
term in this argument, and for the present model we can now easily
see that this fine tuning is automatic at the critical point of
the gauge theory in Eq.~(\ref{eq:znmodel}).

The key is to note that the logarithmic interaction between the
monopoles is equivalent to the statement that the correlators of
$e^{2i\pi n\chi}$ decay as a power law (at the fixed point without
monopoles). In the theory of the monopoles in
Eq.~(\ref{eq:sdualchi}), the $e^{2i\pi\chi}$ field is at the
critical point of the 3D $XY$ model. The RPA theory approximates
this non-trivial interacting critical theory by an equivalent
Gaussian theory, which also happens to give power law correlations
for the $e^{2i\pi\chi}$ field (this is a property of the
$K^3|\chi|^2$ form). In the full theory in
Eq.~(\ref{eq:sdualchi}), the screening of monopole interactions is
associated with corrections higher order in $\lambda$. In the
context of conformal perturbation theory about the critical point
of Eq.~(\ref{eq:sdualchi}), however, it is clear that these higher
order effects in $\lambda$ actually represent shifts in the
position of the critical point, and not any changes in the scaling
dimensions of operators.

Hence for the N\'eel-VBS transition we conclude that the `na\"ive'
computation of monopole scaling dimensions in the large $N$ limit
\cite{MS} is actually correct, and that we should neglect
screening between multiple monopoles in determining this scaling
dimension. The latter effects are more correctly accounted for by
shifting the position of the critical point.

Note that for the $n=1$ case of Eq.~(\ref{eq:znmodel}) considered
in Ref.~\onlinecite{sudbo}, computation of the scaling dimension
of the monopole operator using Eq.~(\ref{eq:sdualchi}) shows that
monopoles are relevant at the critical point. Indeed, they round
out the transition to a crossover, and the monopoles are always in
a screened plasma phase. So the conclusions of
Refs.~\onlinecite{igor,igorss} for this case are correct, but not
for completely sound reasons.


\begin{thebibliography}{99}
\bibitem{ssfoot} It seems likely that the
non-relativistic, spinon Fermi surface state discussed in
Ref.~\protect\onlinecite{igorss} is unlikely to behave like a
critical system, and so will not suppress instantons without fine
tuning.
\bibitem{noteN} At large-$N$ even the single monopole operator is
irrelevant so that even without Berry phases a `deconfined'
critical fixed point obtains.  However this is special to
large-$N$ and does not hold at $N = 1$ (or in the easy plane limit
at $N = 2$).
\bibitem{higgs} As shown in Ref.~\protect\onlinecite{ijmp}, in the
bosonic mean-field theory, the Higgs field $Q_i$ has the same
symmetry transformation properties as the spinon pair field
$\epsilon_{\alpha\beta} z_{\alpha}
\partial_i z_{\beta}$. Note that the requirement of $SU(2)$
invariance leads to a spatial vector index on $Q_i$, and makes it
odd under certain reflections. In the easy plane case, we can form
a single charge 2 Higgs field as $z_{\alpha}^2$, and so the single
Higgs field model in Eq.~(\protect\ref{eq:SJ5}) can apply. In the
$SU(2)$ invariant models on the square lattice of
Ref.~\protect\onlinecite{ijmp}, the two-Higgs field model in
Eq.~(\protect\ref{eq:SJ51}) is necessary.
\end{thebibliography}
\end{document}